\documentclass[twocolumn,showpacs,pra,amsmath,noshowpacs]{revtex4}
\usepackage{epsfig,graphicx,bm}
\usepackage{color}

\def\be{\begin{equation}}
\def\ee{\end{equation}}
\def\e#1{\label{#1}\end{equation}}
\def\bea{\begin{eqnarray}}
\def\eea{\end{eqnarray}}
\def\ea#1{\label{#1}\end{eqnarray}}
\def\bes#1{\begin{subequations}\label{#1}}
\def\ese{\end{subequations}}

\begin{document}
\title{Quantum Bayesian approach to circuit QED measurement with moderate bandwidth}
\author{Alexander N.\ Korotkov}
\affiliation{Department of Electrical and Computer Engineering, University of
California, Riverside, California 92521}
\date{\today}

\begin{abstract}
We consider continuous quantum measurement of a superconducting qubit in the circuit QED setup with a moderate bandwidth of the measurement resonator, i.e., when the ``bad cavity'' limit is not applicable. The goal is a simple description of the quantum evolution due to measurement, i.e., the measurement back-action. Extending the quantum Bayesian approach previously developed for the ``bad cavity'' regime, we show that the evolution equations remain the same, but now they should be applied to the entangled qubit-resonator state, instead of the qubit state alone. The derivation uses only elementary quantum mechanics and basic properties of coherent states, thus being accessible to non-experts.
    \end{abstract}
  \maketitle

\section{Introduction}

The problem of instantaneous wavefunction collapse (reduction) due to measurement \cite{vonNeumann} has been a stumbling block for many physicists since the creation of quantum mechanics. The unavoidable ``spookiness'' \cite{EPR} of the quantum collapse is related to the impossibility to find a traditional physical mechanism responsible for the collapse. Mathematically, the ``spookiness'' can be expressed via violation of the Bell inequalities \cite{Bell-ineq}. Even though this violation \cite{Aspect} is common knowledge nowadays, the mechanism and interpretation of the collapse remain debatable \cite{Wheeler-Zurek-book}.

A natural approach to understanding the physics of the wavefunction reduction is through analysis of the gradual evolution at a shorter time scale, i.e., ``inside'' the collapse. A few decades ago there was an idea that such an evolution can be fully described by decoherence. However, nowadays it is becoming common knowledge that gradual collapse of individual quantum systems is governed by a continuous flow of information during the measurement, thus showing essentially the same ``spookiness'' as the textbook collapse. This understanding was significantly influenced by  experiments with superconducting qubits in the last decade \cite{Katz-2006,Katz-2008,Saclay-2010,Vijay-2012,Hatridge-2013, Murch-2013, deLange-2014, Campagne-2014,Weber-2014}, which demonstrated the actual evolution ``inside'' the collapse.

There are many approaches to the theoretical description of the evolution ``inside'' the collapse, i.e., the description of partial or continuous quantum measurement. In spite of very different mathematical treatments, many of these approaches are essentially equivalent. Probably the most well-known approach is based on positive operator-valued measure (POVM) and Kraus operators \cite{Davies-book,Kraus-book,Holevo-book}. Let us also mention quantum trajectories \cite{Wiseman-1993,Carmichael-1993,Wiseman-book,Doherty-1999,Gambetta-2008}, quantum filtering \cite{Belavkin-1992}, Monte Carlo wavefunction approach \cite{Dalibard-1992}, quantum state diffusion \cite{Gisin-1992}, restricted path integral \cite{Mensky-book,Caves-1986}, quantum Bayesian formalism \cite{Korotkov-1999,Korotkov-2001} (see also \cite{Caves-1986} and Chap.\ 2.2 of \cite{Gardiner-book}) and many other approaches, e.g., \cite{Braginsky-book,Diosi-1988,Zoller-1987,Plenio-1998}. Among these approaches, one of the simplest formalisms is the quantum Bayesian formalism, which is based only on elementary quantum mechanics and common sense.

For solid-state systems, the gradual collapse due to continuous measurement was first described using the quantum Bayesian formalism \cite{Korotkov-1999,Korotkov-2001,Korotkov-2002}, and soon after that was also described by the quantum trajectory approach \cite{Goan-2001,Goan-2001a}. From late 1990s to mid-2000s the analysis was mainly focused on the continuous measurement of a charge qubit by a quantum point contact (QPC) or a single-electron transistor (SET) \cite{Korotkov-1999,Korotkov-2001,Korotkov-2002,Goan-2001,Goan-2001a,Ruskov-2002,Korotkov-Averin, Averin-2003,Jordan-2005,Oxtoby-2005}. The next considered system was based on a partially/continuously measured superconducting phase qubit \cite{Katz-2006,Katz-2008,Korotkov-Jordan-2006,Pryadko-2007,Ruskov-2007,Zhong-2014}; the first experimental demonstration of a partial collapse \cite{Katz-2006} and uncollapsing \cite{Katz-2008} was realized with this system. After the development of circuit QED qubit measurement \cite{Blais-2004,Wallraff-2004} and the transmon \cite{Koch-2007}, much attention was paid to this system since it experimentally allowed truly continuous quantum measurement of qubits \cite{Saclay-2010,Vijay-2012,Hatridge-2013,Murch-2013,deLange-2014,Campagne-2014,Weber-2014}. In this measurement setup the qubit state affects the frequency of a coupled resonator, which in turn is probed by an applied microwave in the homodyne way. For the circuit QED measurement of a qubit the quantum trajectory approach was developed in Ref.\ \cite{Gambetta-2008} and the quantum Bayesian approach was introduced in Ref.\ \cite{Korotkov-2011}. In particular, the quantum Bayesian theory was used in several circuit QED experiments on quantum feedback and quantum trajectories \cite{Vijay-2012,Murch-2013,Weber-2014,Roch-2014}, and several experiments used the quantum trajectory theory \cite{deLange-2014,Riste-2013,Roch-2014}.

While the description of the qubit evolution in the process of circuit QED measurement is generally similar to that for measurement by QPC or SET, there is one considerable difference. The measurement by a QPC or SET is of the broad-band type, while the circuit QED measurement is narrow-band. Correspondingly, instead of one output signal $I(t)$ in the QPC/SET case, there are in general two output signals in the circuit QED case, since a narrow-band signal can be represented as $I(t)\cos (\omega t)+ Q(t) \sin (\omega t)$, where $\omega$ is the carrier frequency. The existence of two  signals (two quadratures) leads to the importance of the question of which amplifier is used in the process of measurement. In the case of a phase-sensitive amplifier, only one quadrature is amplified, and therefore only one signal [say, $I(t)$] is available. This makes the phase-sensitive case similar to the measurement by QPC or SET (however, it is still important exactly which quadrature is amplified). For the case of a phase-preserving amplifier, both output signals $I(t)$ and $Q(t)$ are available (and both are noisier than in the phase-sensitive case \cite{Caves-1982,Devyatov-1986,Clerk-2010}); this makes the description of qubit evolution significantly different from that for the QPC/SET case.

An advantage of the quantum Bayesian formalism in comparison with the quantum trajectory formalism is its simplicity, so that it does not require special theoretical training, and can be used by non-experts. This simplicity is due to a transparent physical meaning, which directly relates the quantum back-action to the information acquired during measurement. In Ref.\ \cite{Korotkov-2011} the quantum Bayesian formalism for the circuit QED measurement of a qubit was developed for the so-called ``bad cavity'' limit, in which the damping (bandwidth) $\kappa$ of the measurement resonator is much larger than the rate of qubit collapse (quantum back-action) due to measurement. In this limit the qubit is practically unentangled with the resonator and experiences two kinds of back-action due to measurement. The ``spooky'' back-action (which can also be called ``quantum'', ``informational'', or ``non-unitary'' \cite{Korotkov-2011}) moves the qubit state along the meridians of the Bloch sphere and is directly related to the continuous information on the qubit state ($|0\rangle$ or $|1\rangle$) obtained during the measurement. This back-action does not have a physical mechanism, similarly to the Einstein-Podolsky-Rosen-Bell example \cite{EPR,Bell-ineq}. The other type, the ``phase'' backaction (called ``realistic'', ``classical'' or ``unitary'' back-action in \cite{Korotkov-2011}) has a physical mechanism: fluctuation of the (ac Stark-shifted) qubit frequency due to a fluctuating number of photons in the resonator. The phase back-action moves the qubit state along the parallels of the Bloch sphere. In spite of the clear physical mechanism, the phase back-action also has some ``spookiness''; for example, it is possible to choose the qubit movement along the parallels or meridians afterwards, by choosing the amplified microwave quadrature \cite{Korotkov-2011} (this prediction has been confirmed experimentally \cite{Murch-2013}).

In the present paper we extend the quantum Bayesian formalism to the case when the ``bad cavity'' limit is not applicable. As we will see, in this case the evolution equations remain practically the same as in the ``bad cavity'' regime \cite{Korotkov-2011}; however, now they should be applied to the entangled qubit-resonator system. In the derivation we will assume that the qubit evolves only due to measurement; in particular, we assume no Rabi oscillations. The Rabi oscillations can be added later phenomenologically; however, such addition is not really correct if the Rabi frequency is comparable to or larger than the resonator damping $\kappa$. In this respect the theory discussed here has the same limitation as the ``polaron frame approximation'' usually used in the quantum trajectory approach \cite{Gambetta-2008} (see also \cite{Wang-2014,Feng-2016}). Actually, our theory is equivalent to the quantum trajectory theory with this approximation. However, the evolution equations are formally different and have a simple and intuitive physical meaning. We expect that our approach may have advantage over the quantum trajectory theory in numerical simulations, similar to the QPC/SET case. (In the latter case the reason for the numerical advantage was that the quantum trajectory equation is essentially the lowest-order approximation in the time step, while the quantum Bayesian evolution is the exact solution in the absence of Rabi oscillations, and this permits using larger time steps even in the presence of Rabi oscillations.)

Our derivation will be based on elementary quantum mechanics. We will also need some basic facts related to coherent states; for completeness, they are discussed in Appendix A. The paper is mainly addressed to non-experts in continuous quantum measurement and non-experts in quantum optics; this is why we include brief discussion of facts well-known to experts and focus on simple logic. We hope that our derivation  is accessible at the advanced-undergraduate level. While we discuss the circuit QED measurement of one qubit, it is straightforward to extend the discussion to the measurement of several qubits, including entanglement by measurement \cite{Riste-2013,Roch-2014}.

The paper is organized in the following way. In Sec.\ II we discuss the system and the model. In Sec.\ III we review the results of Ref.\ \cite{Korotkov-2011} for the ``bad cavity'' regime of circuit QED measurement. The main section of this paper is Sec. IV, in which we derive the quantum Bayesian formalism for circuit QED measurement with a moderate bandwidth. We first introduce a natural idea of ``history tail'', which consists of the microwave field emitted by the measurement resonator, and thus carries information about the resonator state at previous time moments (Sec.\ \ref{sec:history-tail}). Then we develop the Bayesian formalism by applying a natural measurement procedure to pieces of the ``history tail'' of short duration $\Delta t$ (Sec.\ \ref{sec:main-idea}). The textbook collapse due to this measurement leads to evolution of the entangled qubit-resonator state. We first derive the results for phase-sensitive measurement (Sec.\ \ref{sec:imperfect}) and then for phase-preserving measurement (Sec.\ \ref{sec:ph-pres}). The obtained evolution equations for short $\Delta t$ are also converted into the differential form (Sec.\ \ref{sec:differential}) and integrated for an arbitrary long duration (Sec.\ \ref{sec:arb-duration})   We conclude in Sec.\ V. Appendix A reviews basic facts related to coherent states. In Appendix B we derive the formulas for the phase back-action in the ``bad cavity'' regime via a simple language based on vacuum noise.

\section{System and model}\label{sec:model}

\begin{figure}[tb]
  \centering
\includegraphics[width=8.9cm,trim=8cm 9cm 10cm 6cm,clip=true]{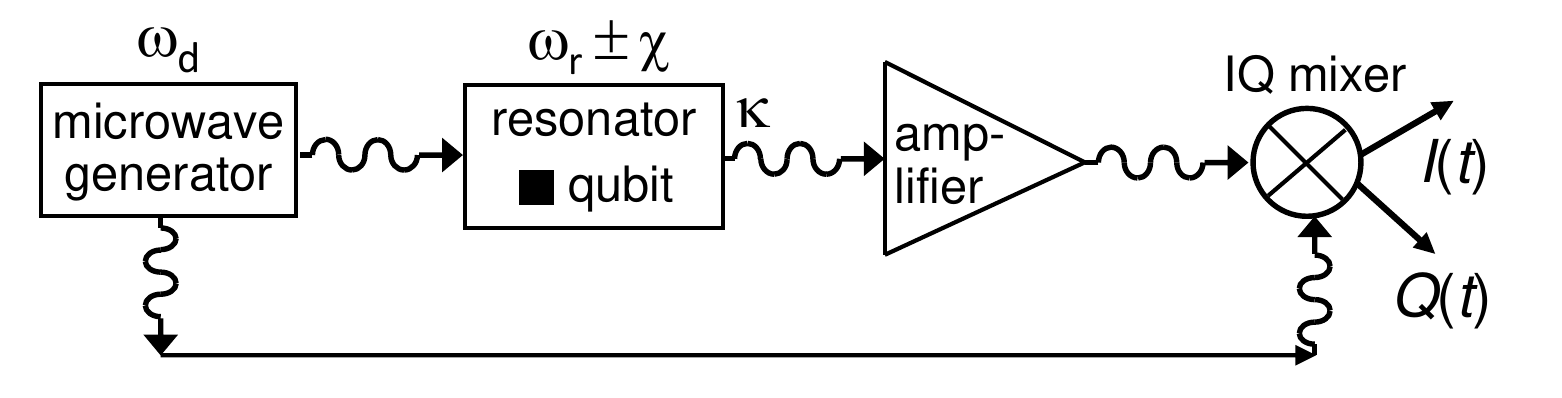}
  \caption{Schematic of the circuit QED setup. Microwave field of frequency
$\omega_{\rm d}$ is transmitted through (or reflected from) the resonator, whose
frequency slightly changes, $\omega_{\rm r}\pm\chi$,
depending on the qubit state. After amplification, the microwave is
sent to the IQ mixer, which produces two quadrature signals: $I(t)$
and $Q(t)$. In the case of a phase-sensitive amplifier we define $I(t)$ as the signal corresponding to the amplified quadrature, while for a phase-preserving amplifier we define $I(t)$ as
the quadrature carrying information about the qubit state.}
  \label{fig:schematic}
\end{figure}

We consider a superconducting qubit (transmon) measured in the circuit QED setup (Fig.\ \ref{fig:schematic}). The idea of the measurement \cite{Blais-2004,Wallraff-2004} is based on the dispersive coupling of the qubit with a microwave resonator, whose frequency slightly changes  depending on whether the qubit is in the state $|0\rangle$ or $|1\rangle$ (both are the eigenstates of qubit energy). This frequency shift affects the phase and amplitude of a probing microwave, which is transmitted through or reflected from the resonator (theoretically, there is no significant difference between the transmission and reflection configurations; however, in practice it is often better to use reflection). The outgoing microwave is amplified, and then the GHz-range signal is downconverted by mixing it with the original microwave tone, so that the low-frequency ($\alt 100$ MHz) output of the IQ mixer provides information about the qubit state. The rate of the information acquisition is limited by the output noise, which is mainly determined by the first amplifying stage (pre-amplifier). In recent years nearly quantum-limited superconducting parametric amplifiers \cite{Bergeal-2010,Castellanos-2008,Vijay-2011,Mutus-2013} became the standard pre-amplifiers, replacing formerly used cryogenic high-electron-mobility transistors (HEMTs), which have a much higher noise level.

The Hamiltonian of the qubit interacting with the resonator in the dispersive approximation \cite{Blais-2004} is
    \be
    H_{\rm q\&r}/\hbar =(\omega_{\rm q}/2) \, \sigma_z + \omega_{\rm r\,} a^\dagger a\ + \chi \, a^\dagger a\, \sigma_z,
    \label{Ham-disp}\ee
where $\omega_{\rm q}$ is the (effective) qubit frequency, $\omega_{\rm r}$ is the (effective) resonator frequency, $\chi$ is the dispersive coupling, $a^\dagger$ and $a$ are the creation and annihilation operators for the resonator (so that $n=a^\dagger a$ is the number of photons in the resonator), and the Pauli operator $\sigma_z$ acts on the qubit state in the energy basis $|1\rangle$ and $|0\rangle$. As we see from this Hamiltonian, the resonator frequency increases by $2\chi$ when the qubit state changes from $|0\rangle$ to $|1\rangle$; conversely, the qubit frequency increases by $2\chi$ per each additional photon in the resonator (ac Stark shift). The typical value of $|\chi|$ is crudely 1 MHz, while the qubit and resonator frequencies are typically between 4 and 9 GHz, with the detuning $|\Delta|$ of crudely 1 GHz, where $\Delta =\omega_{\rm q}-\omega_{\rm r}$.

The microwave drive of the resonator can be described by the standard additional Hamiltonian
    \be
 H_{\rm d}/\hbar =\varepsilon(t) \, e^{-i\omega_{\rm d}t}   a^\dagger + \varepsilon^*(t) \, e^{i\omega_{\rm d}t}  a ,
    \label{Ham-drive}\ee
 where $\omega_{\rm d}$ is the drive frequency and $\varepsilon (t)$ is the properly normalized drive amplitude. [This form is the Rotating Wave Approximation (RWA) of the physical Hamiltonian ${\rm Re}[\varepsilon(t)\,e^{-i\omega_{\rm d}t}](a+a^\dagger )$.] We do not consider the case when the resonator is driven by a squeezed microwave or a squeezed vacuum. We assume that the field in the resonator decays with the rate $\kappa /2$ (so that the energy decays with the rate $\kappa$) due to coupling with transmission lines and possibly due to other mechanisms of decay (at zero temperature). For the ensemble-averaged evolution, the effect of damping with rate $\kappa$ can be described via the standard Lindblad term in the master equation; however, we will not use it, since we are interested in evolution of an individual quantum system rather than an ensemble.

Note that the derivation of the dispersive Hamiltonian (\ref{Ham-disp}) for a transmon is somewhat involved (see, e.g., Ref.\ \cite{Koch-2007} and Appendix of Ref.\ \cite{Sete-2015}) because at least 3 transmon levels should be taken into account to find the coupling $\chi$ (4 levels are needed for the lowest-order dependence of $\chi$ on $n$). The small-$n$ value of the coupling $\chi$ can be approximated \cite{Sete-2015} as
    \be
\chi = \frac{ \omega_{\rm r}}{\omega_{\rm q}} \, \frac{g^2\delta_{\rm q}}{\Delta (\Delta-\delta_{\rm q})},
    \ee
where $g$ is the coupling in Jaynes-Cummings Hamiltonian and $\delta_{\rm q}=\omega_{\rm q}-\omega_{\rm q, 12}$ is the transmon anharmonicity ($\omega_{\rm q, 12}$ is the transition frequency between  transmon levels $|1\rangle$ and $|2\rangle$). With increasing $n$ the value of $\chi$ changes (as well as $\omega_{\rm r}$), and a better description of the evolution should be based on the eigenlevels of the transmon-resonator system, rather than bare levels \cite{Khezri-2016}. In the present paper we do not take these complications into account and use the simple Hamiltonian (\ref{Ham-disp}); however, there is a natural way to include these effects into our formalism phenomenologically.
One more subtlety is that the resonator damping $\kappa$ leads to the qubit energy relaxation  \cite{Esteve-1986,Houck-2008} via the Purcell effect, which we do not take into account. However, in many present-day experiments this effect is suppressed by a Purcell filter \cite{Reed-2010,Jeffrey-2014,Sete-2015}, so description by the simple Hamiltonian (\ref{Ham-disp}) again becomes a good approximation.

In this paper we will be using the rotating frame, based on the drive frequency $\omega_{\rm d}$ for the resonator and the frequency $\omega_{\rm q}$ for the qubit. This essentially means that instead of fast-oscillating coefficients in the lab-frame wavefunction $c_{0,n}(t)\,|0,n\rangle + c_{1,n}(t)\,|1,n\rangle$ (here $n$ is the number of photons in the resonator), we implicitly operate with slower-varying coefficients  $c_{0,n}(t)\, e^{-i\omega_{\rm q} t/2} e^{i\omega_{\rm d} n t}$ and  $c_{1,n}(t)\, e^{i\omega_{\rm q} t/2} e^{i\omega_{\rm d} n t}$. Equivalently, we can change the Hamiltonian (\ref{Ham-disp}) and (\ref{Ham-drive}) to the rotating-frame Hamiltonian
    \be
     H_{\rm rot}/\hbar  =  (\omega_{\rm r}-\omega_{\rm d})\, a^\dagger a
     +  \chi a^\dagger a \, \sigma_z
+  \varepsilon a^\dagger + \varepsilon^* a.
    \label{Ham-2}\ee
Note that in Appendix A  we use tilde signs for the rotating-frame variables, which are omitted in the main text.

Our goal in this paper is to find (in a simple way) the evolution of the qubit-resonator state in the process of measurement. For that we assume that the qubit evolves only due to measurement, so we explicitly assume the absence of a Rabi drive applied to the qubit and absence of qubit energy relaxation. Since the Hamiltonian (\ref{Ham-disp}) is of the quantum non-demolition (QND) type \cite{Braginsky-book}, then if the initial qubit state is $|0\rangle$, it will remain $|0\rangle$ during the whole measurement process. Similarly, the initial qubit state $|1\rangle$ will remain $|1\rangle$. In these two simple cases, evolution of the resonator state is decoupled from the qubit, but the effective resonator frequency $\omega_{\rm r}\pm \chi$ depends on the qubit state (the upper sign is for the qubit state $|1\rangle$). Then the classical evolution of the resonator field $\alpha (t)$ can be described in the standard RWA way as \cite{Walls-Milburn-book}
    \be
    \dot{\alpha}_\pm = -i (\omega_{\rm r} \pm \chi -\omega_{\rm d}) \, \alpha_\pm -\frac{\kappa}{2}\, \alpha_\pm -i\varepsilon ,
    \label{alpha-dot1}\ee
where the rotating frame is based on the drive frequency $\omega_{\rm d}$. The quantum evolution is described by exactly the same equation \cite{Walls-Milburn-book}, with the classical field state  replaced by the coherent state $|\alpha_\pm (t)\rangle$ (see Appendix A). Note that $|\alpha|^2=\bar{n}$ is the average number of photons in the resonator. Besides the notation $\alpha_\pm$, we will interchangeably use a notation that explicitly shows the corresponding qubit state,
    \be
    \alpha_1 \equiv \alpha_+, \,\,\, \alpha_0 \equiv \alpha_-.
    \ee

From Eq.\ (\ref{alpha-dot1}), we see that the resonator field depends on the qubit state. In particular, the steady state is
    \be
    \alpha_{\pm ,\rm st} = \frac{-i\varepsilon}{ i(\omega_{\rm r}\pm \chi -\omega_{\rm d}) + \kappa /2}.
    \label{alpha-st}\ee
(It is easy to see that these complex numbers always belong to the circle in the complex plane, which is centered at $-i\varepsilon /\kappa$ and passes through the origin.)

\begin{figure}[tb]
  \centering
\includegraphics[width=9.0cm,trim=8cm 7.8cm 10cm 5.5cm,clip=true]{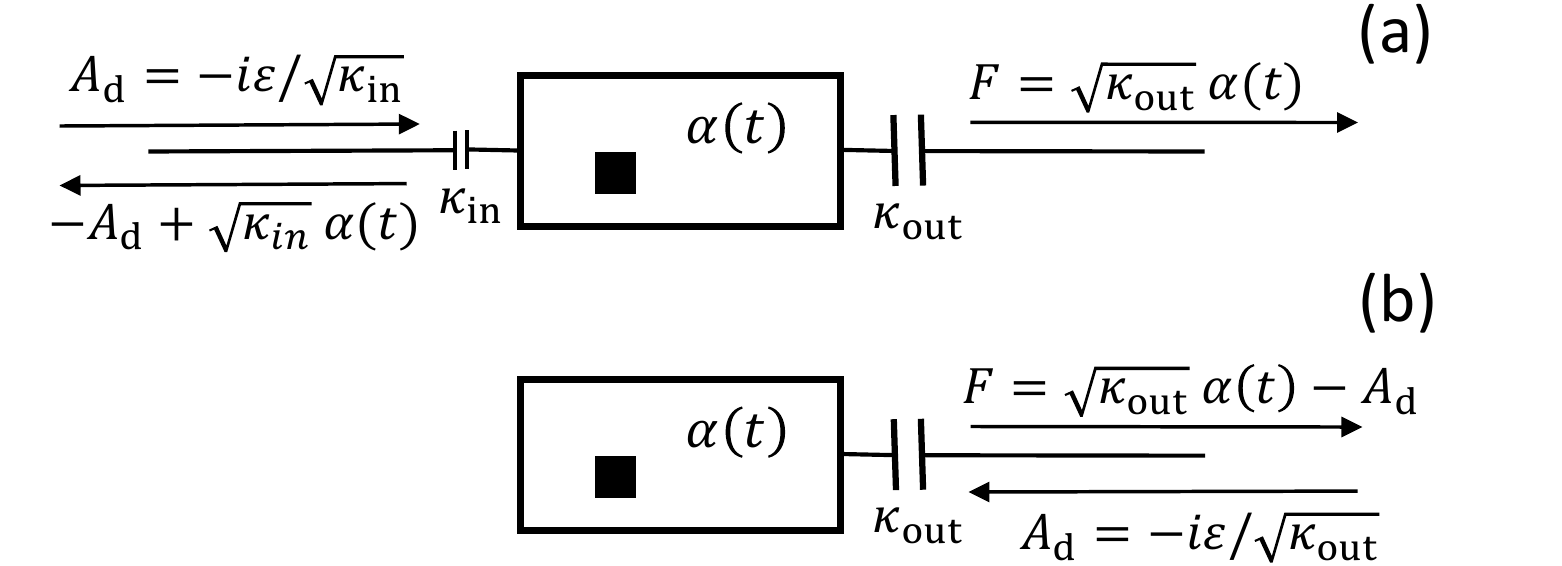}
  \caption{Comparison between (a) transmission and (b) reflection configurations. The incoming drive field $A_{\rm d}$ is mostly reflected, but its small part enters the resonator, contributing to the field change as $\dot{\alpha}=\sqrt{\kappa_{\rm in}}\, A_{\rm d}$ in the transmission case and $\dot{\alpha}=\sqrt{\kappa_{\rm out}}\, A_{\rm d}$ in the reflection case. For the resonator field,  $|\alpha |^2$ is the average number of photons, while for the propagating fields, $|F|^2$ and $|A_{\rm d}|^2$ are the average numbers of photons per second. }
  \label{fig:trans-refl}
\end{figure}

The outgoing field $F$ in the transmission and reflection configurations (Fig.\ \ref{fig:trans-refl}) can be described as \cite{Walls-Milburn-book}
    \be
    F_{\rm trans}= \sqrt{\kappa_{\rm out}} \, \alpha, \,\,\,\,\,
    F_{\rm refl}= \sqrt{\kappa_{\rm out}} \, \alpha + \frac{i \varepsilon }{\sqrt{\kappa_{\rm out}}} ,
    \label{F-out}\ee
where $\kappa_{\rm out}$ is the resonator damping due to coupling with the outgoing transmission line ($\kappa_{\rm out}\leq \kappa$), and in this normalization $|F|^2$ is the average number of propagating photons per second. (Note that the phase of $F$ can be chosen arbitrarily; in our choice the coefficient between $F$ and $\alpha$ is real and positive.) By combining Eqs.\ (\ref{alpha-st}) and (\ref{F-out}) it is easy to see that in the case $\kappa_{\rm out} \approx \kappa$ the reflection configuration operates with smaller fields for the same response (and therefore larger phase response) than the transmission configuration, and in this sense it is preferable from the practical point of view. However, for our purposes in this paper the two configurations are equivalent (the well-defined difference $i\varepsilon/\sqrt{\kappa_{\rm out}}$ can theoretically be simply subtracted). We will implicitly assume the transmission configuration (without loss of generality), while all the results are applicable to both the transmission and reflection configurations.

The outgoing microwave field $F$ is then amplified (either in a phase-preserving or a phase-sensitive way) and sent to the IQ mixer (Fig.\ \ref{fig:schematic}). A phase-sensitive amplifier amplifies only a certain phase (quadrature) $\phi_{\rm a}$ of the microwave field and de-amplifies the $\pi/2$-shifted phase (orthogonal quadrature). For the complex number $F$ this means amplification of only a certain direction on the complex plane along $e^{i\phi_{\rm a}}$. For a faster qubit measurement, the obvious choice is to amplify the quadrature that connects the complex numbers $\alpha_0$ and $\alpha_1$ corresponding to qubit states $|0\rangle$ and $|1\rangle$. We will consider amplification of an arbitrary quadrature, including this optimal case. The IQ mixer produces two low-frequency signals, which correspond to the real and imaginary parts of an amplified $F$; however it is easy to rotate the axes of the complex plane by using the linear combinations of the two outputs. Since only one quadrature is amplified by a phase-sensitive detector, there is no information in the mixer output corresponding to the orthogonal quadrature. Therefore, the phase-sensitive amplifier essentially produces {\it only one output signal} after the mixer, which we will call $I(t)$. Note that the amplified phase $\phi_{\rm a}$ can in principle vary in time; then we also vary the quadrature, corresponding to $I(t)$.

A phase-preserving amplifier equally amplifies any quadrature, so both outputs of the IQ mixer are important. (Note that usual non-parametric amplifiers, including HEMT, are phase-preserving.) In this case we will call $I(t)$ the linear combination of the outputs corresponding to the quadrature connecting $\alpha_0$ and $\alpha_1$, so that $I(t)$ carries information about the measured qubit state $|0\rangle$ or $|1\rangle$. The output signal for the orthogonal quadrature will be called $Q(t)$; it does not carry information about the qubit state, but will still be important for producing phase back-action. Since $\alpha_0 (t)$ and $\alpha_{1} (t)$ evolve before reaching steady values, we will correspondingly vary the quadratures corresponding to $I(t)$ and $Q(t)$.

The main reason why phase-sensitive amplifiers are often preferred for the qubit measurement is that their quantum limitation for the output noise is twice smaller than that for phase-preserving amplifiers \cite{Caves-1982,Devyatov-1986,Clerk-2010}. The output noise of a phase-sensitive amplifier should exceed the ``half quantum'', which exactly corresponds to the width of the ground state of the oscillator, representing the amplified field (so that the energy is $\hbar\omega_{\rm d} /2$). In other words, this is the amplified vacuum noise of the coherent state of the field, and the ideal phase-sensitive amplifier does not add its own noise (the output noise can be smaller if a squeezed state is amplified). The output noise power of a phase-preserving amplifier is at least two ``half quanta'': one comes from the amplified vacuum noise, and one more is added by the amplifier \cite{Caves-1982,Devyatov-1986,Clerk-2010}.

As discussed above, the dynamics of the system is very simple when the qubit is either in the state $|0\rangle$ or $|1\rangle$ during the whole measurement process. The goal of this paper is to describe the evolution when the initial qubit state is a superposition $c_0 |0\rangle +c_1|1\rangle$ or, more generally, an arbitrary density matrix $\rho(0)$.

\section{``Bad cavity'' limit}

In this section we review results of Ref.\ \cite{Korotkov-2011} for the ``bad cavity'' limit, which assumes $\kappa \gg \Gamma$, where $\Gamma$ is the qubit ensemble dephasing due to measurement, discussed below. In this case we can neglect transient evolution of the resonator state, and there is practically no entanglement between the qubit and the measurement resonator, because the two steady states (\ref{alpha-st}) are very close to each other, $|\alpha_{1,\rm st}-\alpha_{0,\rm st}|\ll 1$. Therefore, the evolution of the qubit state can be considered by itself. It is assumed that parameters of the measurement setup ($\kappa$, $\varepsilon$, etc.) do not change in time.  We review here the ``bad cavity'' limit mainly for later comparison with the more general case $\kappa \sim \Gamma$; the derivation in the next section does not rely on results discussed in this section.

Ensemble dephasing of the qubit in the ``bad cavity'' regime is \cite{Gambetta-2006}
    \be
    \Gamma = 2\chi \, {\rm Im} (\alpha_{1,\rm st}^* \, \alpha_{0,\rm st}) = \frac{\kappa }{2} \, |\alpha_{1,\rm st}-\alpha_{0,\rm st}|^2.
    \label{Gamma-bc1}\ee
We see that the condition $\kappa \gg \Gamma$ is equivalent to $|\alpha_{1,\rm st}-\alpha_{0,\rm st}|\ll 1$.
 In the case when $|\chi | \ll \kappa$, the ensemble dephasing \cite{Blais-2004,Gambetta-2006} can be expressed as [see Eq.\ (\ref{alpha-st})]
    \be
    \Gamma = \frac{8\chi^2 \bar{n}}{\kappa} \, \frac{1}{1+ [2(\omega_{\rm r}-\omega_{\rm d})/\kappa ]^2} ,
    \label{Gamma-bc2}\ee
and the ac Stark shift contribution $\delta\omega_{\rm q}$ to the effective qubit frequency $\omega_{\rm q}+\delta \omega_{\rm q}$ is \cite{Blais-2004,Gambetta-2006}
    \be
    \delta \omega_{\rm q}= 2\chi \bar{n}.
    \label{Stark-bc}\ee
If $|\chi |$ is comparable to $\kappa$ (but still $\Gamma \ll \kappa$), then Eqs.\ (\ref{Gamma-bc2}) and (\ref{Stark-bc}) should be modified (see Sec.\ \ref{sec:ensemble-aver}), but the Bayesian formalism reviewed in this section does not change.

    \subsection{Phase-sensitive amplifier}\label{sec:ph-sens-bc}

Evolution of the qubit density matrix $\rho (t)$ due to measurement of an {\it arbitrary duration}  $\tau$ can be described by simple equations  \cite{Korotkov-2011}
    \begin{eqnarray}
&&   \frac{\rho_{11}(t+\tau)}{\rho_{00}(t+\tau)} =
    \frac{\rho_{11}(t)}{\rho_{00}(t)} \, \exp \bigg[
 \frac{\tilde{I}_{\rm m}(\tau) \, \Delta I}{D} \bigg], \,\,\,\qquad
    \label{ph-sens-diag-bc}\\
&&  \frac{\rho_{10}(t+\tau)}{\rho_{10}(t)}= \frac{\sqrt{\rho_{11}(t+\tau)\,
\rho_{00}(t+\tau)}} {\sqrt{\rho_{11}(t)\, \rho_{00}(t)}}\, \,
    \nonumber \\
&& \hspace{1.7cm}
 \times \, \exp [-i K \tilde{I}_{\rm m}(\tau)\, \tau]  \, e^{-\gamma \tau} \, e^{-i\delta \omega_{\rm q}\tau} , \qquad
    \label{ph-sens-off-bc}
    \end{eqnarray}
where
    \be
     \tilde I_{\rm m}(\tau) = I_{\rm m}(\tau) - \frac{I_0+I_1}{2}, \,\,\, I_{\rm m}(\tau)= \frac{1}{\tau} \int_t^{t+\tau} I(t')\, dt',
    \label{ph-sens-Im-bc}\ee
so that $I_{\rm m}$ is the measured output signal $I(t)$ averaged over the time interval $[t,t+\tau]$, while for $\tilde I_{\rm m}$ we also subtract the mean value $(I_0+I_1)/2$, with $I_0$ and $I_1$ being the average output signals, corresponding to the qubit states $|0\rangle$ and $|1\rangle$. The measurement response is
    \begin{eqnarray}
&& \Delta I= I_1-I_0= \Delta I_{\rm max}\cos\phi_{\rm d}, \,\,\,
\qquad
   \label{ph-sens-DeltaI-bc}
   \\
&& \phi_{\rm d} = \phi_{\rm a} -{\rm arg} (\alpha_{1,\rm st}-\alpha_{0,\rm st}) ,
   \label{phi-d-def}\end{eqnarray}
where $\phi_{\rm d}$ is the phase difference between the amplified quadrature $\phi_{\rm a}$ and the optimal quadrature $\phi_{\rm opt}={\rm arg} (\alpha_{1,\rm st}-\alpha_{0,\rm st})$, which gives the largest response $\Delta I_{\rm max}$. The variance of $\tilde I_{\rm m}(\tau)$ due to the amplifier noise is
    \be
    D=\frac{S_I}{2\tau},
    \label{D-def-bc}\ee
where $S_I$ is the single-sided spectral density of the noise [for different definitions of the spectral density, Eq.\ (\ref{D-def-bc}) should be changed correspondingly]. The phase back-action depends on the coefficient $K$, which equals
    \be
    K= \frac{\Delta I_{\rm max}}{S_I} \, \sin\phi_{\rm d}= \frac{\Delta I_{\rm max}}{2D \tau} \, \sin\phi_{\rm d}.
     \label{ph-sens-K-bc}\\
    \ee
The dephasing rate $\gamma$ is due to non-ideality of the measurement,
    \be
    \gamma = \Gamma - \frac{(\Delta I_{\rm max})^2}{4S_I} =\Gamma - \frac{(\Delta I_{\rm max})^2}{8D\tau},
    \ee
where $\Gamma$ is the qubit ensemble dephasing [see Eqs.\ (\ref{Gamma-bc1}) and (\ref{Gamma-bc2})]. The quantum efficiency of the measurement process can be introduced in two different ways,
    \be
    \eta =1-\frac{\gamma}{\Gamma}=\frac{(\Delta I_{\rm max})^2}{8D\tau \Gamma}=\eta_{\rm amp}\eta_{\rm col}, \,\,\, \tilde\eta = \eta \, \cos^2 \phi_{\rm d},
    \ee
where $\eta$ ($0\leq \eta\leq 1$) takes into account quantum efficiency $\eta_{\rm amp}$ of the phase-sensitive amplifier and efficiency $\eta_{\rm col}$ of the microwave signal collection, while $\tilde\eta$ also includes the effect of choosing a non-optimal quadrature for amplification. Here $\eta_{\rm col}=\kappa_{\rm col}/\kappa$ is the ratio of the microwave energy reaching amplifier to the total energy loss by the resonator, so that $\kappa_{\rm col}/\kappa_{\rm out}$ describes the loss in the transmission line before reaching the amplifier. The amplifier efficiency $\eta_{\rm amp}=S_{I,\rm q.l.}/S_I$ is the ratio between the output noise $S_{I,\rm q.l.}$ of an ideal quantum-limited amplification chain to the actual output noise $S_I$. The last term in Eq.\ (\ref{ph-sens-off-bc}) is due to the ac Stark shift $\delta\omega_{\rm q}$ given by Eq.\ (\ref{Stark-bc}) (note that in Ref.\ \cite{Korotkov-2011} the rotating frame was already accounting for this term and the equation was written for the conjugate variable $\rho_{01}$). Note that $\rho_{00}+\rho_{11}=1$ and therefore $\rho_{00}=(1+\rho_{11}/\rho_{00})^{-1}$.

Equations (\ref{ph-sens-diag-bc})--(\ref{ph-sens-Im-bc}) can be used to find the qubit evolution in an experiment by using experimental output signal record $I(t)$; in numerical simulations
$\tilde I_{\rm m}(\tau)$ can be picked randomly from the probability density distribution
    \begin{eqnarray}
    && \hspace{-0.2cm} P(\tilde I_{\rm m})= \rho_{00}(t) \, P(\tilde I_{\rm m}|0) + \rho_{11}(t) \, P(\tilde I_{\rm m}|1) ,
    \label{ph-sens-P(I)}\\
&&  \hspace{-0.2cm}  P(\tilde I_{\rm m}|j) = \frac{1}{\sqrt{2\pi D}} \, \exp \bigg[ -\frac{[\, \tilde I_{\rm m} +(-1)^j(\Delta I/2)\,]^2}{2D} \bigg] , \qquad
    \label{ph-sens-Gauss}\end{eqnarray}
where $j=0,1$ and $P(\tilde I_{\rm m}|j)$ is the standard Gaussian distribution in the case when the qubit is in the state $|j\rangle$. For an infinitesimally small averaging time $\tau$, this is equivalent to using
    \be
    I(t)=\frac{I_0+I_1}{2} + \frac{\Delta I}{2} \, [\rho_{11} (t)-\rho_{00}(t)] +\xi_I(t),
\,\,\, S_{\xi_I}=S_I,
    \label{ph-sens-I(t)}\ee
where $\xi_I(t)$ is the white noise with spectral density $S_I$.

The qubit evolution equations (\ref{ph-sens-diag-bc}) and (\ref{ph-sens-off-bc}) have a very simple physical meaning. The evolution (\ref{ph-sens-diag-bc}) of the diagonal matrix elements of the density matrix is the classical Bayesian update for the probabilities,
    \be
    \rho_{jj}(t+\tau) = \frac{\rho_{jj}(t) \, P(\tilde{I}_{\rm m}|j)}{\rm Norm},
    \ee
where $P(\tilde{I}_{\rm m}|j)$ is the likelihood, given by Eq.\ (\ref{ph-sens-Gauss}). Note that another form of Eq.\ (\ref{ph-sens-diag-bc}) in terms of the non-centered measurement result $I_{\rm m}$ is
    \be
\frac{\rho_{11}(t+\tau)}{\rho_{00}(t+\tau)} =
    \frac{\rho_{11}(t) \exp [-(I_{\rm m}-I_1)^2/2D ]}{\rho_{00}(t)\exp [-(I_{\rm m}-I_0)^2/2D ]} .
    \ee
The evolution (\ref{ph-sens-off-bc}) of the off-diagonal matrix element contains the natural term due to change of the diagonal elements (conservation of relative purity), the phase back-action term, decoherence due to non-ideality, and contribution from the ac Stark shift. The phase back-action has a natural mechanism: when measuring a non-optimal quadrature, $\phi_{\rm d}\neq 0$, the output signal gives us information about the fluctuating number of photons in the resonator, and therefore the fluctuating ac Stark shift. The factor $K$ in Eq.\ (\ref{ph-sens-off-bc}) is the coefficient  characterising this linear relation between the ac Stark shift and output signal fluctuations.

The evolution equations (\ref{ph-sens-diag-bc}) and (\ref{ph-sens-off-bc}) have been derived in Ref.\ \cite{Korotkov-2011} in the following way. The Bayesian evolution (\ref{ph-sens-diag-bc}) of the diagonal matrix elements was essentially postulated from the necessary correspondence between the classical and quantum evolution. This follows from common sense as much as the standard collapse postulate in quantum mechanics. For the off-diagonal elements, the logic of the derivation (sketched below) was essentially the same as in the first derivation \cite{Korotkov-1999} for measurement by a QPC. Using the general inequality $|\rho_{10}|\leq \sqrt{\rho_{11}\rho_{00}}$  and evolution (\ref{ph-sens-diag-bc}) for the diagonal elements, it is easy to derive inequality for the ensemble dephasing, $\Gamma \geq (\Delta I)^2/4S_I$. In the quantum-limited case [in this case $|\Delta \alpha| =1$ is resolved with signal-to-noise ratio of 1 after time $1/\kappa$, and therefore $S_I=S_{\rm min}= (\Delta I_{\rm max})^2 \kappa [1+4(\omega_{\rm r}-\omega_{\rm d})^2/\kappa^2]/(32\chi^2\bar{n})\,$], and for $\phi_{\rm d}=0$, the lower bound of this inequality for $\Gamma$ coincides with the actual value (\ref{Gamma-bc2}). Therefore, in this case the evolution of $\rho_{10}$ should be precisely the first term in Eq.\ (\ref{ph-sens-off-bc}) and possibly a result-independent phase (which is naturally associated with the qubit frequency shift in the last term); otherwise the ensemble dephasing would be larger than in Eq.\ (\ref{Gamma-bc2}). Thus, in the ideal case Eqs.\ (\ref{ph-sens-diag-bc}) and (\ref{ph-sens-off-bc}) have been derived ``logically'', by comparing unavoidable evolution due to acquired information with the ensemble dephasing.

In the non-optimal case ($\phi_{\rm d}\neq 0$), the derivation in Ref.\ \cite{Korotkov-2011} took into account the phase back-action by explicitly analyzing the information on the fluctuating photon number in the resonator provided by the measurement result $\tilde{I}_{\rm m}$. In this way Eq.\ (\ref{ph-sens-K-bc}) for the correlation factor $K$ was obtained, leading to the term with $K$ in Eq.\ (\ref{ph-sens-off-bc}). Finally, the term $e^{-\gamma \tau}$ in Eq.\ (\ref{ph-sens-off-bc}) was obtained by averaging over the extra noise from a non-ideal amplifier \cite{Korotkov-nonid} and averaging over the signal that was lost due to imperfect microwave collection. This is how the qubit evolution equations (\ref{ph-sens-diag-bc}) and (\ref{ph-sens-off-bc}) have been obtained in Ref.\ \cite{Korotkov-2011}.

Actually, the derivation for the phase back-action coefficient $K$ in Ref.\ \cite{Korotkov-2011} was presented only for the case of resonant microwave frequency, $\omega_{\rm d}=\omega_{\rm r}$. In Appendix B we show the derivation, which is still valid in the case of a significant detuning, $|\omega_{\rm d}-\omega_{\rm r}| \agt \kappa$. This derivation is based on an analysis of the effect of vacuum noise entering the resonator from the transmission line. In this analysis the vacuum noise is treated essentially classically, consistent with the Poisson statistics $\bar{n}\pm \sqrt{\bar{n}}$ for the photon number.

Note that averaging of the evolution equations (\ref{ph-sens-diag-bc}) and (\ref{ph-sens-off-bc}) over random $\tilde{I}_{\rm m}$ with the probability distribution (\ref{ph-sens-P(I)}) produces ensemble-averaged equations
    \begin{eqnarray}
    && \rho_{jj}(t+\tau) = \rho_{jj} (t),
    \\
     && \rho_{10}(t+\tau) = \rho_{10} (t)\, e^{-\Gamma\tau} e^{-i\delta\omega_{\rm q}\tau}, \end{eqnarray}
in which there is no dependence on the measured phase $\phi_{\rm d}$ (as required by causality) because
    \be
    \frac{(\Delta I)^2}{4S_I}+\frac{K^2 S_I}{4}=\frac{(\Delta I_{\rm max})^2}{4S_I}=\Gamma -\gamma. \ee

Let us briefly discuss the role of the ``weak response'' condition $|\chi |\ll \kappa$ in the  formalism reviewed in this section. In the case of not too small a number of photons in the resonator, $\bar{n}\agt 1$, this inequality follows from the ``bad cavity'' condition $|\alpha_{1,\rm st}-\alpha_{0,\rm st}|\ll 1$, and therefore is not needed as an additional condition. However, for $\bar{n}_{0,1}\ll 1$ it is possible to have $|\alpha_{1,\rm st}-\alpha_{0,\rm st}|\ll 1$ even when $|\chi | \agt \kappa$. In this case the Bayesian formalism (\ref{ph-sens-diag-bc})--(\ref{ph-sens-Gauss}) is still applicable, but the ensemble dephasing $\Gamma$ and ac Stark shift $\delta \omega_{\rm q}$ are not necessarily given by Eqs.\ (\ref{Gamma-bc2}) and (\ref{Stark-bc}), in particular, because $\bar{n}_0$ and $\bar{n}_1$ may be significantly different, $|\bar{n}_1-\bar{n}_0| \sim \bar{n}_1+\bar{n}_0$. The formulas for $\Gamma$ and $\delta \omega_{\rm q}$ in this case are given in Ref.\ \cite{Gambetta-2006} and also derived in Sec.\ \ref{sec:ensemble-aver} [$\Gamma$ is given by Eq.\ (\ref{Gamma-bc1}), while $\delta\omega_{\rm q}$ is given by Eq. (\ref{delta-omega-sum})].

Note that the Bayesian evolution equations (\ref{ph-sens-diag-bc}) and (\ref{ph-sens-off-bc}) are exactly the same as for the continuous qubit measurement by QPC or SET \cite{Korotkov-1999,Korotkov-2001,Korotkov-2002}. However, the dependence (\ref{ph-sens-DeltaI-bc}) and (\ref{ph-sens-K-bc}) of the response $\Delta I$ and phase back-action coefficient $K$ on the measured quadrature $\phi_{\rm d}$ is a specific feature of the circuit QED (or cavity QED) setup.

\subsection{Phase-preserving amplifier}\label{sec:ph-pres-bc}

As was discussed in Sec.\ II, in the case of a phase-preserving amplifier we choose $I(t)$ to be the output signal, corresponding to the optimal quadrature $\phi_{\rm opt}={\rm arg} (\alpha_{1,\rm st}-\alpha_{0,\rm st})$, while the output $Q(t)$ corresponds to the orthogonal quadrature $\phi_{\rm opt}+\pi/2$. Therefore, $\phi_{\rm d}=0$ for $I(t)$ and $\phi_{\rm d}=\pi/2$ for $Q(t)$.

The qubit state evolution due to a phase-preserving measurement for an {\it arbitrary duration} $\tau$  is described by equations  \cite{Korotkov-2011}
    \begin{eqnarray}
&&   \frac{\rho_{11}(t+\tau)}{\rho_{00}(t+\tau)} =
    \frac{\rho_{11}(t)}{\rho_{00}(t)} \, \exp \bigg[
 \frac{\tilde{I}_{\rm m}(\tau) \, \Delta I}{D} \bigg], \,\,\,\qquad
    \label{ph-pres-diag-bc}\\
&&  \frac{\rho_{10}(t+\tau)}{\rho_{10}(t)}= \frac{\sqrt{\rho_{11}(t+\tau)\,
\rho_{00}(t+\tau)}} {\sqrt{\rho_{11}(t)\, \rho_{00}(t)}}\, \,
    \nonumber \\
&& \hspace{1.7cm}
 \times \, \exp [-i K \tilde{Q}_{\rm m}(\tau)\, \tau]  \, e^{-\gamma \tau} \, e^{-i\delta \omega_{\rm q}\tau} , \qquad
    \label{ph-pres-off-bc}
    \end{eqnarray}
which have exactly the same form as Eqs.\ (\ref{ph-sens-diag-bc}) and (\ref{ph-sens-off-bc}), except $\tilde{I}_{\rm m}(\tau)$ in Eq.\ (\ref{ph-sens-off-bc}) is replaced by $\tilde{Q}_{\rm m}(\tau)$ in Eq.\ (\ref{ph-pres-off-bc}). The measurement result $\tilde{I}_{\rm m}(\tau)$ is given by Eq.\ (\ref{ph-sens-Im-bc}), and similarly
    \be
     \tilde Q_{\rm m}(\tau) = \frac{1}{\tau} \int_t^{t+\tau} Q(t')\, dt'
   - Q_0,
    \label{ph-pres-Qm-bc}\ee
with equal average values, $Q_1=Q_0$, for the two qubit states. Since $I(t)$ and $Q(t)$ are equally amplified, the variances of $\tilde{I}_{\rm m}(\tau)$ and $\tilde{Q}_{\rm m}(\tau)$ due to amplifier noise are both equal to
    \be
    D=\frac{S_I}{2\tau }, \,\,\, S_Q=S_I.
    \ee
The phase back-action is now caused by $\tilde Q_{\rm m}(\tau)$, and the coefficient $K$ is the same as in Eq.\ (\ref{ph-sens-K-bc}) for $\phi_{\rm d}=\pi/2$,
    \be
    K=\frac{\Delta I}{S_I}=\frac{\Delta I}{2D\tau}, \,\,\,\,\, \Delta I=I_1-I_0.
    \label{ph-pres-K}\ee
The dephasing rate $\gamma$ due to non-ideality is
    \be
    \gamma = \Gamma - 2 \, \frac{(\Delta I)^2}{4S_I}=\Gamma - \frac{(\Delta I)^2}{4D\tau},
    \ee
where ensemble dephasing $\Gamma$ is still given by Eqs.\ (\ref{Gamma-bc1}) and (\ref{Gamma-bc2}) (it cannot depend on the detector because of causality), and the extra factor of 2 is related to equal contributions due to fluctuations of $I(t)$ and $Q(t)$. The quantum efficiency can again be defined in two different ways,
    \be
    \eta = 1- \frac{\gamma}{\Gamma}=\eta_{\rm amp}\eta_{\rm col}, \,\,\,
    \tilde\eta= \frac{\eta}{2} = \tilde{\eta}_{\rm amp}\eta_{\rm col},
    \label{eta-phase-pres-def}\ee
where $\eta$ ($0\leq\eta\leq 1$) compares the measurement with the ideal phase-preserving case, while $\tilde\eta$ and $\tilde{\eta}_{\rm amp}$ compare the operation using only $I(t)$ channel with the ideal phase-sensitive case ($\tilde{\eta}\leq\tilde{\eta}_{\rm amp}\leq 1/2$ because of twice larger noise in an ideal phase-preserving amplifier).

Equations (\ref{ph-pres-diag-bc}) and (\ref{ph-pres-off-bc}) describe the qubit evolution when the signals $I(t)$ and $Q(t)$ are obtained from an experiment, while in numerical simulations
$\tilde{I}_{\rm m}(\tau)$ can be generated using Eqs.\ (\ref{ph-sens-P(I)}) and (\ref{ph-sens-Gauss}), while $\tilde{Q}_{\rm m}(\tau)$ can be picked from the Gaussian probability distribution
    \be
    P(\tilde{Q}_{\rm m}) = \frac{1}{\sqrt{2\pi D}} \, \exp \bigg[ - \frac{\tilde{Q}_{\rm m}^2}{2D} \bigg] .
    \ee
For infinitesimally small $\tau$, the signal $I(t)$ can also be generated using Eq.\ (\ref{ph-sens-I(t)}),  and for $Q(t)$ we can use
    \be
    Q(t)=Q_0 + \xi_Q (t), \,\,\, S_{\xi_Q}= S_Q= S_I,
    \ee
with equal spectral densities, $S_{\xi_Q}=S_{\xi_I}$, of uncorrelated noise in $I(t)$ and $Q(t)$ channels.

Equations (\ref{ph-pres-diag-bc}) and (\ref{ph-pres-off-bc}) have been derived in Ref.\ \cite{Korotkov-2011} in three different ways, leading to the same result. In the first derivation, Eq.\ (\ref{ph-pres-diag-bc}) has been again postulated from the necessary correspondence with classical evolution of probability, and the phase back-action coefficient $K$ in Eq.\ (\ref{ph-pres-K}) has been calculated from information on fluctuation of photon number, provided by $Q(t)$. This gives the inequality $\Gamma \geq 2(\Delta I)^2/4S_I$, whose lower bound in the ideal case coincides with the actual value (\ref{Gamma-bc2}). Thus in the ideal case Eqs.\ (\ref{ph-pres-diag-bc}) and (\ref{ph-pres-off-bc}) can be derived ``logically'', while the non-ideal case ($\eta <1$) can be analyzed by averaging over the extra noise of the amplifier ($\eta_{\rm amp}<1$) and over information contained in the lost fraction of the microwave signal ($\eta_{\rm col}<1$).

In the second derivation \cite{Korotkov-2011}, Eqs.\ (\ref{ph-pres-diag-bc}) and (\ref{ph-pres-off-bc}) have been obtained from Eqs.\ (\ref{ph-sens-diag-bc}) and (\ref{ph-sens-off-bc}) by considering a phase-preserving amplifier as a phase-sensitive amplifier with rapidly rotating amplified phase $\phi_{\rm a}$, so that the difference $\phi_{\rm d}$ from the optimal phase is also changing. Then averaging the evolution in Eqs.\ (\ref{ph-sens-diag-bc}) and  (\ref{ph-sens-off-bc}) over the period of phase rotation, we obtain Eqs.\ (\ref{ph-pres-diag-bc}) and (\ref{ph-pres-off-bc}). Finally, the third derivation in Ref.\ \cite{Korotkov-2011} has been based on considering a phase-preserving amplifier as two phase-sensitive amplifiers, which amplify orthogonal quadratures in two microwave channels, obtained from the microwave signal by using a symmetric beam splitter. Then using Eqs.\ (\ref{ph-sens-diag-bc}) and (\ref{ph-sens-off-bc}) for each channel, we again obtain Eqs.\ (\ref{ph-pres-diag-bc}) and (\ref{ph-pres-off-bc}).

Note that since in the ``bad cavity'' regime the qubit is practically not entangled with the measurement resonator, it is easy to include the qubit evolution due to Rabi oscillations, energy relaxation, etc. (this extra evolution should be much slower than $\kappa$, but can be slower, comparable, or faster than $\Gamma$). For that we need to take the derivative of the evolution equations (\ref{ph-sens-diag-bc}) and (\ref{ph-sens-off-bc}) for the phase-sensitive case or Eqs.\ (\ref{ph-pres-diag-bc}) and (\ref{ph-pres-off-bc}) for the phase-preserving case,
and simply add the terms due to other mechanisms of evolution (this is equivalent to interleaving the both types of evolution). As always \cite{Oksendal,Korotkov-2001}, in taking the derivative it is important to specify whether the It\^o or Stratonovich definition of the derivative is used.

\section{Moderate bandwidth}

Now let us discuss the main subject of this paper: the Bayesian formalism for continuous qubit measurement in the circuit QED setup (Fig.\ \ref{fig:schematic}) in the case when the ``bad cavity'' limit is not applicable. Therefore, we now assume that the resonator damping rate $\kappa$ is comparable to the speed of the qubit evolution due to measurement back-action, which can be characterized by the qubit ensemble dephasing $\Gamma$. In this case there is significant entanglement between the qubit and resonator, so we should consider the evolution of the combined qubit-resonator system. Also, since the typical measurement time is comparable to $\kappa^{-1}$, our formalism should focus on the transient evolution. The parameters of the measurement setup ($\varepsilon$, $\phi_{\rm a}$, $\kappa$, etc.) are allowed to change in time (this change should be much slower than $\omega_{\rm d}$ for RWA to be valid, but can be comparable to or even faster than $\kappa$).

In general, this problem is rather complicated, but we use a simplifying assumption: we assume that the qubit evolution is only due to measurement, i.e., there are no Rabi oscillations, qubit energy relaxation, etc. In practice this means that the frequency of Rabi oscillations and rate of energy relaxation should be much smaller than $\kappa$ (then the extra evolution can be added phenomenologically, as discussed in the previous section).

We will assume that the initial state of the qubit-resonator system is unentangled, and the resonator starts in a coherent state $|\alpha_{\rm in}\rangle$,
     \be
|\psi (0)\rangle = (c_0 \, |0\rangle +c_1 \, |1\rangle )\otimes
|\alpha_{\rm in}\rangle,
    \label{psi-in}\ee
where $|c_0|^2+|c_1|^2 =1$ and, for example, $|\alpha_{\rm in}\rangle$ is vacuum. Generalization to a mixed initial state,
     \be
\rho^{\rm q\&r} (0) = \rho_{\rm q, in} \otimes |\alpha_{\rm in}\rangle \langle \alpha_{\rm in}| ,
    \label{rho-in}\ee
or a slightly more general state [see Eq.\ (\ref{rho-qr-def}) below] will be straightforward. In the analysis we will use the rotating frame, corresponding to the Hamiltonian (\ref{Ham-2}). We will first discuss a simple general point of view, which describes the evolution due to measurement, then derive equations for the ensemble-averaged evolution, and then discuss the evolution during an individual realization of the measurement process. Until Sec.\ \ref{sec:imperfect} we will assume the ideal case, in particular $\kappa_{\rm out}=\kappa$ -- see Fig.\ \ref{fig:trans-refl}(a), in which we need to assume $\kappa_{\rm in}\ll \kappa_{\rm out}$.

\subsection{Idea of ``history tail''}\label{sec:history-tail}

Suppose the initial state of the qubit is $|0\rangle$. Since the measurement is of the QND type and the qubit does not evolve by itself, it will remain in the state $|0\rangle$, and since the resonator is initially in a coherent state, its state will remain to be an (evolving) coherent state (see Appendix A). Therefore, the qubit-resonator system will evolve in the rotating frame as
    \be
|\psi (t)\rangle =|0\rangle \, e^{-i\varphi_0 (t)}
|\alpha_0 (t)\rangle,
    \ee
where the coherent state amplitude $\alpha_0 (t)$ and the overall phase $\varphi_0 (t)$ evolve  according to Eqs.\ (\ref{tilde-alpha-dot}) and (\ref{phi-dot-2}) with the resonator
frequency $\omega_{\rm r}-\chi$ and drive (rotating frame) frequency $\omega_{\rm d}$,
    \begin{eqnarray}
&&    \dot{\alpha}_0 = -i (\omega_{\rm r}-\chi-\omega_{\rm d}) \,
 \alpha_0 -\frac{\kappa}{2}\,  \alpha_0 -i\varepsilon ,
    \label{alpha0-dot}\\
&&   \dot\varphi_0 = {\rm Re} (\varepsilon^* \alpha_0 ) .
    \label{phi0-dot}
    \end{eqnarray}
Here the drive amplitude $\varepsilon$ can be time-dependent and the damping $\kappa$ can in general be also time-dependent. Note that the evolution (\ref{phi0-dot}) of the overall phase is often not considered in textbooks, but for us it is very important. Derivation of Eq.\ (\ref{phi0-dot}) from the Schr\"odinger equation with the Hamiltonian (\ref{Ham-2}) is very simple.

\begin{figure}[tb]
  \centering
\includegraphics[width=9cm,trim=4cm 11.5cm 9.5cm 3cm, clip=true]{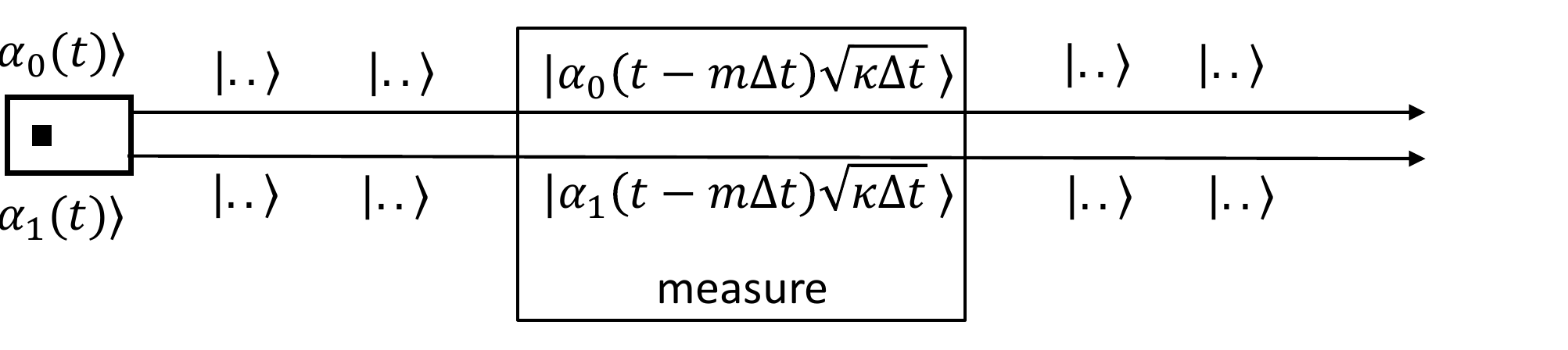}
  \caption{Illustration of the ``history tail'' idea. Since the qubit does not evolve by itself, we can think in terms of superposition [Eq.\ (\ref{evol-01})] of two evolutions of the resonator state [$|\alpha_{0} (t)\rangle$ and $|\alpha_1(t)\rangle$], which leave the ``history record'' in the form of propagating field, leaked from the resonator. We call this propagating field a ``history tail''. The quantum Bayesian formalism developed in this paper is based on measuring small pieces of the history tail in the textbook way. The corresponding state collapse leads to change of the coefficients in the superposition, which is the evolution due to measurement.}
  \label{fig:tail}
\end{figure}

Now let us consider a larger
physical system, which includes the field leaking from the resonator
to the transmission line (actually, we also necessarily need to consider
the incoming field from the transmission line, but we assume
that it is always vacuum). This larger system keeps a record of the
previous evolution in a form of a ``flying away tail'' (a propagating microwave), which we will call a ``history tail'' (Fig.\ \ref{fig:tail}). Let us divide this tail into sufficiently short pieces of duration $\Delta t$ ($\Delta t \ll \kappa^{-1}$); each of them will also be a coherent state, as follows from the property 2.6 discussed in Appendix A for a beam splitter (in our case a leaking ``mirror'' at the end of the resonator). The $m$th piece of the history tail (counting back in time) will be $|\alpha_0(t-m\,\Delta t)\sqrt{\kappa\, \Delta
t}\rangle$, which is the resonator state at time $t-m\Delta t$,
passed through the beam splitter with transmission amplitude
$\sqrt{\kappa \, \Delta t}$ [this value follows from the energy
conservation, $\sqrt{1-e^{-\kappa \, \Delta t}}\approx
\sqrt{\kappa \, \Delta t}$]. Therefore, the wavefunction, including
the history tail is
    \be
    |\Psi (t)\rangle = e^{-i\varphi_0 (t)} |0\rangle \,
|\alpha_0 (t)\rangle  \prod _m |\alpha_0 (t-m\, \Delta t)
\sqrt{\kappa \, \Delta t} \rangle .
    \label{evol-0}\ee
If $\kappa$ depends on time, then the factor $\sqrt{\kappa\, \Delta t}$ in this equation should be replaced with $\sqrt{\kappa (t-m\, \Delta t)\, \Delta t}$. Note that the coherent
states in the tail are unentangled with each other and with the
resonator state (see property 2.6 in Appendix A). Also note that the number of terms in the direct
product (\ref{evol-0}) increases with time; this seems unphysical, but it is only a
matter of notation; we can keep the number of terms constant by
adding vacuum states of the pieces of field incoming from the
transmission line.

If the qubit is initially in the state $|1\rangle$, then it remains
in $|1\rangle$, so that the wavefunction of the system including the
history tail is given by Eq.\ (\ref{evol-0}) with $|0\rangle$
replaced with $|1\rangle$, $\alpha_0$ replaced with $\alpha_1$, and
$\varphi_0$ replaced with $\varphi_1$, where $\alpha_1(t)$ and
$\varphi_1(t)$ evolve according to Eqs.\ (\ref{tilde-alpha-dot}) and
(\ref{phi-dot-2}) with the resonator frequency $\omega_{\rm
r}+\chi$,
   \begin{eqnarray}
&&    \dot{\alpha}_1 = -i (\omega_{\rm r}+\chi-\omega_{\rm d}) \,
 \alpha_1 -\frac{\kappa}{2}\,  \alpha_1 -i\varepsilon ,
    \label{alpha1-dot}\\
&&   \dot\varphi_1 = {\rm Re} (\varepsilon^* \alpha_1 ) .
    \label{phi1-dot}
    \end{eqnarray}

Now let us make a simple but very important logical step in the
derivation. Since the qubit does not evolve by itself, we can consider
two evolutions at the same time: for the qubit in the state
$|0\rangle$ and in the state $|1\rangle$, so that the coefficients
in the initial superposition (\ref{psi-in}) do not change in time (Fig.\ \ref{fig:tail}). This follows from the
general linearity of quantum mechanics and somewhat resembles the
``many worlds'' interpretation. Therefore, for the initial state (\ref{psi-in}) of
the qubit-resonator system we obtain the
wavefunction evolution (including the flying away history tail)
    \begin{eqnarray}
&& \hspace{-0.4cm}   |\Psi (t)\rangle = c_0 \, e^{-i\varphi_0 (t)}
|0\rangle \, |\alpha_0 (t)\rangle \prod _m |\alpha_0 (t-m\, \Delta
t) \sqrt{\kappa \, \Delta t} \rangle
    \nonumber \\
&& \hspace{0.1cm} + c_1 \, e^{-i\varphi_1 (t)} |1\rangle \,
 |\alpha_1 (t)\rangle  \prod _m |\alpha_1 (t-m\, \Delta t)
\sqrt{\kappa \, \Delta t} \rangle , \qquad
    \label{evol-01}\end{eqnarray}
where $c_0$ and $c_1$ are constant in time, while $\alpha_{0}$, $\varphi_0$, $\alpha_1$, and
$\varphi_1$ evolve according to Eqs.\ (\ref{alpha0-dot}), (\ref{phi0-dot}), (\ref{alpha1-dot}) and (\ref{phi1-dot}), starting with $\alpha_0(0)=\alpha_1(0)=\alpha_{\rm in}$ and
$\varphi_0(0)=\varphi_1(0)=0$.

The approach to qubit measurement via the wavefunction evolution
in Eq.\ (\ref{evol-01}) is physically transparent and quite
powerful. In particular, it will easily allow us to describe evolution
of the qubit-resonator system in the process of measurement by
applying the textbook collapse postulate to measurement of the tail pieces (Fig.\ \ref{fig:tail}).
However, let us first discuss the ensemble-averaged evolution.

    \subsection{Ensemble-averaged evolution}\label{sec:ensemble-aver}

If the result of the tail measurement is not taken into account, we
need to average the quantum state over all possible measurement
results, which is equivalent to tracing the state (\ref{evol-01})
over the tail. This leads to a density operator in the
qubit-resonator Hilbert space, $\rho^{\rm
q\&r}=\rho_{00}^{\rm  q\&r}+\rho_{11}^{\rm
q\&r}+\rho_{01}^{\rm  q\&r}+\rho_{10}^{\rm  q\&r}$, in
which the parts diagonal in the qubit subspace are
    \begin{eqnarray}
 \rho_{00}^{\rm  q\&r} (t) = |c_0|^2 \,|0\rangle \langle 0 | \otimes |\alpha_0
 (t)\rangle \langle \alpha_0 (t)| ,
    \label{rho-00-qr}\\
 \rho_{11}^{\rm  q\&r} (t) = |c_1|^2 \, |1\rangle \langle 1 | \otimes |\alpha_1
 (t)\rangle \langle \alpha_1 (t)|,
    \label{rho-11-qr}\end{eqnarray}
while the off-diagonal parts contain the inner product of the tails
for the two different evolutions,
    \begin{eqnarray}
&&  \hspace{-0.5cm} \rho_{10}^{\rm q\&r} (t) = c_1 c_0^* \,
e^{-i[\varphi_1(t)-\varphi_0(t)]}\, |1\rangle \langle 0 | \otimes
|\alpha_1
 (t)\rangle \langle \alpha_0 (t)|
    \nonumber \\
&& \hspace{0.1cm} \times  \prod_m \langle \alpha_0 (t-m\, \Delta t)
\sqrt{\kappa \, \Delta t} \, |\, \alpha_1 (t-m\, \Delta t)
\sqrt{\kappa \, \Delta t} \rangle ,  \qquad
    \label{rho-10-qr-1}\end{eqnarray}
and similarly for $\rho_{01}^{\rm q\&r} =(\rho_{10}^{\rm
 q\&r})^\dagger$.
 The inner product for each time piece $\Delta t$
is given by Eq.\ (\ref{inner-product}) in Appendix A, so that we find
    \begin{eqnarray}
&&  \rho_{10}^{\rm  q\&r} (t) = c_1 c_0^* \,
e^{-i[\varphi_1(t)-\varphi_0(t)]}\, |1\rangle \langle 0 | \otimes
|\alpha_1
 (t)\rangle \langle \alpha_0 (t)| \,
    \nonumber \\
&& \hspace{1cm} \times \exp \left( - \int_0^t
\frac{\kappa}{2}\,|\alpha_1(t')-\alpha_0(t')|^2 \, dt' \right)
    \nonumber \\
&& \hspace{1cm} \times  \exp \left( -i \int_0^t \kappa\, {\rm Im}\,
[\alpha_1^*(t') \, \alpha_0 (t')]\,  dt' \right) .  \qquad
    \label{rho-10-qr}\end{eqnarray}

In this equation the second line obviously describes dephasing with the rate
    \be
    \Gamma_{\rm d}(t) =\frac{\kappa}{2}\,
    |\alpha_1(t)-\alpha_0(t)|^2,
    \label{Gamma-d}\ee
which is directly related to distinguishability of the field emitted
into the transmission line and therefore to the information that
can in principle be obtained from measurement. The third line in Eq.\
(\ref{rho-10-qr}) is the changing phase factor which can be ascribed
to the shift of the qubit frequency in the process of measurement,
    \be
    \delta \omega_{\rm q,1}(t) = \kappa \,  {\rm Im}\,
[\alpha_1^* (t) \, \alpha_0 (t)].
    \label{delta-omega-1}\ee
However, a similar frequency shift comes from the term
$e^{-i[\varphi_1(t)-\varphi_0(t)]}$ in Eq.\ (\ref{rho-10-qr}); using Eqs.\
(\ref{phi0-dot}) and (\ref{phi1-dot}), we obtain the corresponding value
    \be
 \delta \omega_{\rm q,2}(t) = {\rm Re}\, \{ \varepsilon^*(t)\,
 [\alpha_1 (t)- \alpha_0 (t)]\},
    \label{delta-omega-2}\ee
so that the total frequency shift of the qubit (which can be called the ac Stark shift) is
    \be
\delta \omega_{\rm q,s} = \delta \omega_{\rm q,1}+ \delta
\omega_{\rm q,2}.
    \label{delta-omega-12}\ee

Thus, the ensemble-averaged evolution of the qubit-resonator state can be described (neglecting the overall phase) by the wavefunction
    \be
    |\psi (t)\rangle = c_0\, |0\rangle \, |\alpha_0 (t)\rangle +
e^{-i \int_0^t \delta\omega_{\rm q,s}(t')\, dt' } c_1\, |1\rangle
\, |\alpha_1 (t)\rangle,
    \ee
subjected to dephasing $\Gamma_{\rm d}(t)$ between the
two components.

If we also want to trace the state over the resonator, then we have
an additional inner product $\langle \alpha_0
(t)\,|\,\alpha_1(t)\rangle$, which changes the dephasing rate
(\ref{Gamma-d}) by
    \be
  \Delta \Gamma (t) = \frac{d}{dt}
  \left[ \frac{1}{2}\,|\alpha_1(t)-\alpha_0(t)|^2\right]
    \label{Delta-Gamma}\ee
(this change can be positive or negative) and introduces additional
contribution to the qubit frequency shift,
    \be
    \delta\omega_{\rm q,3}(t)= \frac{d}{dt} \, {\rm Im}
    [\alpha_1^*(t)\,\alpha_0(t)] ,
    \label{delta-omega-3}\ee
as follows from Eq.\ (\ref{inner-product}).
 The {\it qubit-only} density matrix elements then become
    \begin{eqnarray}
&&    \rho_{00}^{\rm q}(t)= |c_0|^2, \,\,\,\,\,   \rho_{11}^{\rm q}(t)= |c_1|^2,
        \\
&&   \rho_{10}^{\rm q}(t)= c_1c_0^* \, \exp \bigg[ -\int_0^t
[\Gamma_{\rm d}(t')+\Delta\Gamma (t')] \, dt'\bigg]
    \nonumber \\
&& \hspace{0.9cm} \times \,  \exp \bigg[ -i\int_0^t [\delta\omega_{\rm
q,s}(t')+\delta\omega_{\rm q,3}(t')] \, dt' \bigg] . \qquad
    \end{eqnarray}
[Note that $\rho_{ij}^{\rm q}$ are numbers, while $\rho_{ij}^{\rm q\& r}$ in Eqs.\ (\ref{rho-00-qr})--(\ref{rho-10-qr}) are operators.]
Using Eqs.\ (\ref{alpha0-dot}) and (\ref{alpha1-dot}), it is easy to show that
    \begin{eqnarray}
&&    \Gamma_{\rm d}(t)+\Delta\Gamma (t)= 2\chi \, {\rm Im}
    [\alpha_1^* (t) \, \alpha_0(t)],
    \label{Gamma-sum}\\
&& \delta\omega_{\rm q,s}(t)+\delta\omega_{\rm q,3}(t)= 2\chi \, {\rm Re}
    [\alpha_1^* (t) \, \alpha_0(t)] , \qquad
    \label{delta-omega-sum}\end{eqnarray}
which coincide with the results of Refs.\ \cite{Gambetta-2006,Gambetta-2008}
for the qubit dephasing and ac Stark shift.

Note that the results (\ref{Gamma-d}) and (\ref{Gamma-sum}) for the
dephasing rate coincide in the steady state (because then
$\Delta\Gamma=0$), but they are different during the transient
evolution. The rate (\ref{Gamma-d}) reflects the information loss
due to emitted field, while the rate (\ref{Gamma-sum}) also includes
the effect from changing entanglement between the qubit and the resonator. Similarly,
the results (\ref{delta-omega-12}) and (\ref{delta-omega-sum}) for
the ac Stark shift coincide in the steady state (then $\delta
\omega_{\rm q,3}=0$), but differ during transients. Equation
(\ref{delta-omega-12}) is applicable to the entangled
qubit-resonator state, while Eq.\ (\ref{delta-omega-sum}) assumes
tracing over the resonator state.

Also note that all these results for the dephasing rate and ac Stark
shift are applicable only in the case of a non-evolving qubit (i.e., when the evolution is only due to measurement).
Therefore, they are applicable to the Ramsey sequence (with short pulses), but, strictly
speaking, not applicable to Rabi oscillations, spectroscopic
measurement of the ac Stark shift, etc. For the echo sequence our results are not applicable directly, but the exact results can still
be easily obtained using the same derivation (assuming sufficiently
short pulses applied to the qubit).

In the ``bad cavity'' limit we can neglect the transients and use the steady-state values $\alpha_{1,\rm st}$ and $\alpha_{0,\rm st}$. If additionally $|\chi |\ll \kappa$, then Eqs.\ (\ref{Gamma-sum}) and (\ref{delta-omega-sum}) reduce to Eqs.\ (\ref{Gamma-bc2}) and (\ref{Stark-bc}). If $|\chi |\agt \kappa$, then for the qubit dephasing and ac Stark shift in Sec.\ III  we need to use steady-state versions of Eqs.\ (\ref{Gamma-sum}) and (\ref{delta-omega-sum}) or, equivalently, Eqs.\ (\ref{Gamma-d}) and (\ref{delta-omega-12}).

\subsection{Main idea for the state update}\label{sec:main-idea}

To describe an individual measurement realization with a random result and evolution depending on this result, we will measure the pieces of the ``history tail'' in Eq.\ (\ref{evol-01}) -- see Fig.\ \ref{fig:tail}. Note that each piece can be measured in a different way, so the measurement properties can be changing in time. Moreover, in general the sequence of measurement of the pieces can also be arbitrary (e.g., in a ``delayed choice'' experiment).
We are interested in describing homodyne measurement with a phase-sensitive or phase-preserving amplifier.

Let us start with describing an {\it ideal} (with perfect quantum efficiency) {\it phase-sensitive} homodyne measurement. We will use the following physical model to describe such a measurement:
   \begin{enumerate}
\item
A large coherent-state field $\alpha_{\rm p}$ ($|\alpha_{\rm p}| \gg 1$) from a pump is added to the piece of the tail.
\item
The number of photons $n$ is measured in the resulting state.
\item
For a particular random $n$ obtained in this measurement, the wavefunction is collapsed in the standard textbook way.
  \end{enumerate}

This procedure describes well the optical homodyne measurement (note that the photon number does not actually need to be resolved with single-photon precision since the fluctuations are significant). It is also similar to what is done experimentally in a phase-sensitive superconducting parametric amplifier. For example, in Refs.\ \cite{Vijay-2011,Vijay-2012} the phase-sensitive parametric amplifier works by adding a pump microwave to the microwave leaked from the resonator using a directional coupler (a microwave analog of a beam splitter). Then the resulting microwave is sent to a nonlinear oscillator, whose frequency depends on the oscillation amplitude. This frequency change is then sensed via the corresponding phase change at the mixer. Thus we measure the power of the pump with added signal, i.e., within the time interval $\Delta t$ we essentially measure the corresponding number of photons $n$ (again, single-photon precision in measuring $n$ is not needed). In a more complicated case of sideband pumping (double-pumping) of the parametric amplifier \cite{Murch-2013,Roch-2014,Weber-2014}, the added resonant pump wave is modulated in amplitude; however, the general principle remains practically the same. The case of a parametric pumping at the doubled frequency is different, but it is still practically equivalent to the measurement described by our model.

Note that in order to add the field $\alpha_{\rm p}$, we need a beam splitter (directional coupler) which almost fully passes the signal, so the applied pump field should be much larger than the already large field $\alpha_{\rm p}$. Also note that a small part of the signal in this case will be lost, so that perfect quantum efficiency is impossible. However, we will not consider these details, and will also not consider ways to go around these problems (e.g., by using a balanced homodyne detection).

It is rather simple to analyze the measurement using our model. For describing measurement of $m$th piece of the tail, let us rewrite Eq.\ (\ref{evol-01}) as
    \begin{eqnarray}
&&    |\Psi \rangle = c_0 |\psi_0\rangle |\alpha_{0,\rm t}\rangle + c_1 |\psi_1\rangle |\alpha_{1,\rm t}\rangle ,
        \label{Psi-1}\\
&&    |\alpha_{j,\rm t}\rangle =|\alpha_j (t-m\, \Delta t) \sqrt{\kappa \, \Delta t} \rangle, \,\,\,\,\,  j=0,1,
    \\
&&        |\psi_j\rangle = e^{-i\varphi_j} |j\rangle |\alpha_j\rangle \prod_{k\neq m} |\alpha_j (t-k\Delta t)\sqrt{\kappa\,\Delta t}\rangle , \qquad
    \label{psi-j}
    \end{eqnarray}
where $|\alpha_{j,\rm t}\rangle$ is the measured $m$th piece of the tail, while remaining terms in Eq.\ (\ref{evol-01}) are denoted as the normalized wavefunctions $|\psi_j\rangle$. [Actually, if we measure each piece of the tail immediately as it emerges, then it is sufficient to consider $|\psi_j\rangle = e^{-i\varphi_j}|j\rangle |\alpha_j\rangle$, which contain only the qubit and resonator states and do not contain unmeasured pieces of the tail. However, with Eqs.\ (\ref{Psi-1})--(\ref{psi-j}) we can in general consider a ``delayed choice'' version of the measurement.]

After the first step in the procedure (addition of the pump field
$\alpha_{\rm p}$),  the tail pieces $|\alpha_{j,\rm
t}\rangle$ become $|\alpha_{\rm p}+\alpha_{j,\rm t}\rangle \exp
[-i{\rm Im}(\alpha_{\rm p}^* \alpha_{j,\rm t})]$ [see Eq.\
(\ref{displacement-comp}) for displacement by operator
$\hat{D}(\alpha_{\rm p})$], therefore,  the state (\ref{Psi-1})
becomes
    \begin{eqnarray}
&& |\Psi \rangle = c_0 |\psi_0\rangle\, e^{-i\, {\rm Im}(\alpha_{\rm p}^* \alpha_{0,\rm t})} |\alpha_{\rm p}+\alpha_{0,\rm t} \rangle
    \nonumber \\
&& \hspace{0.7cm} + c_1 |\psi_1\rangle
\, e^{-i\,{\rm Im}(\alpha_{\rm p}^* \alpha_{1,\rm t})} |\alpha_{\rm p} + \alpha_{1,\rm t} \rangle  .
        \label{Psi-2}
    \end{eqnarray}

At the second step of our procedure we need to measure the number of
photons $n$ in the pump-plus-piece-of-tail part of the state
(\ref{Psi-2}). The probability distribution for obtaining a
particular $n$ is
    \begin{eqnarray}
&&    P(n)= |c_0|^2 e^{-|\alpha_{\rm p}+\alpha_{0, \rm t}|^2}
    |\alpha_{\rm p}+\alpha_{0, \rm t}|^{2n}/n!
    \nonumber \\
&& \hspace{1cm}    +
    |c_1|^2 e^{-|\alpha_{\rm p}+\alpha_{1, \rm t}|^2}
    |\alpha_{\rm p}+\alpha_{1, \rm t}|^{2n}/n!  ,
    \label{P(n)-1}\end{eqnarray}
as follows from Eqs.\ (\ref{Psi-2}) and (\ref{prop-2-1}). This
distribution is normalized,  $\sum_n P(n)=1$,   because
$|c_0|^2+|c_1|^2=1$.

The third step of the procedure is the orthodox collapse of the
state (\ref{Psi-2}) onto the particular (random) measurement result
$n$. This means that instead of the states $|\alpha_{\rm
p}+\alpha_{0,\rm t}\rangle$ in Eq.\ (\ref{Psi-2}), we pick only the
amplitude corresponding to $|n\rangle$, and then renormalize the
wavefunction (\ref{Psi-2}), so that it becomes [see Eq.\ (\ref{alpha-def-2})]
    \begin{eqnarray}
&&    |\tilde \Psi\rangle = \left( \tilde c_0 |\psi_0\rangle
    +\tilde c_1 |\psi_1\rangle \right) |n\rangle ,
    \label{tilde-Psi}\\
&&    \tilde{c}_j=\frac{c_j \,
    e^{-i\, {\rm Im}(\alpha_{\rm p}^* \alpha_{j,\rm t})}
e^{-\frac{1}{2}|\alpha_{\rm p}+\alpha_{j, \rm t}|^2}
 (\alpha_{\rm p}+\alpha_{j, \rm t})^n
    }{\rm Norm} , \qquad
    \label{tilde-c-1}\end{eqnarray}
where the normalization $\rm Norm$ ensures that $|\tilde c_0|^2
+|\tilde c_1|^2=1$. Note that the the overall phase of $|\tilde \Psi\rangle$ is not important.

As we see, the ``quantum back-action'' due to the collapse changes
the amplitudes of the pre-measured state (\ref{Psi-1}):
$c_0\rightarrow \tilde c_0$ and $c_1\rightarrow \tilde c_1$. This is the main idea for
the description of the evolution due to measurement in the quantum
Bayesian formalism. The procedure can be applied to measurement
of other pieces of the ``history tail'' in the same way.

\subsection{Gaussian approximation}\label{sec:Gaussian}

Let us transform Eqs.\ (\ref{P(n)-1}) and (\ref{tilde-c-1}) into a more
useful form, using the assumption of a large pump
amplitude, $|\alpha_{\rm p}|\gg 1$ and $|\alpha_{\rm p}|\gg
|\alpha_{j,\rm t}|$. In this case we can use the Gaussian
approximation for the coherent states $|\alpha_{\rm p}+\alpha_{j,\rm
t}\rangle$ [see Eq.\ (\ref{alpha-def-2})],
        \begin{eqnarray}
&&  \hspace{-0.0cm}   |\alpha_{\rm p}+\alpha_{j,\rm t}\rangle \approx \sum_{n=0}^\infty \sqrt{\frac{\exp [-(n-\bar{n}_j)^2/2\sigma^2]}{\sqrt{2\pi\sigma^2}}
     }
     \nonumber \\
&& \hspace{2.6cm}
     \times \exp [i n\, {\rm arg} (\alpha_{\rm p}+\alpha_{j,\rm t}) ]\, |n\rangle, \qquad  \label{alpha-Gauss}\\
&& \hspace{-0.0cm} \bar{n}_j = |\alpha_{\rm p}|^2 + 2\, {\rm Re} (\alpha_{\rm p}^*\alpha_{j,\rm t})+ |\alpha_{j,\rm t}|^2
    \nonumber \\ 
&& \hspace{0.5cm} \approx |\alpha_{\rm p}|^2 + 2\, {\rm Re} (\alpha_{\rm p}^*\alpha_{j,\rm t}) , 
    \label{n-bar-01} \\
&&     \sigma=|\alpha_{\rm p}|\gg 1,
    \end{eqnarray}
where $\bar{n}_j$ is the average number of photons for the state  $|\alpha_{\rm p}+\alpha_{j,\rm t}\rangle$ (we can neglect the last term $|\alpha_{j,\rm t}|^2$ for $\bar{n}_j$ since $|\alpha_{j,\rm t}|\ll |\alpha_{\rm p}|$) and $\sigma=\sqrt{\bar{n}}$ is the standard deviation. Note that we use the same $\sigma$ for both states because $|\bar{n}_1-\bar{n}_0|\ll \bar{n}_j$. Also note that for exact normalization of the Gaussian state (\ref{alpha-Gauss}) at finite $|\alpha_{\rm p}|$ the denominator  $\sqrt{2\pi \sigma^2}$ should be slightly changed; however, this is not important for the derivation.

The probability distribution (\ref{P(n)-1}) for measuring $n$
photons in this case becomes
    \be
    P(n)= \frac{|c_0|^2\, e^{-(n-\bar{n}_0)^2/2\sigma^2}}{\sqrt{2\pi\sigma^2}} +
 \frac{|c_1|^2\, e^{-(n-\bar{n}_1)^2/2\sigma^2}}{\sqrt{2\pi\sigma^2}},
    \label{P(n)-Gauss}\ee
and the updated amplitudes $\tilde c_0$ and $\tilde c_1$ given by Eq.\
(\ref{tilde-c-1}) become
    \be
    \tilde{c}_j=\frac{c_j \, e^{-i\, {\rm Im(\alpha_{\rm p}^* \alpha_{j,\rm t})}}
    e^{-(n-\bar{n}_j)^2/4\sigma^2} e^{in \,
    {\rm arg}(\alpha_{\rm p} +\alpha_{j,\rm t})}}{\rm Norm} .
    \label{tilde-c-2}\ee
We see that if the measurement result $n$ is closer to $\bar{n}_0$ than to
$\bar{n}_1$, then the amplitude $\tilde{c}_0$ increases (by absolute value)
in this update. This is the expected feature of the quantum Bayesian formalism: if the measurement result is more consistent with the qubit state $|0\rangle$, then the amplitude of this state increases.

To simplify the phase factor in Eq.\ (\ref{tilde-c-2}), let us write
$e^{i n\, {\rm arg} (\alpha_{\rm p}+\alpha_{j,\rm t})}$ as $e^{i n\,
{\rm arg} (\alpha_{\rm p})} e^{in\, {\rm arg} (1 +\alpha_{j,\rm
t}/\alpha_{\rm p})}$, and then expand ${\rm arg} (1 +\alpha_{j,\rm
t}/\alpha_{\rm p})$ to the second order, so that
    \be
    e^{i n\, {\rm arg} (\alpha_{\rm p}+\alpha_{j,\rm t})} = e^{i n\, {\rm arg} (\alpha_{\rm p})} e^{in\, {\rm Im}(\alpha_{j,\rm t}/\alpha_{\rm p})
    [1-{\rm Re}(\alpha_{j,\rm t}/\alpha_{\rm p})]}
    \label{phase-1}\ee
(we need the second order because $\alpha_{j,\rm t}\propto
\sqrt{\Delta t}$ and we wish to keep the terms linear in $\Delta
t$). The $j$-independent phase factor $e^{in\, {\rm arg}(\alpha_{\rm p})}$
can be ignored as an overall phase. In the remaining phase in Eq.\
(\ref{phase-1}) let us represent $n$ as $n=\bar{n}_{\rm c}+
(n-\bar{n}_{\rm c})$ with the center point
    \be
\bar{n}_{\rm c}\equiv (\bar{n}_0+\bar{n}_1)/2.
    \ee
From Eq.\ (\ref{n-bar-01}), neglecting the term $|\alpha_{j,\rm
t}|^2$, we find $\bar{n}_{\rm c}=|\alpha_{\rm p}|^2 + {\rm
Re}[\alpha_{\rm p}^*(\alpha_{0,\rm t}+\alpha_{1,\rm t})]$.

If $n=\bar{n}_{\rm c}$, then the phase in Eq.\ (\ref{tilde-c-2})
(neglecting the overall phase) is $-{\rm Im}(\alpha_{\rm p}^*\alpha_{j,\rm t})+\bar{n}_{\rm c}\, {\rm
Im}(\alpha_{j,\rm t}/\alpha_{\rm p})[1-{\rm Re}(\alpha_{j,\rm
t}/\alpha_{\rm p})]$,
which can be written (neglecting the terms of order $\alpha_{\rm
p}^{-1}$) as ${\rm Im}(\alpha_{j,\rm t}/\alpha_{\rm p})\,{\rm
Re}[(\alpha_{0,\rm t}+\alpha_{1,\rm t}-\alpha_{j,\rm t})\alpha_{\rm
p}^*]$. Moving the phase difference to $j=1$ (i.e., considering the
phase for $j=0$ as an unimportant overall phase), we find that the phase evolution in Eq.\ (\ref{tilde-c-2}) can be described by multiplying $c_1$ by $e^{-i\,{\rm
Im}(\alpha_{1,\rm t}^*\alpha_{0,\rm t})}$. This is exactly what we
would expect from the phase of the inner product $\langle
\alpha_{0,\rm t}|\alpha_{1,\rm t}\rangle$, and it is fully consistent with
the result (\ref{delta-omega-1}) for the ac Stark shift contribution. Thus,
for $n=\bar{n}_{\rm c}$ the phase shift produced by the collapse is
the same as the ensemble-averaged phase shift.

When $n\ne \bar{n}_{\rm c}$, there is an additional phase factor
$e^{i (n-\bar{n}_{\rm c})\, {\rm Im} (\alpha_{j,\rm t}/\alpha_{\rm
p})}$ in Eqs.\ (\ref{tilde-c-2}) and (\ref{phase-1}) [we now use
$1-{\rm Re}(\alpha_{j,\rm t}/\alpha_{\rm p})\approx 1$, neglecting a phase correction of order $\alpha_{\rm
p}^{-1}$]. Moving the phase difference to $j=1$, we find that $c_1$
should be additionally multiplied by $e^{-i(n-\bar{n}_{\rm c})\,{\rm
Im}[(\alpha_{0,\rm t}-\alpha_{1,\rm t})/\alpha_{\rm p}]}$.

Thus, the evolution due to measurement [see Eqs.\ (\ref{Psi-1}),
(\ref{tilde-Psi}), and (\ref{tilde-c-2})] can be described as
    \begin{eqnarray}
&& \tilde{c}_0=\frac{c_0\,\exp [-(n-\bar{n}_0)^2/4\sigma^2]}{\rm
Norm} , \qquad
    \label{tilde-c0}\\
&&    \tilde{c}_1= \frac{c_1\,\exp [-(n-\bar{n}_1)^2/4\sigma^2]}{\rm
Norm} \, e^{-i\,\Delta \varphi} ,
    \label{tilde-c1}\\
&&    {\rm Norm} = \sqrt{\sum\nolimits_{j=0,1} |c_j|^2
    \exp [-(n-\bar{n}_j)^2/2\sigma^2] } ,
    \\
&&    \Delta\varphi = -\frac{n-\bar{n}_{\rm c}
}{\sigma}\,\sqrt{\kappa \Delta t} \,\, {\rm Im}\{
[\alpha_1(t_m)-\alpha_0(t_m)]\,e^{-i\phi_{\rm a}}\}
    \nonumber \\
&&\hspace{0.8cm}  +\kappa \Delta t\, {\rm Im}[\alpha_{1
}^*(t_m)\,\alpha_{0}(t_m)], \,\,\,\
    \label{delta-phi} \\
&&  \phi_{\rm a} =  \phi_{\rm p}={\rm arg} (\alpha_{\rm p}) ,
    \end{eqnarray}
where $t_m=t-m\Delta t$ is the time moment when the measured piece
of the ``history tail'' leaked from the resonator, and $\phi_{\rm a}$ is the phase of the pump, which determines the amplified quadrature. We emphasize that the measured piece becomes unentangled with the rest of the wavefunction [see Eq.\ (\ref{tilde-Psi})] and therefore can be disregarded when the measurement of the next piece is analyzed.

Note that the pump phase $\phi_{\rm a}$ affects the response [see Eq.\ (\ref{n-bar-01})]
    \be
\bar{n}_1-\bar{n}_0 = 2\sigma\sqrt{\kappa \Delta t} \,\,  {\rm Re}
\{[\alpha_1(t_m)-\alpha_0(t_m)]\, e^{-i\phi_{\rm a}} \}
    \ee
and also affects the phase shift $\Delta \varphi$ in Eq.\
(\ref{delta-phi}). Thus, the choice of the measured quadrature affects evolution of the system (as in the ``bad cavity'' case \cite{Korotkov-2011}). However, it is simple to show that the state update (\ref{tilde-c0})--(\ref{delta-phi}) averaged over the measurement result (\ref{P(n)-Gauss}) does not depend on the choice of $\phi_{\rm a}$.

\subsection{Continuous phase-sensitive measurement}\label{sec:ph-sens}

The formalism developed in Secs.\ \ref{sec:main-idea} and \ref{sec:Gaussian} allows us to consider measurement of the ``history tail'' pieces in an arbitrary sequence and thus to describe various ``delayed choice'' experiments. However, usually this is not needed, and we can assume measurement of the pieces as soon as they leak from the resonator. In this case it is sufficient to describe the system by an entangled qubit-resonator wavefunction (we still consider an ideal case)
    \be
    |\psi (t)\rangle = c_0(t)\, |0\rangle \, |\alpha_0(t)\rangle + c_1(t)\, |1\rangle \, |\alpha_1(t)\rangle ,
    \label{psi-cont-meas}\ee
with the coefficients $c_0(t)$ and $c_1(t)$ evolving in time due to measurement.
Then the evolution equations are essentially the
same as Eqs.\ (\ref{tilde-c0})--(\ref{delta-phi}), but now there is
no delay in measurement, and we have included the phase factors
$e^{-i\varphi_0}$ and $e^{-i\varphi_1}$ in Eq.\ (\ref{evol-01}) into
the coefficients $c_0$ and $c_1$ in Eq.\ (\ref{psi-cont-meas}) [actually, we include the phase difference $e^{-i(\varphi_1-\varphi_0)}$ into $c_1$, neglecting the overall phase $e^{-i\varphi_0}$].
Also, instead of the measured number of photons $n$, let us introduce the output signal $I_{\rm m}
= n/\Delta t$. Similarly, $I_0 = \bar{n}_0/\Delta t$ and $I_1 =
\bar{n}_1/\Delta t$ are the corresponding average values, and $D = (\sigma /\Delta t)^2$ is the variance of $I_{\rm m}$. Then Eqs.\ (\ref{tilde-c0})--(\ref{delta-phi}) can
be rewritten as
     \begin{eqnarray}
&& c_0(t+\Delta t) = \frac{c_0(t)\,\exp [-(I_{\rm m}-I_0)^2/4D]}{\rm
Norm} , \qquad
    \label{c0-1}\\
&&    c_1(t+\Delta t)= \frac{c_1 (t)\,\exp [-(I_{\rm m}-I_1)^2/4D]}{\rm
Norm} \, e^{-i\,\Delta \varphi} , \quad \,\,\,
    \label{c1-1}\\
&&    {\rm Norm} = \sqrt{\sum\nolimits_{j=0,1} |c_j(t)|^2
    \exp [-(I_{\rm m}-I_j)^2/2D] } , \qquad
    \\
    &&    \Delta\varphi = -
 \frac{I_{\rm m}-(I_0+I_1)/2}{\sqrt{D}}\,\sqrt{\kappa \Delta t} \,\,
 {\rm Im} [(\alpha_1-\alpha_0)\,e^{-i\phi_{\rm a}}]
    \nonumber \\
 && \hspace{1.0cm}
+ \, \delta \omega_{\rm q,s} \, \Delta t ,
    \label{delta-phi-1}\end{eqnarray}
where the qubit ac Stark shift $\delta \omega_{\rm q,s}$ due to leaking field  is given by Eqs.\ (\ref{delta-omega-1})--(\ref{delta-omega-12}).

Note that Eqs.\ (\ref{c0-1})--(\ref{delta-phi-1}) do not change if we multiply $I_{\rm m}$, $I_0$, $I_1$, and $\sqrt{D}$ by an arbitrary factor. Therefore, we can consider them just as experimental output signals (in arbitrary units), so that
    \be
   I_{\rm m}=\frac{1}{\Delta t}\int_t^{t+\Delta t} I(t')\, dt'
    \ee
for a continuous measurement output $I(t)$, and the variance $D$ is related to the single-sided spectral density $S_I$ of the output signal as
    \be
    D= \frac{S_I}{2\Delta t} .
    \label{D-def}\ee

Now let us introduce the angle difference $\phi_{\rm d}$ between the amplified quadrature along $\alpha_{\rm p}$ and the ``information-carrying'' quadrature along $\alpha_1-\alpha_0$,
    \be
    \phi_{\rm d} = \phi_{\rm a} - {\rm arg}[\alpha_1 (t) -\alpha_0(t)] ,
    \ee
and also introduce the maximum response $\Delta I_{\rm max}>0$,
which {\it would} correspond to $\phi_{\rm d}=0$, so that
    \be
    \Delta I = I_1 -I_0 =\Delta I_{\rm max} \cos(\phi_{\rm d}).
    \ee
Then we can write the factor $\sqrt{\kappa\Delta t}\,{\rm Im}
[(\alpha_1-\alpha_0)\,e^{-i\phi_{\rm a}}]$ in Eq.\ (\ref{delta-phi-1}) as
$-\sin (\phi_{\rm d}) \, \Delta I_{\rm max}/(2\sqrt{D})$.

Also counting the measurement signal from the central point
$(I_0+I_1)/2$,
    \be
    \tilde I_{\rm m} = I_{\rm m} - \frac{I_0+I_1}{2},
    \ee
we can rewrite evolution equations (\ref{c0-1})--(\ref{delta-phi-1})
as
     \begin{eqnarray}
&& c_0(t+\Delta t) = \frac{c_0(t)\,e^{-\tilde{I}_{\rm m}
\cos(\phi_{\rm d}) \,\Delta I_{\rm max}/4D} }{\rm Norm} , \qquad
    \label{c0-2}\\
&&    c_1(t+\Delta t)= \frac{c_1 (t) \, e^{\tilde{I}_{\rm m}
\cos(\phi_{\rm d}) \,\Delta I_{\rm max}/4D} }{\rm Norm} \,
e^{-i\,\Delta \varphi} , \quad \,\,\,
    \label{c1-2}\\
    &&    \Delta\varphi =
 \frac{\tilde{I}_{\rm m}\sin(\phi_{\rm d})\, \Delta I_{\rm max}}{2D}
+  \delta \omega_{\rm q,s} \Delta t ,
    \label{delta-phi-2}\end{eqnarray}
where ${\rm Norm}$ ensures $|c_0(t+\Delta t)|^2+|c_1(t+\Delta
t)|^2=1$.

Note that for short $\Delta t$ the variance $D$ of the noisy signal
$\tilde{I}_{\rm m}$ is much larger than $(\Delta I_{\rm max})^2$, and then
$|\tilde{I}_{\rm m}\Delta I_{\rm max}|\ll D$, so that the change of $c_0$ and $c_1$ is small. However, from the structure of Eqs.\ (\ref{c0-2})--(\ref{delta-phi-2}) it is easy to see that they {\it remain valid for an arbitrary long $\Delta t$} if $\Delta I_{\rm max}$, $\phi_{\rm d}$, $\delta\omega_{\rm q,s}$ and noise $S_I$ do not change with time. Therefore, the practical upper limit for the time step $\Delta t$ (e.g., in numerical simulations) is determined by transients, which change the resonator states $\alpha_0(t)$ and $\alpha_1(t)$, and by possible changes of the amplified quadrature phase $\phi_{\rm a}$ .

\subsection{Phase-sensitive measurement with imperfect quantum efficiency}\label{sec:imperfect}

So far we considered an ideal phase-sensitive measurement, so that the evolution description using a wavefunction was sufficient. To describe a measurement with imperfect efficiency, we need to use the language of density matrices (we still assume that the qubit evolves only due to measurement). Then instead of Eq.\ (\ref{psi-cont-meas}), the evolution of the entangled qubit-resonator system is described by the density operator
    \be
    \rho^{\rm q\& r}(t)= \sum_{j,j'=0,1} \rho_{jj'}(t)\, |j\rangle\langle j'| \otimes |\alpha_j (t)\rangle\langle \alpha_{j'}(t)| ,
    \label{rho-qr-def}\ee
where $\alpha_0(t)$ and $\alpha_1(t)$ are given by Eqs.\ (\ref{alpha0-dot}) and (\ref{alpha1-dot}).
We emphasize that the matrix elements  $\rho_{jj'}(t)$ describe the entangled qubit-resonator state, not only the qubit state. (Note that using the form (\ref{rho-qr-def}) for the qubit-resonator state is equivalent to the polaron-frame approximation used in the theory of quantum trajectories \cite{Gambetta-2008}.) In the ideal case the evolution of the matrix elements $\rho_{jj'}$  can be obtained by converting Eqs.\ (\ref{c0-2})--(\ref{delta-phi-2}) into the language of density matrices,
     \begin{eqnarray}
&& \frac{\rho_{11}(t+\Delta t)}{\rho_{00}(t+\Delta t)} = \frac{\rho_{11}(t)}{\rho_{00}(t)} \,\exp [\tilde{I}_{\rm m}
\cos(\phi_{\rm d}) \,\Delta I_{\rm max}/D]  , \qquad
    \label{rho-diag}\\
&& \frac{\rho_{10}(t+\Delta t)}{\rho_{10}(t)} = \frac{\sqrt{\rho_{11}(t+\Delta t)\,\rho_{00}(t+\Delta t)}}{\sqrt{\rho_{11}(t)\,\rho_{00}(t)}}\, e^{-i\Delta\varphi} ,
     \label{rho-offdiag}\end{eqnarray}
where the phase shift $\Delta \varphi$ is still given by Eq.\ (\ref{delta-phi-2}). Another, more intuitive way to describe the evolution of the diagonal elements is by using the uncentered signal $I_{\rm m}$ as in Eq.\ (\ref{c0-1}):
    \be
    \rho_{jj}(t+\Delta t)=\frac{\rho_{jj}(t)\exp[-(I_{\rm m}-I_j)^2/2D]}{{\rm Norm}} .
    \label{rho-jj}\ee

Note that these evolution equations for $\rho_{jj'}$ are exactly the same as Eqs.\ (\ref{ph-sens-diag-bc}) and (\ref{ph-sens-off-bc}) of the quantum Bayesian formalism in the ``bad cavity'' limit, except the ac Stark shift $\delta \omega_{\rm q}$ is now $\delta \omega_{\rm q,s}$, dephasing is so far absent ($\gamma =0$), and, most importantly, Eqs.\ (\ref{rho-diag}) and (\ref{rho-offdiag}) describe the entangled qubit-resonator state (\ref{rho-qr-def}) and are capable of describing transient evolution. During transients there is significant time-dependence in $\alpha_{0}(t)$ and $\alpha_{1}(t)$, which also leads to time-dependence in $\Delta I_{\rm max}(t)$, the quadrature phase difference $\phi_{\rm d}(t)$, the response $I_1-I_0=\cos (\phi_{\rm d})\Delta I_{\rm max}$, and the middle point $(I_1+I_0)/2$. Therefore, during transients the time step $\Delta t$ in Eqs.\ (\ref{rho-diag}) and (\ref{rho-offdiag}) should be much smaller than $\kappa^{-1}$, in contrast to arbitrary $\tau$ in Eqs.\ (\ref{ph-sens-diag-bc}) and (\ref{ph-sens-off-bc}).

Imperfect quantum efficiency $\eta$ of the measurement ($0\leq \eta \leq 1$), similar to the case discussed in Sec.\ \ref{sec:ph-sens-bc}, mainly originates from two mechanisms: imperfect collection efficiency and imperfect amplifier efficiency. First, a fraction of the field leaked from the resonator is lost before reaching the amplifier. Second, the amplifier produces more output noise than the quantum limitation. Therefore, we can define the total quantum efficiency $\eta$ as
    \be
    \eta = \eta_{\rm col}\eta_{\rm amp}, \,\,\,  \eta_{\rm col} =\frac{\kappa_{\rm col}}{\kappa}, \,\,\,
    \eta_{\rm amp} = \frac{S_{I,\rm q.l.}}{S_I} ,
    \ee
where $\kappa_{\rm col}/\kappa$ is the ratio of the ``collected'' microwave energy, which reaches  amplifier, to the total energy lost by the resonator ($\kappa_{\rm col}/\kappa=\kappa_{\rm out}/\kappa \, \times \, \kappa_{\rm col}/\kappa_{\rm out}$), and $S_{I,\rm q.l.}$ is the spectral density of the output noise if a quantum-limited phase-preserving amplifier were used instead of the actual amplifier, which produces a larger noise $S_I$. Let us now discuss the effects produced by imperfect $\eta_{\rm col}$ and $\eta_{\rm amp}$.

An imperfect collection efficiency $\eta_{\rm col}$ can be modelled by adding an asymmetric beam splitter on the path of the leaked field, which splits each piece of the ``history tail'' into two pieces, $|\alpha_{j, \rm t}\rangle \rightarrow |\sqrt{\eta_{\rm col}}\,\alpha_{j, \rm t}\rangle \otimes |\sqrt{1-\eta_{\rm col}} \,\alpha_{j,\rm t}\rangle$, so that the first piece is measured, while the second one remains unmeasured. Since no information can be obtained from the unmeasured piece, we need to trace over it, as in the calculation of the ensemble-averaged evolution, while for the measured piece we use the same procedure as above. The tracing over the unmeasured piece does not change the diagonal matrix elements in Eq.\ (\ref{rho-qr-def}); therefore, the total change of $\rho_{00}$ and $\rho_{11}$ is still given by Eq.\ (\ref{rho-diag}). Note, however, that imperfect collection efficiency reduces the response, $\Delta I_{\rm max}=\sqrt{\eta_{\rm col}}\, \Delta I_{\rm max,ideal}$ and changes the central point $(I_0+I_1)/2$, while the variance $D$ (determined by the amplifier noise) remains unchanged.

For the off-diagonal matrix element $\rho_{10}$, the tracing over the unmeasured piece $|\sqrt{1-\eta_{\rm col}} \,\alpha_{j,\rm t}\rangle$ produces the factor $\langle \sqrt{1-\eta_{\rm col}}\, \alpha_{0,\rm t}|\sqrt{1-\eta_{\rm col}}\,\alpha_{1,\rm t}\rangle=e^{-(1-\eta_{\rm col})\Gamma_{\rm d}(t)\Delta t} e^{-i (1-\eta_{\rm col}) \delta\omega_{\rm q, s}\Delta t}$ [see Eqs.\ (\ref{rho-10-qr-1})--(\ref{delta-omega-12}) in Sec.\ \ref{sec:ensemble-aver}], while the measured piece gives the evolution described by Eq.\ (\ref{rho-offdiag}), with $\delta\omega_{\rm q, s}$ in Eq.\ (\ref{delta-phi-2}) multiplied by $\eta_{\rm col}$. Therefore, the total evolution of $\rho_{10}$ is described by Eq.\ (\ref{rho-offdiag}) with the extra factor $e^{-(1-\eta_{\rm col})\Gamma_{\rm d}(t)\Delta t}$, while  the phase $\Delta \varphi$ is still given by Eq.\  (\ref{delta-phi-2}) [note again that $\Delta I_{\rm max}$ and $(I_0+I_1)/2$ are affected by $\eta_{\rm col}$, but still correspond to the experimentally measured values]. Thus, the only effect of an imperfect collection efficiency $\eta_{\rm col}$ on the system evolution (\ref{rho-diag})--(\ref{rho-offdiag}) is the extra factor $e^{-(1-\eta_{\rm col})\Gamma_{\rm d}(t)\Delta t}$ in Eq.\ (\ref{rho-offdiag}), where $\Gamma_{\rm d}$ is given by Eq.\ (\ref{Gamma-d}).

Imperfect quantum efficiency $\eta_{\rm amp}$ of the amplifier produces additional noise at the output, so that the response $\Delta I_{\rm max}$ and the middle point $(I_0+I_1)/2$ do not change, while the variance $D$ given by Eq.\ (\ref{D-def}) increases because of the increased noise spectral density $S_I$. To take into account the extra noise, for a given measured output value $\tilde I_{\rm m}$, we need to guess what was the ``actual'' value  $\tilde I_{\rm m,a}$ (the probability distribution is given by the classical Bayesian analysis), then apply the evolution (\ref{rho-diag})--(\ref{rho-offdiag}) using the value $\tilde I_{\rm m,a}$, and then average over all possible values of $\tilde I_{\rm m,a}$. This is exactly what was done in Ref.\ \cite{Korotkov-nonid} for a qubit measurement by QPC or SET. Since the evolution equations (\ref{rho-diag}), (\ref{rho-offdiag}), and (\ref{delta-phi-2}) have exactly the same form as what was considered in Ref.\ \cite{Korotkov-nonid}, we can simply use the obtained result: the evolution is still given by Eqs.\  (\ref{rho-diag}), (\ref{rho-offdiag}), and (\ref{delta-phi-2}) with two changes. First, the variance $D$ is the actual (increased) variance; second, there is an extra dephasing factor in Eq.\ (\ref{rho-offdiag}), which can be found from comparison with ensemble-averaged evolution.

Even though the formal derivation of this result is rather lengthy \cite{Korotkov-nonid}, it is easy to understand it. The evolution of the diagonal elements of the density matrix is the evolution of probabilities, and therefore must obey the classical Bayes formula, which directly gives Eq.\ (\ref{rho-diag}). The extra dephasing in Eq.\ (\ref{rho-offdiag}) comes from uncertainty of the phase $\Delta\varphi$ due to uncertainty of the unknown ``actual'' value $\tilde I_{\rm m,a}$. The reduced proportionality factor between $\Delta\varphi$ and the (centered) measurement result $\tilde I_{\rm m}$ in Eq.\ (\ref{delta-phi-2}) due to increased value of $D$ can be understood from the fact that for uncorrelated Gaussian-distributed zero-mean random numbers $x_1$ and $x_2$, the averaging of $x_1$ for a fixed sum $x_1 + x_2$ gives the smaller value, $\langle x_1\rangle =(x_1 +x_2) \, {\rm var}(x_1)/[{\rm var}(x_1)+ {\rm var}(x_2)]$.

Thus, combining both imperfection mechanisms of the quantum efficiency $\eta$, we can describe the evolution of the qubit-resonator system (\ref{rho-qr-def}), measured using a phase-sensitive amplifier. The resulting equations are very similar to Eqs.\ (\ref{ph-sens-diag-bc}) and (\ref{ph-sens-off-bc}) for the ``bad cavity'' case,
     \begin{eqnarray}
&& \hspace{-0.2cm} \frac{\rho_{11}(t+\Delta t)}{\rho_{00}(t+\Delta t)} = \frac{\rho_{11}(t)}{\rho_{00}(t)} \,\exp \bigg[ \frac{\tilde{I}_{\rm m}
\,\Delta I}{D}\bigg]  , \qquad
    \label{rho-diag-2}\\
&& \hspace{-0.2cm} \frac{\rho_{10}(t+\Delta t)}{\rho_{10}(t)} = \frac{\sqrt{\rho_{11}(t+\Delta t)\,\rho_{00}(t+\Delta t)}}{\sqrt{\rho_{11}(t)\,\rho_{00}(t)}}\,
    \nonumber \\
&& \hspace{1.5cm} \times  \exp(-iK \tilde{I}_{\rm m} \, \Delta t) \,\, e^{-\gamma \Delta t}  \,  e^{-i\delta\omega_{\rm q,s}\Delta t} , \qquad
     \label{rho-offdiag-2}\end{eqnarray}
where
    \begin{eqnarray}
    && \tilde{I}_{\rm m} = \frac{1}{\Delta t} \int_{t}^{t+\Delta t} I(t')\, dt' -\frac{I_0+I_1}{2},
    \label{tilde-I-m}\\
&& \Delta I = I_1-I_0 = \Delta I_{\rm max}\cos \phi_{\rm d},
    \\
&& K= \frac{\Delta I_{\rm max}\sin \phi_{\rm d}}{2D\,\Delta t} = \frac{\Delta I_{\rm max}\sin \phi_{\rm d}}{S_I},
       \label{K-2}\\
&& \hspace{-0.0cm} \gamma = \Gamma_{\rm d} - (\Delta I_{\rm max})^2/4S_I =(1-\eta) \Gamma_{\rm d},
    \label{gamma-2}\\
&& \Gamma_{\rm d}=(\kappa /2) \, |\alpha_1 (t)-\alpha_0(t)|^2,
    \label{Gamma-d-2}\\
&& \delta\omega_{\rm q,s}= 2\chi {\rm Re}(\alpha_1^*\alpha_0) -\frac{d}{dt}{\rm Im} (\alpha_1^*\alpha_0),
  \label{Stark-2}\end{eqnarray}
and we repeated several previous formulas here for convenience. Note that most parameters in these equations depend on time during transients, and therefore the time step $\Delta t$ should be sufficiently small. Recall that $S_I$ is the single-sided spectral density of the output noise and $D=S_I/(2\Delta t)$ is the corresponding noise variance of $\tilde I_{\rm m}$.
 Equations (\ref{rho-diag-2})--(\ref{Stark-2}) are the {\it main result} of this paper.

If instead of using experimental output signal $I(t)$ we want to simulate the process, we can pick $\tilde {I}_{\rm m}$ from the probability distribution
    \begin{eqnarray}
&& P(\tilde I_{\rm m}) = \rho_{00}(t)\, \frac{\exp [ - (\tilde I_{\rm m}+ \Delta I /2)^2/2D ] }{\sqrt{2\pi D}}
    \nonumber \\
&& \hspace{1.0cm} + \rho_{11}(t)\, \frac{\exp [ - (\tilde I_{\rm m}- \Delta I/2)^2/2D ] }{\sqrt{2\pi D}} . \qquad
    \label{tilde-I-distr}\end{eqnarray}
For an infinitesimally small $\Delta t$ this is equivalent to using
   \be
    I(t)=\frac{I_0+I_1}{2} + \frac{\Delta I}{2} \, [\rho_{11} (t)-\rho_{00}(t)] +\xi_I(t),
    \label{ph-sens-I(t)-2}\ee
where $\xi_I (t)$ is a white noise with spectral density $S_I$.

It is easy to check that averaging of $\rho_{jj'}(t+\Delta t)$ given by Eqs.\ (\ref{rho-diag-2}) and (\ref{rho-offdiag-2}) over $\tilde I_{\rm m}$ with probability distribution (\ref{tilde-I-distr}) produces the expected ensemble-averaged equations
    \begin{eqnarray}
    && \rho_{00}(t+\Delta t)=\rho_{00}(t), \,\,\,  \rho_{11}(t+\Delta t)=\rho_{11}(t), \qquad
        \label{rho-ea-diag}\\
      && \rho_{10}(t+\Delta t)=\rho_{10}(t) \,e^{-\Gamma_{\rm d}\Delta t} \, e^{-i\delta\omega_{\rm q,s}\Delta t}  .  \qquad
        \label{rho-ea-offdiag}\end{eqnarray}
Similar to what was mentioned in Sec.\ \ref{sec:ph-pres-bc}, the averaging over the fluctuating phase $-K\tilde{I}_{\rm m}\Delta t$ in Eq.\ (\ref{rho-offdiag-2}) produces an ensemble dephasing rate $K^2S_I/4$, while the averaging of the first term in Eq.\ (\ref{rho-offdiag-2}) produces an ensemble dephasing rate $(\Delta I)^2/4S_I$. Both dephasing rates depend on the amplified quadrature via the angle $\phi_{\rm d}$, but their sum, $(\Delta I_{\rm max})^2/4S_I$, does not depend on $\phi_{\rm d}$.

If there is extra (not measurement-related) dephasing rate $\gamma_{\rm int}$ of the qubit-resonator system, e.g., due to intrinsic pure dephasing of the qubit, then it can be easily included into Eq.\ (\ref{rho-offdiag-2}) by adding the factor $e^{-\gamma_{\rm int}\Delta t}$. Alternatively, we can include $\gamma_{\rm int}$ into the ensemble-averaged dephasing, $\Gamma_{\rm d} \rightarrow \Gamma=\Gamma_{\rm d}+\gamma_{\rm int}$, so that evolution equations (\ref{rho-diag-2}) and (\ref{rho-offdiag-2}) remain unchanged, but now $\gamma =\Gamma - (\Delta I_{\rm max})^2/4S_I$. In this case the overall quantum efficiency includes the extra dephasing, $\eta=\eta_{\rm col}\eta_{\rm amp}\eta_{\rm int}$, with $\eta_{\rm int}=\Gamma_{\rm d}/(\Gamma_{\rm d}+\gamma_{\rm int})$ \cite{Korotkov-2011}.

Note that for the evolution equations discussed in this section the initial state should not necessarily be pure, so Eq.\ (\ref{psi-in}) for the initial state can be replaced with Eq.\ (\ref{rho-in}). Moreover, it is sufficient to have an initial state of the form (\ref{rho-qr-def}); the only necessary condition is that each qubit state $|j\rangle$ corresponds to a certain coherent state $|\alpha_j(0)\rangle$.

\subsection{Phase-preserving amplifier}\label{sec:ph-pres}

So far we considered the measurement using a phase-sensitive amplifier. In this section we use the results of the previous section to describe the case when a phase-preserving amplifier is used. We will do it in two ways, which give the same result.

First, let us model the measurement using a phase-preserving amplifier in the following way. Let us pass each piece of the ``history tail'' through a symmetric beam splitter $|\alpha_{j, \rm t}\rangle \rightarrow | \alpha_{j, \rm t}/\sqrt{2}\rangle \otimes |\alpha_{j, \rm t}/\sqrt{2}\rangle$, amplify orthogonal quadratures in these two parts, measure as discussed in Sec.\ \ref{sec:main-idea}, and output the results for both quadratures. From the structure of Eqs.\ (\ref{rho-diag-2})--(\ref{tilde-I-distr}) it is easy to see that it does not matter how these two orthogonal quadratures are chosen if the amplification conditions in both channels are the same (the same $D$ and $\Delta I_{\rm max}$, which also means the same quantum efficiency). Note that for both channels we should simultaneously use either the first or the second Gaussian in Eq.\ (\ref{tilde-I-distr}), though no correlation is needed in the infinitesimal limit (\ref{ph-sens-I(t)-2}).  It is natural to choose one quadrature (we call it $I$) along the informational direction $\alpha_1(t)-\alpha_0(t)$, while the other quadrature (we call it $Q$) is shifted by $\pi/2$, so that $\phi_{\rm d}=0$ for the $I$-quadrature and  $\phi_{\rm d}=\pi/2$ for the $Q$-quadrature. Thus, the directions of the $I$ and $Q$ quadratures are changing in time, but they are practically constant during the time step $\Delta t$. Note that if $D$ for both quadratures is kept the same as for a phase-sensitive amplifier, then the response $\Delta I$ is  a factor $\sqrt{2}$ smaller than $\Delta I_{\rm max}$ for the phase-sensitive case (because of the beam splitter). Equivalently, if $\Delta I$ is kept the same as in the phase-sensitive case (e.g., by an additional classical amplification by the factor $\sqrt{2}$), then $D$ for both quadratures is twice larger than for the phase-sensitive case (in the ideal case this corresponds to the fact that the noise of a phase-preserving amplifier is twice as large as for a phase-sensitive amplifier).

Therefore, for the phase-preserving case we can simply use Eqs.\ (\ref{rho-diag-2})--(\ref{tilde-I-distr}) twice, for the optimal quadrature ($\phi_{\rm d}=0$) and for the orthogonal quadrature ($\phi_{\rm d}=\pi/2$), assuming the same output noise, $S_I=S_Q$, for both output quadratures $I(t)$ and $Q(t)$. The $I$-quadrature has the maximum response, $\Delta I=I_1-I_0=\Delta I_{\rm max}$, while the $Q$-quadrature has no response, $Q_1=Q_0$. Thus, after the time step $\Delta t$ the density matrix (\ref{rho-qr-def}) of the qubit-resonator system changes as
     \begin{eqnarray}
&& \hspace{-0.2cm} \frac{\rho_{11}(t+\Delta t)}{\rho_{00}(t+\Delta t)} = \frac{\rho_{11}(t)}{\rho_{00}(t)} \,\exp \bigg[ \frac{\tilde{I}_{\rm m}
\,\Delta I}{D}\bigg]  , \qquad
    \label{rho-diag-3}\\
&& \hspace{-0.2cm} \frac{\rho_{10}(t+\Delta t)}{\rho_{10}(t)} = \frac{\sqrt{\rho_{11}(t+\Delta t)\,\rho_{00}(t+\Delta t)}}{\sqrt{\rho_{11}(t)\,\rho_{00}(t)}}\,
    \nonumber \\
&& \hspace{1.0cm} \times \exp (-i\tilde{Q}_{\rm m} \Delta I/2D) \,\, e^{-\gamma \Delta t} \, e^{-i\delta\omega_{\rm q,s}\Delta t}  , \qquad
       \label{rho-offdiag-3}
        \end{eqnarray}
where $\tilde{I}_{\rm m}$ is given by Eq.\ (\ref{tilde-I-m}), while
      \begin{eqnarray}
      && \tilde{Q}_{\rm m}= \frac{1}{\Delta t} \int_t^{t+\Delta t} Q(t')\, dt' -Q_0,
      \\
&& \hspace{-0.0cm} \gamma = \Gamma_{\rm d} - 2\, \frac{(\Delta I)^2}{4S_I}=  \Gamma_{\rm d} -  \frac{(\Delta I)^2}{4D\Delta t},
    \label{gamma-3}\\
&& D=S_I/(2\Delta t)=S_Q/(2\Delta t),
  \label{D-ph-pres}\end{eqnarray}
the ensemble-averaged dephasing $\Gamma_{\rm d}$ is given by Eq.\ (\ref{Gamma-d}),
and the ac Stark shift $\delta\omega_{\rm q,s}$ is given by Eq.\ (\ref{Stark-2}) or Eqs.\ (\ref{delta-omega-1})--(\ref{delta-omega-12}). An equivalent form for Eq.\ (\ref{rho-diag-3}) in terms of the non-centered signal $I_{\rm m}$ is given by Eq.\ (\ref{rho-jj}). The factor of 2 in Eq.\ (\ref{gamma-3}) appears because averaging over the result in
each channel produces the contribution $(\Delta I)^2/4S_I$ into the total ensemble dephasing $\Gamma_{\rm d}$.

Another way to derived Eqs.\ (\ref{rho-diag-3}) and (\ref{rho-offdiag-3}) from Eqs. (\ref{rho-diag-2}) and (\ref{rho-offdiag-2}) is to assume a slightly shifted pump frequency for a phase-sensitive amplifier, so that the angle $\phi_{\rm d}$ rotates sufficiently fast, and for both quadratures $I(t)$ and $Q(t)$ we collect only the values averaged over $\phi_{\rm d}$. Then we have a natural formation of two quadratures in Eqs.\ (\ref{rho-diag-2}) and (\ref{rho-offdiag-2}): $\tilde I_{\rm m}^{\rm ps}\cos (\phi_{\rm d})\rightarrow \tilde I_{\rm m}^{\rm pp}$ and $\tilde I_{\rm m}^{\rm ps}\sin (\phi_{\rm d})\rightarrow \tilde Q_{\rm m}^{\rm pp}$, where the superscripts indicate the phase-sensitive (ps) or phase-preserving (pp) case.
The variance of the noise in each quadrature is $D^{\rm pp}=D^{\rm ps}/2$ (because $\overline{ \cos^2\phi_{\rm d}}=1/2$) and the response in the information-carrying quadrature is $\Delta I^{\rm pp}=\Delta I_{\rm max}^{\rm ps}/2$ (because the phase-sensitive response $\Delta I_{\rm max}^{\rm ps}\cos\phi_{\rm d}$ should be multiplied by $\cos\phi_{\rm d}$ to project onto the proper quadrature). Therefore $\Delta I_{\rm max}^{\rm ps}/D^{\rm ps}=\Delta I^{\rm pp}/D^{\rm pp}$, and Eqs.\ (\ref{rho-diag-2}) and (\ref{rho-offdiag-2}) directly transform into Eqs.\ (\ref{rho-diag-3}) and (\ref{rho-offdiag-3}). For the dephasing we get $\gamma = \Gamma_{\rm d} - (\Delta I_{\rm max}^{\rm ps})^2/4S_I^{\rm ps}=\Gamma_{\rm d} - 2\times (\Delta I^{\rm pp})^2/4S_I^{\rm pp}$ since $S_I^{\rm pp}=S_I^{\rm ps}/2$  and   $\Delta I^{\rm pp}=\Delta I_{\rm max}^{\rm ps}/2$, thus  reproducing Eq.\ (\ref{gamma-3}).

In numerical simulations the probability distribution for $\tilde I_{\rm m}$ is still given by Eq.\ (\ref{tilde-I-distr}), while for $\tilde Q_{\rm m}$  it is
    \be
P(\tilde Q_{\rm m}) = \frac{\exp [ - \tilde Q_{\rm m}^2/2D ] }{\sqrt{2\pi D}} .
     \label{tilde-Q-distr-2}
    \ee
For infinitesimal $\Delta t$ these distributions are equivalent to using Eq.\ (\ref{ph-sens-I(t)-2}) for $I(t)$ and
    \be
    Q(t)=Q_0 + \xi_Q (t), \,\,\, S_{\xi_Q}= S_Q= S_I,
    \ee
for $Q(t)$, with uncorrelated white noises in the two channels. Averaging of $\rho_{jj'}(t+\Delta t)$ in Eqs.\ (\ref{rho-diag-3}) and (\ref{rho-offdiag-3}) over random $\tilde I_{\rm m}$ and $\tilde Q_{\rm m}$ using the probability distributions (\ref{tilde-I-distr}) and (\ref{tilde-Q-distr-2}) produces the ensemble-averaged evolution equations (\ref{rho-ea-diag}) and (\ref{rho-ea-offdiag}). The ensemble-averaged evolution should remain the same as in the phase-sensitive case because of causality.

Similar to Eq.\ (\ref{eta-phase-pres-def}), the quantum efficiency for a phase-preserving measurement can be defined in two ways,
    \be
    \eta = 1- \gamma /\Gamma_{\rm d}, \,\,\, \tilde\eta = \eta /2,
    \ee
where the first definition is based on the comparison with ideal phase-preserving measurement,
while in the second definition we compare the information in $I$-channel only with the ideal phase-sensitive case. We emphasize that monitoring of a pure quantum state is still possible with a phase-preserving amplifier if $\eta=1$, in spite of the fundamental limitation $\tilde\eta \leq 1/2$.

\subsection{Differential equations for evolution} \label{sec:differential}

We intentionally wrote the evolution equations (\ref{rho-diag-2}), (\ref{rho-offdiag-2}), (\ref{rho-diag-3}) and (\ref{rho-offdiag-3}) for a finite $\Delta t$ because this form is more transparent physically, suitable for numerical simulations, and also unambiguous. The differential form for an infinitesimal $\Delta t$ is significantly more ambiguous because it depends on a chosen definition of the derivative (as should be for nonlinear stochastic differential equations \cite{Oksendal}).

If we define the derivative in the symmetric way $\dot{f}(t)\equiv \lim_{\Delta t\rightarrow 0} [f(t+\Delta t/2)-f(t-\Delta t/2)]/\Delta t$ (the so-called Stratonovich form), then the standard calculus rules apply,  and the differential equations for the  evolution can be derived from Eqs.\ (\ref{rho-diag-2}), (\ref{rho-offdiag-2}), (\ref{rho-diag-3}) and (\ref{rho-offdiag-3}) in a straightforward way (keeping linear order in $\Delta t$). Thus, for the phase-sensitive measurement we obtain the Stratonovich-form evolution as (see \cite{Korotkov-2002})
    \begin{eqnarray}
&& \dot{\rho}_{11}=\rho_{11}\rho_{00} \,\frac{2\cos{\phi_{\rm d}}\Delta I_{\rm max}}{S_I} \left[ I(t)-\frac{I_0+I_1}{2} \right] , \qquad
        \label{dot-rho-11-1}\\
&&   \dot{\rho}_{10}= -(\rho_{11}-\rho_{00})\, \frac{\cos{\phi_{\rm d}}\Delta I_{\rm max}}{S_I}  \left[ I(t)-\frac{I_0+I_1}{2} \right]
    \nonumber \\
&& \hspace{0.9cm} -i \,\frac{\sin{\phi_{\rm d}}\Delta I_{\rm max}}{S_I}   \left[ I(t)- \frac{I_0+I_1}{2} \right] \rho_{10}
    \nonumber \\
&& \hspace{0.9cm} -\gamma \rho_{10} -i\,\delta\omega_{\rm q,s}\rho_{10},  \quad
    \label{dot-rho-10-1}\\
&& \gamma =\Gamma_{\rm d} -(\Delta I_{\rm max})^2/4S_I ,
    \\
&& I(t) = \rho_{00}(t) \, I_0+\rho_{11}(t) \, I_1+\xi_I(t), \,\,\, S_{\xi_I}=S_I,
    \label{I(t)}\end{eqnarray}
where for convenience we repeated equations for $\gamma$ and $I(t)$. We emphasize that $I_0$, $I_1$, $\Delta I_{\rm max}$, and $\phi_{\rm d}$ may significantly depend on time during transients.

For the phase-preserving measurement we similarly obtain the Stratonovich-form equations
    \begin{eqnarray}
&& \dot{\rho}_{11}=\rho_{11}\rho_{00} \, \frac{2\Delta I}{S_I}  \left[ I(t)-\frac{I_0+I_1}{2} \right], \qquad
        \label{dot-rho-11-2}\\
&&   \dot{\rho}_{10}= -(\rho_{11}-\rho_{00})\, \frac{\Delta I}{S_I}  \left[ I(t)-\frac{I_0+I_1}{2} \right]
    \nonumber \\
&& \hspace{0.9cm} -i \,\frac{\Delta I}{S_I} \,  [ Q(t)-  Q_0 ] \,\rho_{10}
    \nonumber \\
&& \hspace{0.9cm} -\gamma \rho_{10} -i\,\delta\omega_{\rm q,s}\rho_{10} , \quad
    \label{dot-rho-10-2}
        \\
&& \gamma =\Gamma_{\rm d} -2\times (\Delta I)^2/4S_I ,
    \\
&& I(t) = \rho_{00}(t) \, I_0+\rho_{11}(t) \, I_1+\xi_I(t),
    \label{I(t)-2}\\
    &&
    Q(t) =   Q_0 + \xi_Q(t), \,\,\,\,\, S_{\xi_Q}=S_{\xi_I}=S_I.
    \label{Q(t)}\end{eqnarray}

If we define the derivative in the ``forward'' way, $\dot{f}(t) \equiv \lim_{\Delta t\rightarrow 0} [f(t+\Delta t)-f(t)]/\Delta t$ (the so-called It\^o form), then the usual calculus rules are no longer correct, and the derivation of the differential equations from Eqs.\ (\ref{rho-diag-2}), (\ref{rho-offdiag-2}), (\ref{rho-diag-3}) and (\ref{rho-offdiag-3}) should retain the second order in $\Delta t$. Alternatively, we can use the standard rules of the transformation from the Stratonovich form into the It\^o form \cite{Oksendal,Korotkov-2002}, applied to Eqs.\ (\ref{dot-rho-11-1})--(\ref{Q(t)}). The resulting It\^o-form equations for the phase-sensitive  measurement are
    \begin{eqnarray}
&& \hspace{-0.4cm} \dot{\rho}_{11}=\rho_{11}\rho_{00} \,\frac{2\cos{\phi_{\rm d}}\Delta I_{\rm max}}{S_I} \, [ I(t)- (\rho_{00}I_0+\rho_{11}I_1)], \qquad \,
        \label{dot-rho-11-3}\\
&& \hspace{-0.4cm}  \dot{\rho}_{10}= -(\rho_{11}-\rho_{00})\, \frac{\cos{\phi_{\rm d}}\Delta I_{\rm max}}{S_I} \, [ I(t)- (\rho_{00}I_0+\rho_{11}I_1)]
     \nonumber \\
&& \hspace{-0.4cm} \hspace{0.9cm} -i \,\frac{\sin{\phi_{\rm d}}\Delta I_{\rm max}}{S_I} \,   [ I(t)- (\rho_{00}I_0+\rho_{11}I_1)]\, \rho_{10}
   \nonumber \\
&& \hspace{-0.4cm} \hspace{0.9cm} -\Gamma_{\rm d} \rho_{10} -i\,\delta\omega_{\rm q,s}\rho_{10} ,
    \end{eqnarray}
and the It\^o-form equations for the phase-preserving  measurement are
    \begin{eqnarray}
&& \dot{\rho}_{11}=\rho_{11}\rho_{00} \, \frac{2\Delta I}{S_I}  [ I(t)- (\rho_{00}I_0+\rho_{11}I_1)] , \qquad
        \\
&&   \dot{\rho}_{10}= -(\rho_{11}-\rho_{00})\, \frac{\Delta I}{S_I}  [ I(t)- (\rho_{00}I_0+\rho_{11}I_1)]
    \nonumber \\
&& \hspace{0.9cm}
 -i \,\frac{\Delta I}{S_I} \,  [ Q(t)- Q_0 ] \,\rho_{10}
     \nonumber \\
&& \hspace{0.9cm}
 -\Gamma_{\rm d} \rho_{10} -i\,\delta\omega_{\rm q,s}\rho_{10} , \qquad
    \label{dot-rho-10-4}
    \end{eqnarray}
while $I(t)$ and $Q(t)$ for numerical simulations are still given by Eqs.\ (\ref{I(t)}), (\ref{I(t)-2}), and (\ref{Q(t)}). The It\^o-form equations (\ref{dot-rho-11-3})--(\ref{dot-rho-10-4}) have two differences compared with the Stratonovich equations: (i) the combination $I(t)-(I_0+I_1)/2$ is replaced with the ``pure noise'' combination $I(t)- (\rho_{00}I_0+\rho_{11}I_1)=\xi_I (t)$ and (ii) the dephasing rate $\gamma$ is replaced with ensemble dephasing $\Gamma_{\rm d}$.

Note that It\^o and Stratonovich equations have identical solutions when the corresponding definitions of the derivative are used. The drawback of the It\^o form is the loss of intuition based on the standard calculus, because the standard calculus rules are not valid in the It\^o form. However, the advantage is that the ensemble-averaged equations can be obtained by simply replacing the noises $\xi_I(t)$ and $\xi_{Q}(t)$ with zero. The quantum trajectory formalism \cite{Wiseman-1993,Carmichael-1993,Wiseman-book,Doherty-1999,Gambetta-2008} is based on the It\^o form, while the quantum Bayesian formalism \cite{Korotkov-1999,Korotkov-2001,Korotkov-2002} usually uses the Stratonovich form (some formalisms use both forms \cite{Gough-2016}).

We emphasize that while the evolution equations in the differential form are useful in analytical analysis, for numerical calculations a relatively large time step $\Delta t$ is often preferable. For finite time steps, the formalism discussed in Secs.\ \ref{sec:imperfect} and \ref{sec:ph-pres} is more useful than the differential equations. The use of non-infinitesimal $\Delta t$ also avoids possible confusion between Stratonovich and It\^o forms.

\subsection{Evolution for an arbitrary duration}\label{sec:arb-duration}

Now let us discuss evolution of the qubit-resonator system for an arbitrarily long duration $\tau$. As in the previous sections, we assume that the qubit does not evolve due to Rabi oscillations, energy relaxation, etc. It is not obvious what the solution of the differential equations discussed in Sec.\ \ref{sec:differential} is. However, the structure  of equations for a small time step $\Delta t$ derived in Secs.\ \ref{sec:imperfect} and \ref{sec:ph-pres} permits very simple integration for an arbitrary $\tau$. This simple solution is also expected from the picture of the ``history tail'' in Fig.\ \ref{fig:tail}.

    \subsubsection{Phase-sensitive case}

The evolution equations (\ref{rho-diag-2}) and (\ref{rho-offdiag-2}) for the qubit-resonator system (\ref{rho-qr-def}) can be easily integrated within the time interval $[t, t+\tau]$,
     \begin{eqnarray}
&& \hspace{-0.2cm} \frac{\rho_{11}(t+\tau )}{\rho_{00}(t+\tau )} = \frac{\rho_{11}(t)}{\rho_{00}(t)} \,\exp (R_{\rm m}^\parallel ), \,\,  \qquad
    \label{rho-diag-tau}\\
&& \hspace{-0.2cm} \frac{\rho_{10}(t+\tau )}{\sqrt{\rho_{11}(t+\tau )\,\rho_{00}(t+\tau )} } = \frac{\rho_{10}(t)}{\sqrt{\rho_{11}(t)\,\rho_{00}(t)}}\, \exp(-iR_{\rm m}^\perp)
    \nonumber \\
&& \hspace{0.2cm} \times  \exp \bigg[ -\int_t^{t+\tau}\gamma (t')\, dt' -i \int_t^{t+\tau}\delta\omega_{\rm q,s} (t')\, dt'\bigg] , \qquad \,\,\,
     \label{rho-offdiag-tau}\end{eqnarray}
where
    \begin{eqnarray}
&& R_{\rm m}^\parallel= \int_t^{t+\tau} \tilde{I} (t') \, \frac{2\, \Delta I(t')}{S_I}\,\, dt' , \qquad
    \\
 && R_{\rm m}^\perp= \int_t^{t+\tau } \tilde{I}(t') \, \frac{\Delta I_{\rm max}(t')\sin [\phi_{\rm d}(t')]}{S_I} \,\, dt'
    \\
 && \tilde{I}(t') = I(t') -\frac{I_0(t')+I_1(t')}{2} ,
\label{tilde-I(t')}\end{eqnarray}
and the time-dependent dephasing $\gamma (t)$ and ac Stark shift $\delta\omega_{\rm q,s}(t)$ are given by Eqs.\ (\ref{gamma-2})--(\ref{Stark-2}). Note that because parameters are time-dependent, there is no simple relation between the effective measurement results $R_{\rm m}^\parallel$ and $R_{\rm m}^\perp$, which produce ``spooky'' and phase back-actions. The choice of notations $\parallel$ and $\perp$ relate to quadratures that are parallel or perpendicular to the informational quadrature.

If we need to generate measurement results numerically, then $R_{\rm m}^\parallel$ can be picked from the probability distribution $P(R_{\rm m}^\parallel )$, which consists of two Gaussians, as  usual in the Bayesian formalism,
    \begin{eqnarray}
&& P(R_{\rm m}^\parallel ) = \rho_{00}(t) \,  P(R_{\rm 0}^\parallel ) + \rho_{11}(t)\,  P(R_{\rm 1}^\parallel ) ,
    \label{P(R)-1}\\
   &&  P(R_{\rm j}^\parallel ) = (2\pi D_R^\parallel )^{-1/2} \exp [-(R_{\rm j}^\parallel -\bar{R}_{\rm j}^\parallel )^2 /2D_R^\parallel ] , \qquad
   \label{P(R)-2}\\
  && \bar{R}_{\rm 1}^\parallel =- \bar{R}_{\rm 0}^\parallel = \int_{t}^{t+\tau} \frac{[\Delta I(t')]^2}{S_I} \, dt', \,\,\, D_R^\parallel = 2 \bar{R}_{\rm 1}^\parallel .
    \label{P(R)-3}\end{eqnarray}
The validity of this formula can be checked by analyzing a composition of two evolutions for $\tau_1$ and $\tau_2$, and by checking consistency with formulas in Sec.\ \ref{sec:imperfect} for small $\tau$. From Eqs.\ (\ref{P(R)-1})--(\ref{P(R)-3}) we see that the qubit will eventually be collapsed onto the state $|0\rangle$ or $|1\rangle$ (unless $\Delta I=0$), as expected for a measured qubit with no additional evolution. Note that Eqs.\ (\ref{P(R)-1})--(\ref{P(R)-3}) for  $P(R_{\rm m}^\parallel )$ can be written in this simple way because Eq.\ (\ref{rho-diag-tau}) is essentially the classical Bayes rule. Unfortunately, $R_{\rm m}^\perp$ cannot be generated in a similar way. Therefore, we need to numerically generate the whole record $I(t')$.

The output realization $I(t')$ within the interval $[t, t+\tau]$ can be generated by dividing $\tau$ into small pieces $\Delta t$ and using Eq.\ (\ref{tilde-I-distr}). The probability of a realization $I(t')$ will then be
    \begin{eqnarray}
&& \hspace{-0.3cm} P\{I(t')\} \propto \rho_{00}(t) \, \exp \bigg[ - \int_t^{t+\tau} \frac{[I(t')-I_0(t')]^2}{S_I} \, dt' \bigg]
    \nonumber \\
    && \hspace{0.5cm} + \rho_{11}(t) \, \exp \bigg[ - \int_t^{t+\tau} \frac{[I(t')-I_1(t')]^2}{S_I} \, dt' \bigg] , \qquad
    \label{I(t')-1}\end{eqnarray}
with an appropriate overall normalization. Alternatively, the probability distribution can be obtained by applying Eq.\ (\ref{ph-sens-I(t)-2}), i.e., taking into account the randomness ``locally'' instead of ``globally'', which produces
    \begin{eqnarray}
&& \hspace{0cm} P\{I(t')\} \propto \exp \bigg[ - \int_t^{t+\tau} \frac{[I(t')-I_{\rm av}(t')]^2}{S_I} \, dt' \bigg],  \qquad
      \label{I(t')-2}\\
    && \hspace{0cm} I_{\rm av}(t')= \rho_{00}(t') \, I_0(t') +  \rho_{11}(t') \, I_1(t') , \qquad
    \label{I(t')-3}\end{eqnarray}
where $\rho_{00}(t')$ and $\rho_{11}(t')$ should be calculated using Eq.\ (\ref{rho-diag-tau}) for the previous period $[t, t']$. Even though this gives the same probability distribution, it is easier to use the ``global'' method (\ref{I(t')-1}).

   \subsubsection{Phase-preserving case}

Integrating Eqs.\ (\ref{rho-diag-3}) and (\ref{rho-offdiag-3}), we obtain the evolution during the time interval $[t, t+\tau]$,
     \begin{eqnarray}
&& \hspace{-0.2cm} \frac{\rho_{11}(t+\tau )}{\rho_{00}(t+\tau )} = \frac{\rho_{11}(t)}{\rho_{00}(t)} \,\exp (R_{\rm m}^I ), \,\,  \qquad
    \label{rho-diag-tau-2}\\
&& \hspace{-0.2cm} \frac{\rho_{10}(t+\tau )}{\sqrt{\rho_{11}(t+\tau )\,\rho_{00}(t+\tau )} } = \frac{\rho_{10}(t)}{\sqrt{\rho_{11}(t)\,\rho_{00}(t)}}\, \exp(-iR_{\rm m}^Q )
    \nonumber \\
&& \hspace{0.2cm} \times  \exp \bigg[ -\int_t^{t+\tau}\gamma (t')\, dt' -i \int_t^{t+\tau}\delta\omega_{\rm q,s} (t')\, dt'\bigg] , \qquad \,\,\,
     \label{rho-offdiag-tau-2}\end{eqnarray}
where
    \begin{eqnarray}
&& R_{\rm m}^I = \int_t^{t+\tau} \tilde{I} (t') \, \frac{2\, \Delta I(t')}{S_I}\,\, dt' , \qquad
    \\
 && R_{\rm m}^Q= \int_t^{t+\tau } \tilde{Q}(t') \, \frac{\Delta I(t')}{S_I} \,\, dt' ,
    \\
 && \tilde{Q}(t') = Q(t') - Q_0(t'), \,\,\, Q_1(t')=Q_0(t'), \qquad
    \end{eqnarray}
$\tilde{I}(t')$ is given by Eq.\ (\ref{tilde-I(t')}), $\gamma (t')$ is given by Eq.\ (\ref{gamma-3}), and  $\delta\omega_{\rm q,s}(t)$ is given by Eq.\ (\ref{Stark-2}).

We emphasize that the outputs $I(t)$ and $Q(t)$ correspond to the informational and non-informational quadratures, which change in time. In terms of the ``fixed'' experimental quadratures $I^{\rm fix}(t)$ and $Q^{\rm fix}(t)$ from the IQ mixer they are
    \begin{eqnarray}
    && I(t)= I^{\rm fix} (t) \cos [\phi_{\rm opt}(t)] + Q^{\rm fix} (t) \sin [\phi_{\rm opt}(t)],
    \qquad
    \\
    && Q(t)= Q^{\rm fix} (t) \cos [\phi_{\rm opt}(t)] - I^{\rm fix} (t) \sin [\phi_{\rm opt}(t)],
    \end{eqnarray}
where $\phi_{\rm opt}(t)={\rm arg} [\alpha_1(t)-\alpha_0(t)]$ corresponds to the informational quadrature.

If the measurement results are not taken from an experiment, but have to be generated numerically, then it is always possible to generate $R_{\rm m}^I$ and $ R_{\rm m}^Q$ without explicitly generating the signals $I(t)$ and Q(t). For $R_{\rm m}^I$ we can still use Eqs.\ (\ref{P(R)-1})--(\ref{P(R)-3}), just replacing the superscript $\parallel$ with $I$. The probability distribution for $R_{\rm m}^Q$ is the zero-mean Gaussian,
    \begin{eqnarray}
&& P(R_{\rm m}^Q) = (2\pi D_R^Q )^{-1/2} \exp [-(R^Q)^2 /2D_R^Q ] , \qquad
   \label{P(R)-4}\\
  &&  D_R^Q =D_R^I =2 \int_{t}^{t+\tau} \frac{[\Delta I(t')]^2}{S_I} \, dt' .
    \label{P(R)-5}\end{eqnarray}
The probability distribution for a realization of $I(t')$ is still given by Eqs.\ (\ref{I(t')-1}) or (\ref{I(t')-2}), while the similar probability distribution for $Q(t')$ is
    \be
P\{Q(t')]\} \propto \exp \bigg[ - \int_t^{t+\tau} \frac{[Q(t')-Q_0(t')]^2}{S_I} \, dt' \bigg] .
    \ee 

The evolution equations derived in this paper describe the evolution of an entangled qubit-resonator state (\ref{rho-qr-def}). However, there is an important special case when we can discuss the state of the qubit alone. If the measurement is of a relatively short duration and the microwave drive is switched off after that, then several decay times $\kappa^{-1}$ later (or after the rapid driven reset procedure \cite{McClure-2016}) the resonator field is practically vacuum for both qubit states. In this case our formulas give the resulting qubit state, unentangled from the resonator state.

\section{Conclusion}

In this paper we have developed a simple quantum Bayesian formalism for the qubit measurement in the circuit QED setup with a moderate bandwidth of the measurement resonator, so that transients are important. The simplification comes from three assumptions: (i) we assume that the qubit evolves only due to measurement (in particular, there are no Rabi oscillations or qubit energy relaxation), (ii) we assume that the measurement resonator is driven by a classical, i.e., coherent field (in particular, no squeezed fields are applied), and (iii) the resonator is initially in a coherent state (e.g., vacuum). In this case the entangled qubit-resonator state developing in the process of measurement can be described as the density operator, Eq.\ (\ref{rho-qr-def}), in which each of the two qubit states corresponds to its own coherent state of the resonator. Therefore, the entangled qubit-resonator state at any moment of time is fully characterized by only 4 numbers: $\rho_{00}$, $\rho_{11}$, $\rho_{10}$, $\rho_{01}$, and two field amplitudes of the resonator: $\alpha_0$ and $\alpha_1$. The field amplitudes evolve according to the classical equations (\ref{alpha0-dot}) and (\ref{alpha1-dot}). The elements of the $2\times 2$ matrix $\rho_{ij}$ evolve according to Eqs.\ (\ref{rho-diag-2}) and (\ref{rho-offdiag-2}) if a phase-sensitive amplifier is used in the measurement or according to Eqs.\ (\ref{rho-diag-3}) and (\ref{rho-offdiag-3}) if a phase-preserving amplifier is used. These evolution equations in differential form (in both Stratonovich and It\^o forms) are presented in Sec.\ \ref{sec:differential}. Integrated equations for an arbitrary long evolution are discussed in Sec.\ \ref{sec:arb-duration}.  The equations depend on parameters that are directly measurable in an experiment.

The evolution equations for $\rho_{ij}$ [Eqs.\  (\ref{rho-diag-2}), (\ref{rho-offdiag-2}), (\ref{rho-diag-3}), (\ref{rho-offdiag-3})] have exactly the same form as in the ``bad cavity'' limit \cite{Korotkov-2011} and have a simple physical meaning. We see that the diagonal elements $\rho_{00}$ and $\rho_{11}$ evolve as probabilities, i.e., they follow the classical Bayes rule, which updates the probabilities according to the information on the qubit state acquired from the measurement result. Therefore, this ``spooky'' back-action is sensitive to the ``informational'' quadrature of the microwave field. The evolution of $\rho_{10}$ (and $\rho_{01}=\rho_{10}^*$) necessarily depends on the evolution of $\rho_{00}$ and $\rho_{11}$ (at least because $|\rho_{10}|^2\leq \rho_{11}\rho_{00}$). Besides that, there are three more effects producing evolution of $\rho_{10}$: (i) phase back-action, which depends on the measurement result sensitive to the ``non-informational'' quadrature of the microwave field, (ii) dephasing due to non-ideality of the measurement (essentially loss of potential information), and (iii) ac Stark shift of the qubit frequency. As discussed in Appendix B, the phase back-action can be physically interpreted as being due to fluctuations of the ac Stark shift because of a fluctuating number of photons in the resonator.

Even though the evolution equations (\ref{rho-diag-2}), (\ref{rho-offdiag-2}), (\ref{rho-diag-3}), and (\ref{rho-offdiag-3}) are the same as in the ``bad cavity'' regime \cite{Korotkov-2011}, the time step $\Delta t$ is no longer arbitrary, since the parameters entering the equations (response $\Delta I$, amplified phase difference $\phi_{\rm d}$, ensemble dephasing $\Gamma_{\rm d}$, etc.) change during the transients, and therefore $\Delta t$ should be smaller than the time scale of this change. We emphasize that in the case of non-changing parameters these equations are exact for an arbitrary long $\Delta t$. This may be beneficial for numerical simulations in comparison with the quantum trajectory formalism \cite{Wiseman-1993,Gambetta-2008} based on a Wiener process, which assumes infinitesimal $\Delta t$. In particular, our evolution equations can be easily integrated for an arbitrarily long duration [Eqs.\ (\ref{rho-diag-tau}), (\ref{rho-offdiag-tau}), (\ref{rho-diag-tau-2}), and (\ref{rho-offdiag-tau-2})].

We note that the evolution equations in the phase-sensitive case are also exactly the same as for a qubit measurement by QPC or SET \cite{Korotkov-2002}, except now we consider a significantly entangled qubit-resonator state, with classically evolving resonator fields. The case of a phase-preserving amplifier is different because there are two output signals, $I(t)$ and $Q(t)$, instead of only one signal $I(t)$. Nevertheless, the evolution equations are almost the same, and the only significant difference is that the phase back-action is governed by the non-informational quadrature $Q(t)$, while the ``spooky'' back-action (evolution of $\rho_{00}$ and $\rho_{11}$) is governed by the informational quadrature $I(t)$.

The derivation in this paper has been based on elementary quantum mechanics and basic facts related to coherent states. In general, the idea is similar to the idea of ``microscopic'' derivation \cite{Korotkov-2001} used to describe a qubit measurement by QPC or SET. We solve exactly the quantum evolution due to interaction between the qubit and resonator (which is very simple because the qubit does not evolve by itself and measurement is of the QND type), and then apply the textbook collapse postulate to the pieces of microwave field, leaking from the resonator.

The formalism developed in this paper is equivalent to the ``polaron frame approximation'' used in the quantum trajectory formalism \cite{Gambetta-2008}, even though our language is significantly different. We hope that our derivation is physically transparent and therefore more easily understandable. Also, as mentioned above, our formalism may have advantages in numerical calculations.

For an evolving qubit (e.g., due to Rabi oscillations) it is tempting to simply include additional evolution into the differential equations for evolution derived in Sec.\ \ref{sec:differential}.
However, this is formally incorrect because in this case the approach based on coherent states is no longer applicable (though this is still possible in the ``bad cavity'' limit \cite{Korotkov-2011}). The reason is the following. When the additional evolution of the qubit is comparable to or faster than $\kappa$, the resonator state $|\alpha_{0}(t)\rangle$  or $|\alpha_{1}(t)\rangle$ may correspond to the ``wrong'' qubit state produced by this evolution. Since for a resonator the evolution of a superposition of coherent states (a ``cat state'') cannot be easily described with coherent states, the simple approach based on coherent states fails.

Therefore, for measurement of an evolving qubit the simple formalism discussed in this paper is not applicable and should be replaced with a more complicated formalism. The derivation of the quantum Bayesian formalism for measurement of an evolving qubit is similar ideologically (using the measurement of the ``history tail''), but much more cumbersome technically. The result is equivalent to ``full'' quantum trajectory formalism \cite{Wiseman-1993,Gambetta-2008}, but uses an explicit Fock-space evolution in the Schr\"odinger picture instead of the language of superoperators. We will discuss this formalism in another paper.

The formalism developed in this paper can be easily generalized to measurement of a multi-level transmon or measurement of several qubits, which evolve only due to measurement. Such a generalization is useful to describe the process of entanglement of superconducting qubits by measurement \cite{Riste-2013,Roch-2014,Motzoi-2015,Chantasri-2016}. For $N$ qubits the state of the system can be described in the way similar to Eq.\ (\ref{rho-qr-def}), so that each of $2^N$ qubit basis states corresponds to particular coherent states of the resonators, obtained via the classical field evolution. Therefore, we only need to describe the evolution of $2^N\times 2^N$ matrix of coefficients, for which we can easily use the quantum Bayesian approach to update the coefficients, depending on the measurement results. This will also be the subject of a future publication.

\acknowledgments

The author thanks Justin Dressel, Eyob Sete, Mark Dykman, Farid Khalili, and Konstantin Likharev for useful discussions. The author also thanks Justin Dressel and Juan Atalaya for critical reading of the manuscript. The work was supported by ARO grant no. W911NF-15-1-0496.

\appendix

\section{Coherent states}

In this appendix we review  basic facts related to coherent states. Most of them are very well known in the quantum optics community. However, some of these facts [e.g., Eq.\ (\ref{phi-dot-2})] are usually not discussed in optical textbooks. In contrast to the notation used in the main text, in this appendix we will use hat symbols for operators.

\subsection{Definition of a coherent state}

As known from undergraduate quantum mechanics, for an oscillator
with frequency $\omega_{\rm r}$ and mass $m$, the ground state in
the $x$-representation is
    \be
    |0\rangle = \psi_{\rm gr}(x) = \left( \frac{m\omega_{\rm r}}{\pi\hbar} \right)^{1/4} \exp\left(-\frac{m\omega_{\rm r}}{2\hbar}\, x^2\right).
    \ee
If we want to describe the classical state of this oscillator with
coordinate $x_{\rm c}$ and momentum $p_{\rm c}$ (still taking into
account the uncertainty of the ground state), we need to shift the
ground-state wavefunction by $x_{\rm c}$, producing $\psi_{\rm
gr}(x-x_{\rm c})$, and also apply the momentum shift by adding the
factor $e^{ip_{\rm c} x/\hbar}$. This produces the so-called ``coherent
state'' $|\alpha\rangle$, which is widely used in optics:
    \begin{eqnarray}
    && |\alpha\rangle \equiv \ \psi_{\rm gr}(x-x_{\rm c}) \, \exp (i p_{\rm c} x/\hbar)\, \exp (-ip_{\rm c}x_{\rm c}/2\hbar), \qquad
    \label{alpha-def}\\
    && \hspace{-0.0cm} \alpha \equiv  \frac{x_{\rm c}}{2\sigma_x}
    + i \, \frac{p_{\rm c}}{2\sigma_p} =
    x_{\rm c}\sqrt{\frac{m\omega_{\rm r}}{2\hbar}}
    +i p_{\rm c}\, \frac{1}{\sqrt{2\hbar m\omega_{\rm r}}} ,
    \label{Re-Im}\end{eqnarray}
where $\sigma_x=\sqrt{\hbar/2m\omega_{\rm r}}$ and
$\sigma_p=\hbar/2\sigma_p=\sqrt{\hbar m\omega_{\rm r}/2}$ are the
ground-state uncertainties. The normalization by doubled
uncertanties $\sigma_x$ and $\sigma_p$ in Eq.\ (\ref{Re-Im}) as well as the overall phase
factor $e^{-ip_{\rm c}x_{\rm c}/2\hbar}$ in Eq.\ (\ref{alpha-def})
are to some extent arbitrary, but this conventional choice simplifies most of the formulas discussed
below. Note that the phase $e^{-ip_{\rm c}x_{\rm
c}/2\hbar}$ is exactly in between what we would obtain by first
shifting $x$, and then $p$ [in this case we would obtain $\psi_{\rm
gr}(x-x_{\rm c})\, e^{ip_{\rm c}x/\hbar}$] and, instead, first
shifting $p$ and then $x$ [in this case we would obtain $\psi_{\rm
gr}(x-x_{\rm c})\, e^{ip_{\rm c}(x-x_{\rm c})/\hbar}$].

Equation (\ref{alpha-def}) can be rewritten in a more standard form \cite{Walls-Milburn-book,Gerry-book}
    \begin{eqnarray}
&&    |\alpha \rangle  =  e^{-\frac{1}{2}|\alpha|^2}\sum_{n=0}^\infty \frac{\alpha^n}{\sqrt{n!}} \, |n\rangle
    \label{alpha-def-2}\\
&& \hspace{0.5cm} =  e^{-\frac{1}{2}|\alpha|^2}\sum_{n=0}^\infty \frac{(\alpha \, \hat{a}^\dagger)^n}{n!}\,|0\rangle  =  e^{-\frac{1}{2}|\alpha|^2} \, e^{\alpha \hat{a}^\dagger}\, |0\rangle , \qquad
        \label{alpha-def-3}\end{eqnarray}
where $\hat{a}^\dagger = (2\hbar m\omega_{\rm r})^{-1/2} (-i\hat{p}
+ m\omega_{\rm r} \hat{x})$ is the  raising
(creation) operator, $\hat{a}^\dagger |n\rangle =\sqrt{n+1}\,
|n+1\rangle$. The equivalence of Eqs.\ (\ref{alpha-def}) and (\ref{alpha-def-2})
can be verified by explicitly checking that  Eq.\ (\ref{alpha-def}) satisfies
the relations $d|\alpha\rangle/d ({\rm Re}\,\alpha )=[-{\rm Re}(
\alpha ) +\hat{a}^\dagger]\, |\alpha\rangle$ and $d |\alpha\rangle/d
({\rm Im}\,\alpha )=[-{\rm Im}( \alpha ) +i\hat{a}^\dagger]\,
|\alpha\rangle$, which follow from Eq.\ (\ref{alpha-def-3}).
 Note a possible confusion between the notations for
the stationary states $|n\rangle$ and the coherent state $|\alpha
\rangle$ (for example, $|\alpha \rangle$ with $\alpha=1$ is not the
first excited level $|1\rangle$); to avoid the confusion, we can use
Greek letters for coherent states and Roman letters or integer
numbers for the stationary states (Fock states). For the ground
state the notations coincide, $|\alpha =0\rangle =|0\rangle$.

If the oscillator state rotates with frequency $\omega$ (for example, due
to drive with this frequency), $x_{\rm c}(t) =x_{\rm c,
amp}\cos(\omega t+\phi_0)$, $p_{\rm c}(t)= -m\omega x_{\rm c, amp}
\sin(\omega t+\phi_0)$, then from Eq.\ (\ref{Re-Im}) we find
$\alpha(t)=e^{-i(\omega t+\phi_0)}x_{\rm c, amp}/2\sigma_x$. In this
case it is useful to introduce the rotating frame by defining
$\tilde \alpha \equiv e^{i\omega t}\alpha$, so that $\tilde\alpha$
does not change in time. In the general case $\tilde\alpha$ changes
with time slowly, while $\alpha(t) = e^{-i\omega
t}\tilde\alpha(t)$ rapidly oscillates. The rotating frame frequency
$\omega$ can be chosen arbitrarily; in the case with a drive, the
most natural choice is the drive frequency $\omega_{\rm d}$ (because
then $\tilde\alpha$ does not change in the steady state); in the
absence of the drive, a natural choice is the oscillator frequency
$\omega_{\rm r}$. Note that the time dependence for the stationary
states is $e^{-in \omega_{\rm r} t}|n\rangle$ (counting the energy
from the ground state energy), so for a ``non-evolving'' oscillator
(i.e., evolving only naturally), from Eq.\ (\ref{alpha-def-2}) we
find $\alpha (t)=\alpha(0)\, e^{-i\omega_{\rm r} t}$.

Note that in the main text we always use the rotating frame based on the drive frequency $\omega_{\rm d}$ and omit the tilde sign in the notation of the rotating-frame $\alpha$. In contrast, in this appendix we explicitly write $\tilde \alpha$ for the rotating frame.

So far we considered a textbook mechanical oscillator. If we
consider a microwave resonator, then the role of $x$ and $p$ is
played by properly normalized voltage and current (at some point in
the resonator) or by flux and charge; the effective mass $m$ can
also be appropriately introduced. The formalism does not change. In quantum
optics it is often preferred not to introduce coordinates and
effective mass explicitly, and instead to start with the commutation relation
$[\hat{a},\hat{a}^\dagger]=1$, then producing Fock states $|n\rangle$ from vacuum $|0\rangle$ with the creation operator.

\subsection{Some properties}

{\bf 1.} From Eq.\ (\ref{alpha-def-2}) it is easy to see that
    \be
    \hat{a}\, |\alpha\rangle =\alpha \, |\alpha\rangle,
    \label{prop-1}\ee
since $\hat{a}\,|n\rangle =\sqrt{n}\, |n-1\rangle$ for the lowering
(annihilation) operator  $\hat{a} = (2\hbar m\omega_{\rm r})^{-1/2}
(i\hat{p} + m\omega_{\rm r} \hat{x})=(\hat{a}^\dagger)^\dagger$. The property (\ref{prop-1}) is
sometimes used as a definition of the coherent state
$|\alpha\rangle$. Note, however, that it does not specify the
overall phase and normalization, while the overall phase if often
important in analysis (when more than one coherent state is
involved). Also note that $\hat{a}^\dagger |\alpha\rangle$ does not have a simple formula, though $\langle \alpha | \hat{a}^\dagger |\alpha\rangle=\alpha^*$ from conjugation of $\langle \alpha | \hat{a} |\alpha\rangle=\alpha$.

{\bf 2.} From Eq.\ (\ref{alpha-def-2}), the probability to measure $n$ photons in the state $|\alpha \rangle$ is
    \be
    P(n) =e^{-|\alpha|^2}|\alpha^2|^n/n!,
    \label{prop-2-1}\ee
which is the Poissonian distribution with average $|\alpha^2|$. This proves that
the wavefunction (\ref{alpha-def-2}) is normalized and shows that
the mean photon number is
    \be
    \bar{n}=|\alpha|^2.
    \ee

{\bf 3.} The inner product of two coherent states $|\alpha\rangle$
and $|\beta\rangle$ can be easily calculated using Eq.\
(\ref{alpha-def-2}), giving the result \cite{Walls-Milburn-book,Gerry-book}
        \be
 \langle \alpha |\beta \rangle=e^{-\frac{1}{2}(|\alpha|^2+|\beta |^2)}
   e^{\alpha^*\beta} = e^{-\frac{1}{2}|\alpha-\beta |^2}
   \,e^{ -i\,{\rm Im}(\alpha\beta^*)} .
    \label{inner-product}\ee
Note that a shift of the coherent states by the same value changes
the inner product, $\langle \alpha +\gamma \, |\, \beta +\gamma
\rangle \neq   \langle \alpha |\beta \rangle$, since this changes the
phase factor.

{\bf 4.} It is useful to introduce the (unitary) displacement operator $\hat{D}$ \cite{Walls-Milburn-book,Gerry-book},
    \be
    \hat{D}(\alpha) \equiv \exp(\alpha \hat{a}^\dagger -\alpha^*
    \hat{a}),
    \,\,\,\,\,
    \hat{D}(\alpha)\, |0\rangle = |\alpha\rangle .
    \ee
A composition of two displacement operators has a phase factor \cite{Walls-Milburn-book,Gerry-book} similar to the phase factor in Eq.\ (\ref{inner-product}),
    \be
\hat{D}(\alpha) \, \hat{D}(\beta)=   \hat{D}(\alpha + \beta ) \,
\exp [-i\, {\rm Im}(\alpha^*\beta)] ,
    \label{displacement-comp}\ee
as follows from the Baker-Campbell-Hausdorff formula $e^{\hat{A}+\hat{B}}=e^{-c/2}\,
e^{\hat{A}}\, e^{\hat{B}} = e^{c/2}\, e^{\hat{B}}\, e^{\hat{A}}$ for
$[\hat{A},\hat{B}]=c$. Also note the useful relations
    \begin{eqnarray}
     && \hat{D}^\dagger(\alpha) \,\hat{a} \,\hat{D}(\alpha) =\hat{a}+\alpha, \,\,\,\,
      \hat{D}^\dagger(\alpha) = \hat{D}(-\alpha) ,  \qquad
      \\
&&       \hat{D}(\alpha) = e^{-\frac{1}{2}|\alpha|^2} e^{\alpha \hat{a}^\dagger} e^{-\alpha^* \hat{a}} .
    \end{eqnarray}

{\bf 5.} Let us introduce the (Hermitian) quadrature operators $\hat{x}_q$ and $\hat{p}_q$ as \cite{Gerry-book}
    \be
    \hat{x}_q=\frac{\hat{a}+\hat{a}^\dagger}{2}=\frac{\hat{x}}{2\sigma_x}, \,\,\,
    \hat{p}_q=\frac{\hat{a}-\hat{a}^\dagger}{2i}=\frac{\hat{p}}{2\sigma_p}, \,\,\,
    [\hat{x}_q, \hat{p}_q]=\frac{i}{2}.
    \label{quadrature-def}\ee
Note that the quadrature operators are often defined as $\sqrt{2} \, \hat{x}_q$ and $\sqrt{2} \, \hat{p}_q$; then their commutator is $i$; another possible definition \cite{Walls-Milburn-book} is $2 \hat{x}_q$ and $2\hat{p}_q$; then the commutator is $2i$.
The definition (\ref{quadrature-def}) gives simpler formulas for the average values for the coherent states,
    \be
    \langle \alpha |\hat{x}_q|\alpha\rangle = {\rm Re}(\alpha), \,\,\,  \langle \alpha |\hat{p}_q|\alpha\rangle = {\rm Im}(\alpha),
    \ee
which follow from the relation $\hat{a}= \hat{x}_q+i \hat{p}_q$. The variance in this case is
    \be
     \langle \alpha |\hat{x}_q^2|\alpha\rangle - \langle \alpha |\hat{x}_q|\alpha\rangle^2 = \langle \alpha |\hat{p}_q^2|\alpha\rangle - \langle \alpha |\hat{p}_q|\alpha\rangle^2 =\frac{1}{4}.
    \ee
 The quadrature operator at an angle $\phi$ can be defined as
    \be
 \hat{x}_q(\phi)=\frac{\hat{a}e^{-i\phi}+\hat{a}^\dagger e^{i\phi}}{2}=  \hat{x}_q  \cos \phi + \hat{p}_q \sin \phi .
    \ee

{\bf 6.} An important property of a coherent state is that it splits
into two {\it unentangled} coherent states after passing through a beam
splitter, in full analogy with a classical optical wave or microwave.
Actually, so far we defined a coherent state only for a resonator,
and it is not obvious how to introduce it for a propagating wave.
We will not discuss how to do it rigorously \cite{Yurke-1984,Gardiner-1985,Clerk-2010}, just implying
that a piece of propagating wave can be described in a way, similar
to a resonator description.

There is a rather simple rigorous way to describe transformation of
an arbitrary quantum state passing through a beam splitter (see,
e.g., \cite{Gerry-book,Sete-2015}). The idea is essentially
to write classical field relations, but for the annihilation
operators (conjugated relations are for the creation operators),
then express the initial state via vacuum and creation operators of
the input arms, and then substitute these input-arms operators with their
expressions via output-arms operators. This gives the resulting
output state.

Applying this procedure to a beam splitter with
transmission and reflection amplitudes $(t_1,t_2,r_1,r_2)$ and input
state $|\alpha\rangle \otimes |0\rangle$, we obtain the output state
$|t_1\alpha\rangle \otimes |r_1\alpha\rangle$, exactly as we would
expect for a classical field. Technically, this follows from the formula  $|\alpha \rangle  = e^{-\frac{1}{2}|\alpha|^2}
\, e^{\alpha \hat{a}^\dagger_{\rm in}}\, |0\rangle$ [see Eq.\ (\ref{alpha-def-3})] and relation
$\hat{a}^\dagger_{\rm in} = t_1\hat{a}^\dagger_{\rm out}+r_1
\hat{b}^\dagger_{\rm out}$, with commuting output-arms operators
$\hat{a}^\dagger_{\rm out}$ and $\hat{b}^\dagger_{\rm out}$, so that
$e^{\alpha \hat{a}^\dagger_{\rm in}} = e^{\alpha t_1
\hat{a}^\dagger_{\rm out}} e^{\alpha r_1 \hat{b}^\dagger_{\rm
out}}$. Note that if we apply coherent fields to both input arms,
$|\alpha\rangle \otimes |\beta\rangle$, then the resulting output
state is also an unentangled product of classically-expected
coherent states, $|t_1\alpha+r_2\beta\rangle \otimes |r_1\alpha
+t_2\beta\rangle$, without an overall phase.

\subsection{Driven microwave resonator with leakage}

We can think about field leakage from a microwave resonator to a
transmission line through a ``mirror'' (coupler) as transmission
through a beam splitter. Therefore, from the discussed above
property, if the initial state in the resonator is a coherent state
$|\alpha\rangle$, then it remains a coherent state $|\alpha
(t)\rangle$, with $\alpha (t)$ given by the classical field evolution,
    \be
    \alpha (t)=\alpha(0)\, e^{-i\omega_{\rm r} t} \, e^{-\kappa t/2},
    \label{kappa-only}\ee
where $\kappa$ is the energy dissipation rate and $\omega_{\rm r}$ is the resonator frequency.

We emphasize that this property is highly unusual for a quantum system (thus indicating that coherent states are classical to a significant extent). Dissipation
usually leads to decoherence, so that an initially pure quantum
state becomes a mixed state. In this case we have an exception: a
pure state remains pure during the whole evolution. This makes
quantum analysis very simple for an evolution involving coherent
states. Note that Eq.\ (\ref{kappa-only}) is still applicable when
the energy loss rate $\kappa$ has a contribution from intrinsic
energy relaxation (at zero temperature).

Now let us for a moment neglect the energy relaxation, and instead
consider a classical drive with frequency $\omega_{\rm d}$ and
(complex) amplitude $\varepsilon (t)$ (in some normalization). This
is usually described by the Hamiltonian
    \be
    \hat{H}= \hbar \omega_{\rm r} \hat{a}^\dagger \hat{a}
    + \hbar \varepsilon e^{-i\omega_d t} \hat{a}^\dagger
    + \hbar \varepsilon^* e^{i\omega_d t} \hat{a},
    \ee
which already assumes Rotating Wave Approximation, requiring
$|\omega_{\rm d}-\omega_{\rm r}|\ll \omega_{\rm r}$ and sufficiently
slowly changing drive $\varepsilon (t)$. Using this Hamiltonian, we
can find the evolution of an arbitrary quantum state of the resonator
$|\psi (t)\rangle =\sum_n c_n(t)\, |n\rangle$ via the Schr\"odinger
equation $\dot c_n=-i\omega_{\rm r} c_n -i\varepsilon e^{-i\omega_d
t} \sqrt{n}\,c_{n-1} -i \varepsilon^* e^{i\omega_d t} \sqrt{n+1}\,
c_{n+1}$. It is easy to see by solving this equation that if the
initial state is a coherent state, then it remains a coherent state,
though with a {\it nontrivial overall phase} $\varphi(t)$,
    \be
    |\psi (t)\rangle = e^{-i\varphi (t)} \, |\alpha (t)\rangle,
    \ee
so that the evolution is described by two equations,
    \begin{eqnarray}
&&    \dot\alpha = -i \omega_{\rm r} \alpha -i\varepsilon
e^{-i\omega_{\rm d} t} ,
    \label{alpha-dot}\\
&&   \dot\varphi = {\rm Re} (\varepsilon^* e^{i\omega_{\rm d} t}
\alpha ) .
    \label{phi-dot-1}\end{eqnarray}

Now let us combine the drive $\varepsilon$ and dissipation $\kappa$.
Since both of them keep the state coherent (with an overall phase),
their combination will also keep it coherent (with an overall
phase). Introducing the rotating frame based on the drive frequency,
    \be
    \tilde \alpha (t) \equiv e^{i\omega_{\rm d} t} \alpha (t),
    \ee
from Eqs.\ (\ref{kappa-only}), (\ref{alpha-dot}), and
(\ref{phi-dot-1}) we obtain
    \begin{eqnarray}
&&    \dot{\tilde\alpha} = -i (\omega_{\rm r}-\omega_{\rm d}) \,
\tilde \alpha -\frac{\kappa}{2}\,  \tilde \alpha -i\varepsilon ,
    \label{tilde-alpha-dot}\\
&&   \dot\varphi = {\rm Re} (\varepsilon^* \tilde\alpha ) .
    \label{phi-dot-2}\end{eqnarray}
Equation (\ref{tilde-alpha-dot}) is the standard result for the
evolution of a resonator under the drive and dissipation, while Eq.\
(\ref{phi-dot-2}) is usually not discussed in quantum optics, even though it
is very important for quantum dynamics involving more than one
coherent state (for example, for measurement of a qubit in the
circuit QED setup).

Note that Eqs.\ (\ref{tilde-alpha-dot})--(\ref{phi-dot-2}) rely on the fact that
for coherent states the dissipation $\kappa$ does not introduce decoherence
and only brings the term $-\kappa \tilde\alpha /2$ into Eq.\
(\ref{tilde-alpha-dot}). We have derived this fact by considering the problem of a coherent
state passing through a beam splitter. Another (lengthier) way to
prove it, is to consider the Lindblad equation for the density
matrix and to show that (surprisingly) a pure initial state remains
pure if initially it was a coherent state. One of the ways to show
it, is to separate the Lindblad evolution into ``jump'' and ``no
jump'' scenarios (e.g., \cite{Korotkov-2013,Plenio-1998}). Then the
``jump'' scenario (application of operator $\hat{a}$) brings no
evolution because of Eq.\ (\ref{prop-1}), so all the evolution comes
from the ``no jump'' scenario (essentially the Bayesian update), which keeps a coherent state coherent, with decreasing $\alpha (t)$. This is why {\it there is no randomness} \cite{Korotkov-2013}, normally
leading to decoherence. Note that the derivation via the Lindblad equation cannot easily reproduce important equation (\ref{phi-dot-2}), because the overall phase is lost in the density
matrix language.

\section{Derivation of phase back-action via vacuum noise}

\begin{figure}[tb]
  \centering
\includegraphics[width=9cm,trim=9cm 7.5cm 10cm 9cm, clip=true]{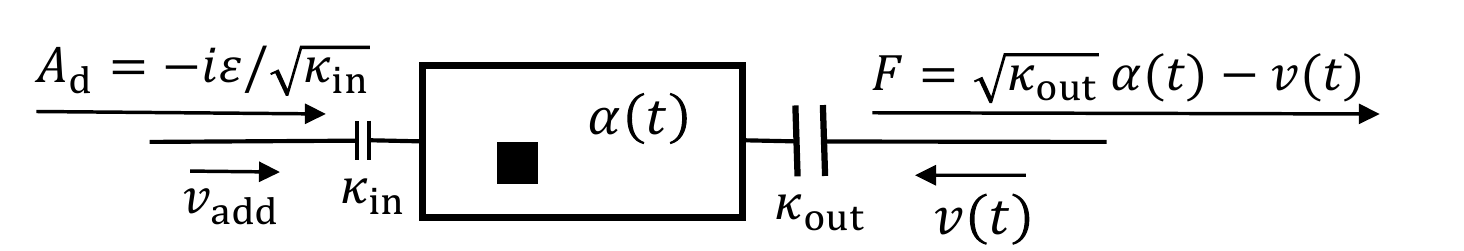}
  \caption{Illustration of the effect of vacuum noise. The vacuum noise $v(t)$ incident from the output side affects the resonator state $\alpha(t)$ via the coupling $\kappa_{\rm out}$. Therefore, $v(t)$ contributes to the outgoing field $F$ twice: due to direct reflection and due to the field leaking from the resonator later. We mostly consider the case $\kappa_{\rm out}\approx \kappa$, $\kappa_{\rm in}\ll \kappa$, so that we can neglect the effect of the vacuum noise $v_{\rm add}(t)$ incident from the input side, which adds to the drive field $A_{\rm d}=-i \varepsilon /\sqrt{\kappa_{\rm in}}$. (The outgoing field from the input port is not important and not shown.) }
  \label{fig:vacuum}
\end{figure}

In this appendix we derive the results for {\it phase back-action} in the process of qubit measurement using the picture of vacuum noise, which is incident on the resonator from the transmission line (Fig.\ \ref{fig:vacuum}). We assume the ``bad cavity'' limit and phase-sensitive amplification. The vacuum noise is treated in a simple classical way.

Let us start with assuming for simplicity that the resonator damping $\kappa$ is only due to coupling with the transmission line carrying the outgoing wave, $\kappa_{\rm out}=\kappa$; in particular, this requires $\kappa_{\rm in}\ll \kappa_{\rm out}$ (later this assumption will be removed). Then the vacuum noise enters the resonator only from the output line (Fig.\ \ref{fig:vacuum}), and the wave equations for the resonator field $\alpha$ and the outgoing field $F$ in the rotating frame based on the drive frequency $\omega_{\rm d}$ are
    \begin{eqnarray}
    && \dot{\alpha} = -i (\omega_{\rm r} -\omega_{\rm d})\, \alpha -\frac{\kappa}{2}\, \alpha -i\varepsilon +\sqrt{\kappa} \, v(t), \qquad
    \label{AppB-alpha-dot}\\
    && F=\sqrt{\kappa} \, \alpha -v(t),
    \label{AppB-F}\end{eqnarray}
where $v(t)$ is the vacuum noise, which is normalized in the same way as $F$. In this normalization $|\alpha|^2$ is the average number of photons in the resonator, while $|F|^2$ is the average number of propagating photons per second. Note that the reflection coefficient in Eq.\ (\ref{AppB-F}) is $-1$, while the transmission through the ``mirror'' is characterized by the coupling $\sqrt{\kappa}$ \cite{Walls-Milburn-book}, as well as in Eq.\ (\ref{AppB-alpha-dot}). The drive term $-i\varepsilon$ can also be written via the properly normalized incoming field $A_{\rm d}$ as $-i\varepsilon = \sqrt{\kappa_{\rm in}}\, A_{\rm d}$. Also note that for the two qubit states we have slightly different resonator frequencies, $\omega_{\rm r}\rightarrow \omega_{\rm r}\pm \chi$; however, in this appendix we will mostly use notation $\omega_{\rm r}$ for brevity and because the resonator frequency shift is not important for the phase back-action, which is our focus here.

In quantum optics the vacuum noise is treated as an operator \cite{Gardiner-book,Clerk-2010,Collett-1984,Gardiner-1985} with correlator $\langle \hat{v}(t)\, \hat{v}^\dagger (t')\rangle=\delta(t-t')$, and Eqs.\ (\ref{AppB-alpha-dot}) and (\ref{AppB-F}) are written for annihilation operators in the Heisenberg representation. However, in our simple derivation we will treat the noise $v(t)$ classically (i.e., as a complex number) and consider evolution of classical fields (which corresponds to the Schr\"odinger picture). It is simple to see that the photon shot noise is properly reproduced if we assume that {\it for any quadrature} (so that $v_{\rm qu}$ is real)
    \be
    \langle v_{\rm qu}(t) \, v(t')_{\rm qu}\rangle =\frac{1}{4}\, \delta (t-t'),
    \label{AppB-v-qu-corr}\ee
which is equivalent to
    \be
    \langle v(t) \, v(t')^*\rangle =\frac{1}{2}\, \delta (t-t'), \,\,\,  \langle v(t) \, v(t')\rangle =0,
    \label{AppB-v-corr}\ee
if $v(t)=v_{\rm qu1}(t)+iv_{\rm qu2}(t)$ is treated as a complex number, describing both quadrature components (obviously, $\langle v\rangle =0$). For example, this relation can be obtained by considering a propagating wave $F_0+v(t)$ with constant $F_0$. Then the fluctuating photon number $\int_0^t |F_0+v(t')|^2\, dt'$ within duration $t$ should have the same variance $\langle \, |\int_0^t 2\,{\rm Re} [F_0^* v(t')]\, dt' |^2 \rangle$ as the mean $|F_0|^2 t$. Therefore,
    \be
        \langle \, | \int_0^t v_{\rm qu}(t') \, dt'\, |^2 \rangle = \frac{t}{4}
    \label{AppB-v-qu-int}\ee
for the quadrature $v_{\rm qu}$ along $F_0$, and Eq.\ (\ref{AppB-v-qu-corr}) follows from (\ref{AppB-v-qu-int}). Note that Eq.\ (\ref{AppB-v-qu-corr}) can be interpreted as following from the standard operator correlator, using the correspondence $v_{\rm qu}=(\hat{v}+\hat{v}^\dagger)/2$.

As another check of this noise formalism, let us derive the correlator for the fluctuating number of photons in the resonator from Eq.\ (\ref{AppB-v-qu-corr}). Using Eq.\ (\ref{AppB-alpha-dot}), we find the fluctuation
    \be
    \delta \alpha (t)= \int_{-\infty}^t e^{-[\kappa /2+i(\omega_{\rm r}-\omega_{\rm d})](t-t')}\, \sqrt{\kappa }\, v(t') \, dt',
    \label{AppB-dalpha}\ee
due to the noise $v(t)$.  For a fixed stationary value $\alpha_{\rm st}$, this leads to photon number fluctuation $\delta n=\alpha_{\rm st}^*\delta \alpha +\alpha_{\rm st}\delta\alpha^*$. Then using Eq.\ (\ref{AppB-dalpha}), performing the double-integration using Eq.\ (\ref{AppB-v-corr}), and denoting  $|\alpha_{\rm st}|^2=\bar{n}$, we find
    \be
    \langle \delta n(t)\, \delta n(t+\tau)\rangle = \bar n \cos [(\omega_{\rm r}-\omega_{\rm d})\tau] \exp \big( -\frac{\kappa}{2} \, |\tau| \big),
    \label{AppB-dn-corr}\ee
which is the standard result for the photon number correlator \cite{Clerk-2010}. Note that the photon number fluctuation decays with the rate $\kappa /2$ instead of naively expected $\kappa$. It is also interesting to note that at time $t'$ only the quadrature $v_{\rm qu}$ along $\alpha_{\rm st}e^{i(\omega_{\rm r}-\omega_{\rm d})(t-t')}$ with the fluctuations (\ref{AppB-v-qu-corr}) contributes to the correlator (\ref{AppB-dn-corr}), while the orthogonal quadrature does not contribute. It is equally possible to say that the contribution comes only from the quadrature $v_{\rm qu}$ along $\alpha_{\rm st}e^{i(\omega_{\rm r}-\omega_{\rm d})(t+\tau-t')}$, while the orthogonal quadrature does not contribute. Also note that from Eq.\ (\ref{AppB-dalpha}) we obtain
    \be
    \langle \, |\delta \alpha |^2\rangle = 1/2,
    \ee
corresponding to the variance of $1/4$ for any quadrature.

Now let us consider the qubit measurement, assuming the ``bad cavity'' regime, as in Sec.\ III. The fluctuation $v(t)$ leads to the fluctuating ac Stark shift
    \be
    \delta \omega_{\rm q} (t) = 2\chi\,\delta n = 4\chi\,{\rm Re} [\alpha_{\rm st}^* \delta \alpha (t)]
    \ee
with $\delta \alpha (t)$ given by Eq.\ (\ref{AppB-dalpha}), and to the fluctuating outgoing field
    \be
    \delta F(t) = -v(t)+\sqrt{\kappa}\, \delta \alpha (t).
    \ee
By integrating these effects over the time period $[t, t+\tau]$ with $\tau \gg \kappa^{-1}$, so that the exponential dependence in Eq.\ (\ref{AppB-dalpha}) has sufficient time to fully decay, we find
    \begin{eqnarray}
&&   \int_t^{t+\tau}  \delta \omega_{\rm q} (t')\, dt' = 4\chi \, {\rm Re}
    \bigg[ \alpha_{\rm st}^* \, \frac{\sqrt{\kappa}}{\kappa/2 +i(\omega_{\rm r}-\omega_{\rm d})}
    \nonumber \\
    && \hspace{2.3cm} \times  \,  \int_t^{t+\tau} v(t')\, dt' \bigg] ,
    \label{AppB-dwq-int}\\
&&  \int_t^{t+\tau}  \delta F (t')\, dt' = \frac{\kappa/2 -i(\omega_{\rm r}-\omega_{\rm d})}{\kappa /2 +i(\omega_{\rm r}-\omega_{\rm d})}
    \nonumber \\
    && \hspace{2.3cm} \times \, \int_t^{t+\tau} v(t')\, dt' .
    \label{AppB-dF-int}\end{eqnarray}
We see that these fluctuating integrals are {\it proportional} to each other. Obviously, the first integral  determines the phase back-action on the qubit state, while the second integral is related to the measurement result. This is how we can relate the phase back-action to the measurement result.

Using Eq.\ (\ref{alpha-st}) for the steady-state values $\alpha_{0, \rm st}$ and $\alpha_{1,\rm st}$ corresponding to the qubit states $|0\rangle$ and $|1\rangle$, and assuming $|\chi | \ll \sqrt{\kappa^2 +4(\omega_{\rm r}-\omega_{\rm d})^2}$, we find
    \be
\alpha_{1,\rm st}-\alpha_{0, \rm st}= \alpha_{\rm st} \, \frac{-2i\chi}{i(\omega_{\rm r}-\omega_{\rm d}) +\kappa /2},
    \ee
and therefore from Eqs.\ (\ref{AppB-dwq-int}) and (\ref{AppB-dF-int}) we obtain
    \begin{eqnarray}
&& \int_t^{t+\tau}  \delta \omega_{\rm q} (t')\, dt' =  2 \, {\rm Re} \bigg[ -i (\alpha_{1,\rm st}-\alpha_{0, \rm st})^* \sqrt{\kappa}
    \nonumber \\
 && \hspace{2.7cm} \times \, \int_t^{t+\tau} \delta F(t')\, dt' \bigg].
    \label{AppB-dwq-int2}\end{eqnarray}
This relation shows that the phase back-action is determined by the output quadrature which is {\it orthogonal} to the informational quadrature along $\alpha_{1,\rm st}-\alpha_{0, \rm st}$. Note that the vacuum fluctuations $v(t)$, which produce the output fluctuations along the informational quadrature, do not affect the qubit state, so the corresponding evolution (\ref{ph-sens-diag-bc}) of the qubit state (diagonal matrix elements) is only due to ``spooky'' back-action and cannot be explained as an effect of $v(t)$.

Let us first consider an ideal phase-sensitive amplification  of the ``orthogonal'' (non-informational) quadrature, so that $\phi_{\rm d}=\pi/2$ [see Eq.\ (\ref{phi-d-def})]. In this case we need to associate the output noise with the effect of $v(t)$ fluctuations (no added noise due to amplifier), and therefore
        \be
\int_t^{t+\tau} \delta F_{\rm qu}(t')\, dt'  = \frac{\tilde{I}_{\rm m}}{\sqrt{D}}
       \sqrt{ \bigg\langle \bigg[ \int_t^{t+\tau} \delta F_{\rm qu}(t')\, dt' \bigg] ^2 \bigg\rangle },
    \label{AppB-dFqu-int}\ee
where $\tilde{I}_{\rm m}$ is the measurement result [Eq.\ (\ref{ph-sens-Im-bc})], $D$ is its variance, and real $\delta F_{\rm qu}$ is the fluctuation along the measured quadrature. Note that the left hand side of this relation is for a particular realization of the noise $\delta F_{\rm qu}$, while the last term in the right hand side assumes averaging over all noise realizations. Since $\delta F_{\rm qu}(t)$ should have the usual vacuum noise statistics, we can use Eq.\ (\ref{AppB-v-qu-int}), which gives $\langle [ \int_t^{t+\tau} \delta F_{\rm qu}(t')\, dt' ] ^2 \rangle=\tau /4$. Following Eq.\ (\ref{AppB-dFqu-int}), we can do the similar conversion for the response,
        \be
\sqrt{\kappa} \, |\alpha_{1,\rm st} -\alpha_{0,\rm st} | \, \tau = \frac{\Delta I}{\sqrt{D}}
       \sqrt{ \tau/4 }.
    \label{AppB-response}\ee
Finally, multiplying Eqs.\ (\ref{AppB-dFqu-int}) and (\ref{AppB-response}) and noticing that this product corresponds to the right hand side of Eq.\ (\ref{AppB-dwq-int2}) multiplied by $2/\tau$, we obtain
    \be
\int_t^{t+\tau}  \delta \omega_{\rm q} (t')\, dt' = \frac{\tilde{I}_{\rm m}\Delta I}{2D},
    \label{AppB-dwq-int3}\ee
which is exactly the result for phase back-action \cite{Korotkov-2011} presented in Sec.\ \ref{sec:ph-sens-bc}, when $\phi_{\rm d}=\pi /2$ -- see Eqs.\ (\ref{ph-sens-off-bc}), (\ref{D-def-bc}), and (\ref{ph-sens-K-bc}). The non-fluctuating part of the ac Stark shift can be simply added.

If we consider an ideal phase-sensitive amplification of an arbitrary quadrature, $\phi_{\rm d}\neq \pi/2$, then the derivation for the fluctuating phase shift $\int_t^{t+\tau}  \delta \omega_{\rm q} (t')\, dt'$ is similar, except the amplified quadrature $\delta F_{\rm qu}(t)$ is no longer along $\alpha_{1,\rm st} -\alpha_{0,\rm st}$, and therefore from Eq.\ (\ref{AppB-dwq-int2}) we obtain an extra factor $\sin (\phi_{\rm d})$, which appears in Eq.\ (\ref{ph-sens-K-bc}) but is absorbed by $\Delta I$ in Eq.\ (\ref{AppB-dwq-int3}). However, it is not obvious if $\tilde I_{\rm m}$ in Eq.\ (\ref{AppB-dFqu-int}) should be counted from  $(I_0+I_1)/2$ or from $\rho_{00}I_0 +\rho_{11} I_1$, and correspondingly if the phase back-action term in Eq.\  (\ref{ph-sens-off-bc}) should be $\exp(-iK \tilde{I}_{\rm m}\tau)$ or $\exp\{-iK [\tilde{I}_{\rm m}- (\rho_{11}-\rho_{00})\Delta I/2 ]\tau \}$.
We can find the answer by requiring that the phase shift due to the phase back-action term in Eq.\ (\ref{ph-sens-off-bc}) is zero on average. Counterintuitively, the phase shift of the averaged $\rho_{10}(t+\tau)$ in Eq.\ (\ref{ph-sens-off-bc}) is zero when the phase back-action term is $\exp(-iK \tilde{I}_{\rm m}\tau)$, even though $\langle \exp(-iK \tilde{I}_{\rm m}\tau)\rangle$ obviously has a non-zero phase if $\rho_{11}\neq \rho_{00}$. This occurs due to a compensating effect from the first term in Eq.\ (\ref{ph-sens-off-bc}), which contains $\rho_{00}$ and $\rho_{11}$: for example, if $\rho_{11} > \rho_{00}$, then a positive $\tilde{I}_{\rm m}$ occurs more often, but produces smaller $|\rho_{10}(t+\tau)|$ than for a negative $\tilde{I}_{\rm m}$. (This somewhat counterintuitive compensation is related to the difference between the It\^o  and Stratonovich approaches.)

Thus, using the approach of the vacuum noise we derived the phase back-action term in Eq.\ (\ref{ph-sens-off-bc}) in the case of ideal phase-sensitive measurement. Let us briefly discuss how in this approach we can take into account non-ideality due to additional resonator damping (e.g., because of coupling to other transmission lines) and the loss of the microwave signal before it reaches amplifier (which is still ideal). Then Eqs.\ (\ref{AppB-alpha-dot}) and (\ref{AppB-F}) can be replaced with
\begin{eqnarray}
    && \dot{\alpha} = -i (\omega_{\rm r} -\omega_{\rm d})\, \alpha -\frac{\kappa}{2}\, \alpha -i\varepsilon +\sqrt{\kappa_{\rm out}} \, v(t)
        \nonumber \\
    && \hspace{0.7cm} + \sqrt{\kappa -\kappa_{\rm out}} \, v_{\rm add,1}(t),  \qquad
    \label{AppB-alpha-dot-2}\\
    && F=\sqrt{\kappa_{\rm col}/\kappa_{\rm out}} \,\, [\sqrt{\kappa_{\rm out}} \, \alpha -v(t)]
        \nonumber \\
    && \hspace{0.7cm}  + \sqrt{1-\kappa_{\rm col}/\kappa_{\rm out}}\, v_{\rm add,2},
\label{AppB-F-2}\end{eqnarray}
where $\sqrt{\kappa -\kappa_{\rm out}} \, v_{\rm add,1}(t)$ is the vacuum noise entering the resonator from other transmission lines, the ratio $\kappa_{\rm col}/\kappa_{\rm out}$ characterizes the energy loss between the resonator and amplifier (which can be modeled via a beam splitter), and because of this loss (at zero temperature) an additional vacuum noise $v_{\rm add,2}$ contributes to the field $F$, which reaches the amplifier. The noises $v$, $v_{\rm add, 1}$, and $v_{\rm add,2}$ are uncorrelated and all satisfy Eq.\ (\ref{AppB-v-corr}); then the noise of $F$ has the same statistics. The calculation becomes more complicated, but it still can be done explicitly. It shows that the correlation (\ref{AppB-dwq-int3}) between the ac Stark shift and the measurement result fluctuations is reduced by the factor $\sqrt{\kappa_{\rm col}/\kappa}$, which is the same factor as for the reduction of $\Delta I$. Therefore, Eq.\ (\ref{AppB-dwq-int3}) and the corresponding Eq.\ (\ref{ph-sens-K-bc}) remain valid. Analysis of imperfection due to a non-ideal amplifier can be performed as in Ref.\ \cite{Korotkov-nonid}; in this case Eqs.\ (\ref{ph-sens-off-bc}) and (\ref{ph-sens-K-bc}) still remain valid.

Note that even though this approach based on vacuum noise gives a natural description of the physical mechanism responsible for the phase back-action, it still cannot explain why in the ideal case with $\phi_{\rm d}=0$ there are no fluctuations of the photon number in the resonator. The fact that in the ideal case only the observed quadrature fluctuates (and the orthogonal quadrature does not fluctuate) is a ``spooky'' property of quantum measurement and cannot have a realistic interpretation.

Derivation of the phase back-action coefficient for the phase-preserving measurement can be done in a similar way. Alternatively, as discussed in Sec.\ \ref{sec:ph-pres-bc}, the results for the phase-preserving case can be obtained from the results for the phase-sensitive measurement.


\begin{thebibliography}{99}

\bibitem{vonNeumann} J. von Neumann, {\it Mathematical Foundations of Quantum Mechanics}
(Princeton University Press, Princeton, NJ, 1955).

\bibitem{EPR} A. Einstein, B. Podolsky, and N. Rosen, Phys. Rev. {\bf 47}, 777 (1935).

\bibitem{Bell-ineq} J. S. Bell, Physics (Long Island City, N.Y.) {\bf 1}, 195 (1964);
   J. F. Clauser, M. A. Horne, A. Shimony, and R. A. Holt, Phys. Rev. Lett. {\bf 23}, 880 (1969).

\bibitem{Aspect} A. Aspect, J. Dalibard, and G. Roger, Phys. Rev. Lett. {\bf 49}, 1804 (1982).

\bibitem{Wheeler-Zurek-book} {\it Quantum Theory of Measurement}, edited by J. A. Wheeler and W. H. Zurek (Princeton University Press, Princeton, NJ, 1983).

\bibitem{Katz-2006} N. Katz, M. Ansmann, R. C. Bialczak, E. Lucero, R. McDermott, M. Neeley, M. Steffen, E. M. Weig, A. N. Cleland, J. M. Martinis, and A. N. Korotkov, Science {\bf 312}, 1498 (2006).

\bibitem{Katz-2008} N. Katz, M. Neeley, M. Ansmann, R. C. Bialczak, M. Hofheinz, E. Lucero, A. O'Connell, H. Wang, A. N. Cleland, J. M. Martinis, and A. N. Korotkov, Phys. Rev. Lett. {\bf 101}, 200401 (2008).

\bibitem{Saclay-2010} A. Palacios-Laloy, F. Mallet, F. Nguyen, P. Bertet, D. Vion, D. Esteve, and A. N. Korotkov, Nature Phys. {\bf 6}, 442 (2010).

\bibitem{Vijay-2012} R. Vijay, C. Macklin, D. H. Slichter, S. J. Weber, K. W. Murch, R. Naik, A. N. Korotkov, and I. Siddiqi, Nature {\bf 490}, 77 (2012).

\bibitem{Hatridge-2013} M. Hatridge, S. Shankar, M. Mirrahimi, F. Schackert, K. Geerlings, T. Brecht, K. M. Sliwa, B. Abdo, L. Frunzio, S. M. Girvin, R. J. Schoelkopf, and M. H. Devoret, Science {\bf 339}, 178 (2013).

\bibitem{Murch-2013} K. W. Murch, S. J. Weber,	C. Macklin, and I. Siddiqi, Nature {\bf 502}, 211 (2013).

\bibitem{deLange-2014} G. de Lange, D. Riste, M. J. Tiggelman, C. Eichler, L. Tornberg, G. Johansson, A. Wallraff, R. N. Schouten, and L. DiCarlo, Phys. Rev. Lett. {\bf 112}, 080501 (2014).

\bibitem{Campagne-2014} P. Campagne-Ibarcq, L. Bretheau, E. Flurin, A. Auffeves, F. Mallet, and B. Huard, Phys. Rev. Lett. {\bf 112}, 180402 (2014).

\bibitem{Weber-2014} S. J. Weber, A. Chantasri, J. Dressel, A. N. Jordan, K. W. Murch, and I. Siddiqi, Nature {\bf 511}, 570 (2014).

\bibitem{Davies-book} E. B. Davies, {\it Quantum theory of open systems} (Academic, London, 1976).

\bibitem{Kraus-book} K. Kraus, {\it States, effects, and Operations: fundamental notions
of quantum theory} (Springer, Berlin, 1983).

\bibitem{Holevo-book} A. S. Holevo, {\it Probabilistic and statistical aspects of quantum theory} (Elsevier, New York, 1982).

\bibitem{Wiseman-1993} H. M. Wiseman and G. J. Milburn, Phys. Rev. A {\bf 47}, 642 (1993).

\bibitem{Carmichael-1993} H. J. Carmichael, {\it An open system approach to quantum optics},
Lecture Notes in Physics (Springer, Berlin, 1993).

\bibitem{Wiseman-book} H. M. Wiseman and G. J. Milburn, {\it Quantum measurement and control} (Cambridge University Press, Cambridge, 2010).

\bibitem{Doherty-1999} A. C. Doherty and K. Jacobs, Phys. Rev. A {\bf 60}, 2700 (1999). 

\bibitem{Gambetta-2008} J. Gambetta, A. Blais, M. Boissonneault,
A. A. Houck, D. I. Schuster, and S. M. Girvin, Phys. Rev. A {\bf
77}, 012112 (2008).

\bibitem{Belavkin-1992} V. P. Belavkin, J. Miltivariate Anal. {\bf 42}, 171 (1992).

\bibitem{Dalibard-1992} J. Dalibard, Y. Castin, and K. Molmer, Phys. Rev. Lett. {\bf 68}, 580 (1992).
\bibitem{Gisin-1992} N. Gisin and I. C. Percival, J. Phys. A {\bf 25}, 5677 (1992).

\bibitem{Mensky-book} M. B. Mensky, {\it Continuous quantum measurements and path integrals} (IOP  Publishing, Bristol, 1993).

\bibitem{Caves-1986} C. M. Caves, Phys. Rev. D {\bf 33}, 1643 (1986).

\bibitem{Korotkov-1999} A. N. Korotkov, Phys. Rev. B {\bf 60}, 5737 (1999).

\bibitem{Korotkov-2001} A. N. Korotkov, Phys. Rev. B {\bf 63}, 115403 (2001).

\bibitem{Gardiner-book} C. W. Gardiner, {\it Quantum Noise} (Springer, Heidelberg, 1991).

\bibitem{Braginsky-book} V. B. Braginsky and F. Ya. Khalili, {\it Quantum Measurement} (Cambridge
University Press, Cambridge, 1992).

\bibitem{Diosi-1988} L. Diosi, Phys. Lett. A 129, 419 (1988).

\bibitem{Zoller-1987} P. Zoller, M. Marte, and D. F. Walls, Phys. Rev. A {\bf 35}, 198 (1987).

\bibitem{Plenio-1998} M. B. Plenio and P. L. Knight, Rev. Mod. Phys. {\bf 70}, 101 (1998).

\bibitem{Korotkov-2002} A. N. Korotkov, arXiv:cond-mat/0209629,  in {\it Quantum Noise in Mesoscopic Physics}, edited by Yu. V. Nazarov (Kluwer, Netherlands, 2003), p. 205.

\bibitem{Goan-2001}  H. S. Goan, G. J. Milburn, H. M. Wiseman, and H. B. Sun, Phys. Rev. B {\bf 63}, 125326 (2001).

\bibitem{Goan-2001a} H. S. Goan and G. J. Milburn, Phys. Rev. B {\bf 64}, 235307 (2001).

\bibitem{Ruskov-2002} R. Ruskov and A. N. Korotkov, Phys. Rev. B {\bf 66}, 041401(R) (2002); R. Ruskov and A. N. Korotkov, Phys. Rev. B {\bf 67}, 241305(R) (2003).

\bibitem{Korotkov-Averin} A. N. Korotkov and D. V. Averin, Phys. Rev. B {\bf 64}, 165310 (2001).

\bibitem{Averin-2003}  D. V. Averin,  in {\it Quantum Noise in Mesoscopic Physics}, edited by Yu. V. Nazarov (Kluwer, Netherlands, 2003), p. 229; arXiv:cond-mat/0301524.

\bibitem{Jordan-2005} A. N. Jordan and M. B\"uttiker, Phys. Rev. Lett. {\bf 95}, 220401 (2005).

\bibitem{Oxtoby-2005} N. P. Oxtoby, P.  Warszawski, H. M. Wiseman, H. B. Sun, and R. E. S. Polkinghorne, Phys. Rev. B {\bf 71}, 165317 (2005).


\bibitem{Korotkov-Jordan-2006} A. N. Korotkov and A. N. Jordan, Phys. Rev. Lett. {\bf 97}, 166805 (2006).

\bibitem{Pryadko-2007} L. P. Pryadko and A. N. Korotkov, Phys. Rev. B {\bf 76}, 100503(R) (2007).

\bibitem{Ruskov-2007} R. Ruskov, A. Mizel, and A. N. Korotkov, Phys. Rev. B {\bf 75}, 220501(R) (2007).

\bibitem{Zhong-2014} Y. P. Zhong, Z. L. Wang, J. M. Martinis, A. N. Cleland, A. N. Korotkov, and H. Wang, Nature Comm. {\bf 5}, 3135 (2014).


\bibitem{Blais-2004} A. Blais, R.-S. Huang, A. Wallraff, S. M. Girvin, and
R. J. Schoelkopf, Phys. Rev. A {\bf 69}, 062320 (2004).

\bibitem{Wallraff-2004} A. Wallraff, D. I. Schuster, A. Blais, L. Frunzio, P. S. Huang, J. Majer, S. Kumar, S. M. Girvin, and R. J. Schoelkopf, Nature {\bf 431}, 162 (2004).

\bibitem{Koch-2007}  J. Koch, T. M. Yu, J. Gambetta, A. A. Houck, D. I. Schuster, J. Majer, A. Blais, M. H. Devoret, S. M. Girvin, and R. J. Schoelkopf, Phys. Rev. A {\bf 76}, 042319 (2007).


\bibitem{Korotkov-2011} A. N. Korotkov, arXiv:1111.4016; in {\it Quantum machines}, Lecture notes of July 2011 Les Houches summer school, edited by M. Devoret et al. (Oxford University Press, New York, 2014).

\bibitem{Riste-2013}  D. Riste, M. Dukalski, C. A. Watson, G. de Lange, M. J. Tiggelman, Y. M. Blanter, K. W. Lehnert, R. N. Schouten, and L. DiCarlo, Nature {\bf 502}, 350 (2013).

\bibitem{Roch-2014} N. Roch, M. E. Schwartz, F. Motzoi, C. Macklin, R. Vijay, A. W. Eddins, A. N. Korotkov, K. B. Whaley, M. Sarovar, and I. Siddiqi, Phys. Rev. Lett. {\bf 112}, 170501 (2014).

\bibitem{Caves-1982} C. M. Caves, Phys. Rev. D {\bf 26}, 1817 (1982).

\bibitem{Devyatov-1986} I. A. Devyatov, L. S. Kuzmin, K. K. Likharev, V. V. Migulin, and A. B. Zorin, J. Appl. Phys. {\bf 60}, 1808 (1986).

\bibitem{Clerk-2010}  A. A. Clerk, M. H. Devoret, S. M. Girvin, F. Marquardt, and R. J. Schoelkopf, Rev. Mod. Phys. {\bf 82}, 1155 (2010).

\bibitem{Wang-2014} P. Y. Wang, L. P. Qin, and X.-Q. Li, New J. Phys. {\bf 16}, 123047 (2014).

\bibitem{Feng-2016} W. Feng, P. F. Liang, L. P. Qin, and X.-Q. Li, Sci. Rep. {\bf 6}, 20492 (2016).


\bibitem{Bergeal-2010}  A. Bergeal, F. Schackert, M. Metcalfe, R. Vijay, V. E. Manucharyan, L. Frunzio, D. E. Prober, R. J. Schoelkopf, S. M. Girvin, and M. H. Devoret, Nature {\bf 465}, 64 (2010).

\bibitem{Castellanos-2008}  M. A. Castellanos-Beltran, K. D. Irwin, G. C. Hilton, L. R. Vale, and K. W. Lehnert, Nature Phys. {\bf 4}, 929 (2008).

\bibitem{Vijay-2011}  R. Vijay, D. H. Slichter, and I. Siddiqi, Phys. Rev. Lett. {\bf 106}, 110502 (2011).

\bibitem{Mutus-2013} J. Y. Mutus, T. C. White, E. Jeffrey, D. Sank, R. Barends, J. Bochmann, Y. Chen, Z. Chen, B. Chiaro, A. Dunsworth, J. Kelly, A. Megrant, C. Neill, P. J. J. O'Malley, P. Roushan, A. Vainsencher, J. Wenner, I. Siddiqi, R. Vijay, A. N. Cleland, and J. M. Martinis, Appl. Phys. Lett. {\bf 103}, 122602 (2013).

\bibitem{Sete-2015} E. A. Sete, E. Mlinar, and A. N. Korotkov, Phys. Rev. B {\bf 91}, 144509 (2015).

\bibitem{Khezri-2016} M. Khezri, E. Mlinar, J. Dressel, and A. N. Korotkov, arXiv:1606.04204.

\bibitem{Esteve-1986}  D. Esteve, M. H. Devoret, and J. M. Martinis, Phys. Rev. B {\bf 34}, 158 (1986).

\bibitem{Houck-2008} A. A. Houck, J. A. Schreier, B. R. Johnson, J. M. Chow, J. Koch, J. M. Gambetta, D. I. Schuster, L. Frunzio, M. H. Devoret, S. M. Girvin, and R. J. Schoelkopf, Phys. Rev. Lett. {\bf 101}, 080502 (2008).

\bibitem{Reed-2010} M. D. Reed, B. R. Johnson, A. A. Houck, L. DiCarlo, J. M. Chow, D. I. Schuster, L. Frunzio, and R. J. Schoelkopf, Appl. Phys. Lett. {\bf 96}, 203110 (2010).

\bibitem{Jeffrey-2014} E. Jeffrey, D. Sank, J. Y. Mutus, T. C.White, J. Kelly, R. Barends, Y. Chen, Z. Chen, B. Chiaro, A. Dunsworth, A. Megrant, P. J. J. O'Malley, C. Neill, P. Roushan, A. Vainsencher, J. Wenner, A. N. Cleland, and J. M. Martinis, Phys. Rev. Lett. {\bf 112}, 190504 (2014).

\bibitem{Walls-Milburn-book} D. F. Walls and G. J. Milburn, {\it Quantum Optics} (Springer, Berlin, 2008).

\bibitem{Gambetta-2006} J. Gambetta, A. Blais, D. I. Schuster, A. Wallraff, L. Frunzio, J. Majer, M. H. Devoret, S. M. Girvin, and R. J. Schoelkopf, Phys. Rev. A {\bf 74}, 042318 (2006).

\bibitem{Korotkov-nonid} A. N. Korotkov, Phys. Rev. B {\bf 67}, 235408 (2003).

\bibitem{Oksendal} B. \O ksendal, {\it Stochastic differential equations} (Springer, Berlin, 1998).

\bibitem{Gough-2016} J. Gough, arXiv:1605.02621.

\bibitem{McClure-2016} D. T. McClure, H. Paik, L. S. Bishop, M. Steffen, J. M. Chow, and J. M. Gambetta, Phys. Rev Applied {\bf 5}, 011001 (2016).

\bibitem{Yurke-1984} B. Yurke and J. S. Denker, Phys. Rev. A {\bf 29}, 1419 (1984).


\bibitem{Gardiner-1985} C. W. Gardiner and M. J. Collett, Phys. Rev. A {\bf 31}, 3761 (1985).


\bibitem{Gerry-book} C. Gerry and P. Knight, {\it Introductory Quantum Optics} (Cambridge University Press, Cambridge, 2006).


\bibitem{Motzoi-2015} F. Motzoi, K. B. Whaley, and M. Sarovar, Phys. Rev. A {\bf 92}, 032308 (2015).

\bibitem{Chantasri-2016} A. Chantasri, M. E. Kimchi-Schwartz, N. Roch, I. Siddiqi, and A. N. Jordan, arXiv:1603.09623.

\bibitem{Korotkov-2013} A. N. Korotkov, arXiv:1309.6405, Appendix B.

\bibitem{Collett-1984} M. J. Collett and C. W. Gardiner, Phys. Rev. A {\bf 30}, 1386 (1984).

\end{thebibliography}
\end{document}